\documentclass[a4paper,11pt]{article}
\pdfoutput=1
\usepackage{jheppub}
\usepackage{amsmath,amssymb,mathtools,shuffle,comment,amsfonts}
\usepackage{empheq}
\usepackage{graphicx}
\usepackage{xcolor}
\usepackage{booktabs}
\usepackage{array}
\usepackage{bm}
\usepackage{slashed}
\usepackage{tikz,diagbox}
\usetikzlibrary{calc}
\usepackage{setspace}
\usepackage{mathrsfs}
\usepackage{enumitem}
\usetikzlibrary{decorations.markings, arrows.meta, calc, decorations.pathmorphing,angles, quotes, arrows.meta}

\usepackage{hyperref}
\hypersetup{
  linktocpage,
  colorlinks  = true,
  urlcolor    = blue,
  linkcolor   = blue,
  citecolor   = blue
}

\newcommand{\AdS}{\mathrm{AdS}}

\newcommand{\dd}{\mathrm{d}}
\newcommand{\Li}{\operatorname{Li}}
\newcommand{\Dbar}{\bar D}

\newcommand{\M}{\mathcal M}
\newcommand{\G}{\mathcal G}

\title{An infrared-safe \texorpdfstring{$S$}{S}-matrix for \texorpdfstring{$\mathcal N=4$}{N=4} super Yang--Mills}

\author[a]{Maria Nocchi}
\author[b]{and Leonardo Rastelli}

\affiliation[a]{Simons Center for Geometry and Physics,
Stony Brook University, Stony Brook, NY 11794, USA}

\affiliation[b]{{\spaceskip=0.92\fontdimen2\font plus \fontdimen3\font minus \fontdimen4\font
C. N. Yang Institute for Theoretical Physics,
Stony Brook University, Stony Brook, NY~11794,~USA}}

\emailAdd{maria.nocchi@stonybrook.edu}
\emailAdd{leonardo.rastelli@stonybrook.edu}

\abstract{
Scattering amplitudes of massless gauge theories are not observables:
they are infrared divergent, and the standard regulators are tied
either to perturbation theory or the planar limit. We study
$\mathcal N=4$ super Yang--Mills theory on rigid AdS$_4$ with Dirichlet
boundary conditions, where the role of the amplitude is played by
correlation functions of boundary operators. These are finite, and
they are defined at any coupling and for any gauge group. Among them,
the four-point function of the flavor-current supermultiplet is the counterpart of the
$2\to2$ amplitude of the elementary supergluon states, and arguably the simplest
non-trivial observable of the theory. In this paper, we compute it at
weak coupling to one loop. The tree-level correlator follows directly from the
microscopic theory, and the one-loop correlator, which we obtain from
AdS unitarity, takes a remarkably simple form in Mellin space. In the
flat-space limit, it reproduces the one-loop scattering amplitude, with
the infrared divergences re-emerging as logarithms.
}

\begin{document}

\maketitle

%%%%%%%%%%%%%%%%%%%%%%%%%%%%%%%%%%%%%%%%%%%%%%%%%%%%%%%%%%%%%%%%
\section{Motivation and summary}
\label{sec:introduction}
%%%%%%%%%%%%%%%%%%%%%%%%%%%%%%%%%%%%%%%%%%%%%%%%%%%%%%%%%%%%%%%%

What is the simplest non-trivial observable of four-dimensional
$\mathcal N=4$ super Yang--Mills (SYM) theory at finite coupling and
finite $N$?
By non-trivial we mean sensitive to the unprotected dynamics:
quantities that are protected, or computable by supersymmetric
localization, are not what we are after.
Correlation functions
of local operators, such as the four-point function of the
stress-tensor multiplet, are the obvious candidates
\cite{Beem:2013qxa,Binder:2019jwn,Chester:2021aun,Chester:2023ehi},
but they are intricate objects: already at zero coupling, the
intermediate spectrum is highly degenerate. The $2\to2$ scattering
amplitude of the elementary massless states is intrinsically simpler,
but it is not an observable.\footnote{Infrared-safe observables can
also be constructed inclusively, by summing over degenerate states:
inclusive cross-sections \cite{Bork:2009ps,Bork:2009nc}, and event
shapes such as energy correlators, which are defined at any coupling
through correlation functions of local operators
\cite{Hofman:2008ar,Belitsky:2013xxa,Belitsky:2013bja}. Here we look
instead for an exclusive observable, the counterpart of the $2\to2$
amplitude itself.} Soft and collinear divergences, organized
by gauge invariance and factorization into universal structures
\cite{Catani:1998bh,Becher:2009qa,Gardi:2009qi,Bern:2005iz}, must be regulated
before the amplitude can be defined at all.

The standard regulators do not turn the amplitude into a
non-perturbative object. Dimensional regularization
\cite{tHooft:1972tcz} is defined order by order in perturbation
theory. Moving onto the Coulomb branch gives masses to some of the
states and regulates planar, color-ordered amplitudes
\cite{Alday:2009zm,Henn:2010bk,Caron-Huot:2021usw}, but at finite $N$ the massless fields of the
unbroken gauge group still run in loops and the divergences return.

In this paper we study a different possibility: placing the theory on
a rigid $\AdS_4$ background \cite{Aharony:2010ay}. The idea that negative curvature tames
the infrared goes back to Callan and Wilczek \cite{Callan:1989em}, and
it is common lore in the study of the flat-space limit
\cite{Fitzpatrick:2011jn,Hijano:2020szl,Duary:2022pyv}
that the AdS radius acts as an infrared regulator. Once boundary
conditions are specified, the natural observables are correlation
functions of operators at the conformal boundary \cite{Paulos:2016fap}. They are finite,
they are defined for any coupling and any gauge group without
reference to a planar limit, and their large-radius limit reproduces
flat-space scattering \cite{Penedones:2010ue,Fitzpatrick:2011hu,Li:2021snj}.
Since $\mathcal N=4$ SYM is conformal, there is a sharper way to say
this. The theory on $\AdS_4$ is Weyl-equivalent to the theory on the
flat half-space, and its boundary correlators do not depend on the
radius $L$. The radius enters only when the dimensionless boundary
data are converted into flat-space momenta, so that the scattering
amplitude is recovered from the regime of large Mellin variables.
Nothing is cut off: the boundary correlator is itself the infrared-safe
observable, and the divergences of the amplitude reappear as
logarithms in that regime.

The price is a choice of boundary conditions, but there is a canonical
one. Among the half-BPS boundary conditions classified in
\cite{Gaiotto:2008sa,Gaiotto:2008ak}, the Dirichlet condition
introduces no boundary degrees of freedom and preserves the whole
gauge group $G$ as a global symmetry of the boundary theory, a
general feature of gauge theories on $\AdS_4$ with Dirichlet
conditions \cite{Aharony:2012jf}. The
associated conserved current sits in a short multiplet of the
preserved $3d$ $\mathcal N=4$ superconformal algebra, whose
superconformal primary is the moment map $\mu$, a scalar of dimension
one in the adjoint of $G$. Its four-point function is the boundary
counterpart of $2\to2$ scattering of the elementary colored states.
We propose it as the simplest non-trivial non-perturbative observable
of $\mathcal N=4$ SYM. At zero coupling, it is a generalized free
field, with a single single-particle multiplet and non-degenerate
two-particle states, in contrast with the free limit of local
correlators. At finite coupling, it defines consistent boundary
conformal data \cite{Liendo:2012hy,Liendo:2016ymz}, although not a
local three-dimensional Conformal Field Theory (CFT), since there is no boundary stress tensor (see \cite{DiPietro:2019hqe,Behan:2020nsf} for related examples).

This makes the correlator a natural target for the conformal
bootstrap. A numerical study at finite coupling, leveraging
localization results for the protected conformal data, will be
reported elsewhere \cite{BoundaryProtectedOPE:inprep}.
Boundary correlators of Yang--Mills theory and related models in AdS
have been studied in connection with confinement
\cite{Aharony:2012jf,Copetti:2023sya,Ciccone:2024guw}, including with
the numerical bootstrap \cite{DiPietro:2025ozw}.
Here we focus on the weak-coupling expansion of the correlator,
which we carry to one loop. There are two kinds of lessons. The first is conceptual: we
show explicitly how the infrared structure of the flat-space
amplitude emerges from the boundary correlator. The divergent terms
have the universal color structure of one-loop infrared divergences,
and a natural subtraction prescription can be compared with
dimensional regularization. The second is
technical: we adapt the AdS unitarity method to a three-dimensional
setting in which the reduced correlator is non-local in position
space, and find that the one-loop answer is nonetheless remarkably
simple in Mellin space.

\subsection{Main results}

We begin by deriving the tree-level moment-map
four-point function directly from the microscopic
$\mathcal N=4$ SYM theory with Dirichlet boundary conditions.
Three ingredients contribute: bulk gauge-field exchange,
the quartic scalar interaction, and a scalar exchange
induced by the Nahm supersymmetric boundary condition.
We show that these contributions combine exactly into
the $d=3,\,p=2$ tree-level supergluon Mellin amplitude
of \cite{Alday:2021odx}, which was fixed by superconformal
symmetry, crossing, factorization, and the exchanged
spectrum. Schematically,
\begin{equation}
    \mathcal M^{\rm tree}_{\mathcal N=4\,{\rm SYM},
    \,{\rm AdS}_4}
    =
    \underbrace{\mathcal M_{\rm gauge}}_{\text{vector exchange}}
    +
    \underbrace{
    \mathcal M_{\rm quartic}+\mathcal M_{\rm Nahm}
    }_{\text{scalar sector}}
    =
    \mathcal M^{\rm tree}_{(p=2)\,\rm supergluon}\, .
\end{equation}
The Nahm term is the one that is easy to miss: it comes
from the non-linear scalar boundary condition rather than from a
bulk vertex, and it is subleading in the flat-space limit.
It is nevertheless required at finite AdS radius by
supersymmetry.

The passage to one loop is where the spectrum of the
microscopic theory becomes essential. A direct evaluation
of loop Witten diagrams is generally cumbersome.
Instead, we use AdS unitarity
\cite{Aharony:2016dwx,Alday:2017xua,Aprile:2017bgs,Alday:2017vkk,Meltzer:2019nbs}, which reconstructs
the non-local loop amplitude from lower-order conformal
data. In our theory, the two-particle cut is obtained
by gluing two copies of the tree-level moment-map
amplitude. The closest precedent is the $\AdS_5$ toy
model of \cite{Alday:2021ajh}, obtained by truncating
the $\AdS_5\times S^3$ supergluon theory to its lowest
$p=2$ Kaluza--Klein (KK) sector. In that model, the
one-loop cut is likewise obtained from the $p=2$
sector alone, whereas the full $\AdS_5\times S^3$
theory requires the tree-level $\langle22pp\rangle$
amplitudes for all allowed $p$. By contrast, our
setup naturally isolates the loop dynamics of a
supersymmetric gauge theory in AdS without additional
compactification or gravitational sectors.

A remarkable finding is that the one-loop
double-discontinuity, obtained directly from the
\textit{full} tree-level correlator, admits a simple
\textit{reduced} Mellin representation involving only
simultaneous poles. This mirrors structures found
in higher-dimensional holographic one-loop correlators
\cite{Alday:2019nin,Alday:2021ajh,Huang:2023ppy}.
The answer can be organized in terms of
a single simultaneous-pole kernel,
\begin{equation}
    B(s,t)
    =
    \sum_{m,n\geq1}
    \frac{c_{mn}}{(s-2m)(t-2n)},
\end{equation}
whose coefficients admit the closed form
\begin{equation}
    c_{mn}
    =
    \frac{1}{9\pi^4}
    \left[
    \frac{
    \sqrt{\pi}\,
    \Gamma(m)\Gamma(n)
    \Gamma\!\left(m+n+\frac12\right)}
    {
    \Gamma\!\left(m+\frac12\right)
    \Gamma\!\left(n+\frac12\right)
    \Gamma(m+n+1)}
    -
    \frac{1}{mn}
    \right],
\end{equation}
which is manifestly symmetric in $m$ and $n$.
The complete reduced amplitude takes the familiar color-box form:
it is the sum over the three channels of the kernel $B$, each dressed
by a one-loop color-box tensor, see
\eqref{eq:complete-reduced-one-loop-amplitude}. The full R-symmetry
dependence is restored by the finite-difference operators that
implement the $3d$ $\mathcal N=4$ superconformal Ward identities in
Mellin space \cite{Chester:2025jxg}, see \eqref{eq:full-ward-lift}. We also resum the elementary Ward-lifted block exactly:
\eqref{eq:flat-exact-W} gives its highest R-symmetry component,
while the identities \eqref{eq:W0W1-identity} determine
the remaining two.

We conclude by studying the flat-space limit of the
reconstructed one-loop correlator. Since the boundary
correlator does not depend on the AdS radius $L$, this is
the regime of large Mellin variables: $L$ enters only
through the conversion of Mellin variables into
flat-space momenta. The discrete tower of AdS Mellin
poles becomes dense and, upon resummation, gives rise to
the logarithmic branch cuts of the four-dimensional
massless box. The infrared divergences of the flat-space
amplitude reappear as logarithms of $L$, which diverge at
fixed flat-space kinematics as $L\to\infty$. In this sense,
the AdS radius acts as an infrared regulator for a
four-dimensional massless gauge theory, for any gauge
group and without reference to the planar limit. The
divergent terms have the universal color structure of
one-loop infrared divergences: the double logarithm is
proportional to the tree-level amplitude, and the single
logarithms take the color-dipole form. A natural AdS
minimal-subtraction scheme is then defined, and the
resulting finite remainder is compared with the
dimensionally regulated flat-space box
\cite{Bern:1992em,Duplancic:2000sk} and with the Higgs
regulator \cite{Alday:2009zm}.

%%%%%%%%%%%%%%%%%%%%%%%%%%%%%%%%%%%%%%%%%%%%%%%%%%%%%%%%%%%%%%%%
\section{Physical setup and conventions}
\label{sec:setup}
%%%%%%%%%%%%%%%%%%%%%%%%%%%%%%%%%%%%%%%%%%%%%%%%%%%%%%%%%%%%%%%%
We start from four-dimensional $\mathcal N=4$ super Yang--Mills theory
with gauge group $G$ in flat space. Since the theory is conformal,
placing it on $\AdS_4$ is equivalent, up to a Weyl transformation, to
placing it on the flat half-space. In Poincar\'e coordinates,
\begin{equation} \label{ads4}
    ds^2_{\AdS_4}
    =\frac{L^2}{z^2}
    \left(
        dz^2 + dx_i dx^i
    \right),
\end{equation}
so that $\AdS_4$ is conformal to the region $z>0$ of flat space.
Throughout, the $\AdS_4$ metric is a fixed background and gravity is
non-dynamical.

A quantum field theory on AdS is not completely specified
until boundary conditions are chosen at the conformal boundary.
In Lorentzian signature, this boundary is timelike, so AdS
is not globally hyperbolic and initial data alone do not
determine the evolution.
A well-posed dynamical problem therefore requires an
admissible choice of boundary conditions.\footnote{
In Euclidean signature, which we use below, these specify
the admissible asymptotic behavior of the fields.
} Since Poincar\'e $\AdS_4$ is conformally
equivalent to the flat half-space, conformally invariant boundary
conditions for $\mathcal N=4$ SYM on $\AdS_4$ can equivalently be
described as conformal boundary conditions for the same theory on
$\mathbb R^4_+$. 

The four-dimensional $\mathcal N=4$ vector multiplet contains the gauge field, fermions, and six real adjoint scalars $\Phi^M$, $M=1,\ldots,6$, transforming in the ${\bf 6}$ of the $SO(6)_R$ R-symmetry. Boundary conditions preserving the maximal supersymmetry compatible with the presence of a boundary were classified in \cite{Gaiotto:2008sa,Gaiotto:2008ak}. A half-BPS boundary admits the three-dimensional superconformal algebra $OSp(4|4)$, with sixteen fermionic generators, corresponding to eight Poincar\'e and eight conformal supercharges. Its bosonic subalgebra is
\begin{equation}
    \underbrace{SO(3,2)}_{
        \substack{\text{$3d$ conformal group}\\ \text{= \,$\AdS_4$ isometries}}
    }
    \times
    \underbrace{SO(4)_R}_{\text{R-symmetry}}\, .
\end{equation}
Thus the boundary condition selects the R-symmetry subgroup
\begin{equation}
    SO(6)_R
    \longrightarrow
    SO(4)_R
    \simeq
    SU(2)_H\times SU(2)_C .
\end{equation}
Accordingly, the ${\bf 6}$ of $SO(6)_R$ branches as
\begin{equation}
    {\bf 6}
    \longrightarrow
    ({\bf 3},{\bf 1})\oplus({\bf 1},{\bf 3}),
\end{equation}
and the six scalars naturally split into two triplets,
\begin{equation}
    X^a\in({\bf 3},{\bf 1}),
    \qquad
    Y^{\dot a}\in({\bf 1},{\bf 3}),
    \qquad
    a,\dot a=1,2,3 .
\end{equation}
Our convention is that the first and second factors refer to $SU(2)_H$ and $SU(2)_C$, respectively. The labels $H$ and $C$ follow the standard three-dimensional $\mathcal N=4$ terminology, where the two R-symmetry factors are associated with the Higgs and Coulomb branches. Equivalently, the triplets may be represented as symmetric rank-two tensors,
\begin{equation}
    X^{AB}=X^{BA},
    \qquad
    Y^{\dot A\dot B}=Y^{\dot B\dot A},
    \qquad
    A,\dot A=1,2 ,
\end{equation}
transforming under $SU(2)_H$ and $SU(2)_C$, respectively.

\paragraph{Dirichlet boundary conditions and the flavor-current multiplet.}
Among the half-BPS boundary conditions described above, we choose the simplest Dirichlet/D-type boundary condition, with no additional boundary degrees of freedom. This means that the local operators in the $3d$ boundary operator algebra arise from the allowed boundary data of the bulk fields, as explained below.

For a bulk gauge theory, the D-type choice also specifies how the bulk gauge symmetry is realized at the boundary. Gauge transformations that vanish at the boundary remain gauge redundancies, whereas transformations approaching a constant non-trivial value at \(z=0\) act as the global \(G\) flavor symmetry on boundary operators. In the flat-half-space frame, the D-type boundary conditions for the bulk fields take the form\footnote{The fermionic boundary conditions complete
\eqref{eq:D-type-bc} into a supersymmetric boundary condition.}
\begin{equation}
\begin{split}
    & A_i^I\big|_{z=0}=a_i^I(x), \\[4pt]
    & Y^{\dot a}\big|_{z=0}=0, \qquad \left(
    D_zX^a-\frac{i}{2}\epsilon^{abc}[X^b,X^c]
    \right)\Big|_{z=0}=0 .
    \label{eq:D-type-bc}
\end{split}
\end{equation}
Here, \(a_i^I(x)\) is a non-dynamical background source for the conserved current $J_i^I$ associated with the boundary flavor symmetry. More specifically, in radial gauge, \(A_z=0\), the bulk equations of motion give the near-boundary expansion\footnote{The four-dimensional gauge connection has vanishing Weyl weight,
so \(A_\mu\) is unchanged under the Weyl transformation between the
flat-half-space and AdS frames. The same near-boundary expansion may
therefore be used in either frame.}
\begin{equation}
    A_i^I(z,x)
    =
    a_i^I(x)+z\,b_i^I(x)+\cdots .
\end{equation}
Since \(F_{zi}^I=\partial_z A_i^I\), the coefficient \(b_i^I(x)\) is precisely the boundary limit of the radial field strength. Up to normalization, this mode represents microscopically the corresponding
dimension-two boundary current,
\begin{equation}
    J_i^I\sim F_{zi}^I\big|_{z=0}. \label{JfromF}
\end{equation}
The two bulk scalar triplets instead obey Dirichlet boundary conditions for $Y^{\dot{a}}$, while $X^a$ satisfies a supersymmetric generalized Neumann/Nahm condition \cite{Gaiotto:2008sa,Dedushenko:2020vgd}. We choose the trivial Nahm background, as a pole $X^a\sim 1/z$ generally breaks the boundary flavor symmetry \(G\) to a subgroup.  

Microscopically, the boundary value of $X$ defines a scalar operator
\begin{equation}
    \mu^{aI}\sim X^{aI}\big|_{z=0},
    \label{mufromX}
\end{equation}
while the Dirichlet triplet $Y$ is fixed to vanish at $z=0$ and therefore does not define an independent boundary operator. The first non-vanishing boundary mode is instead its normal derivative,
\begin{equation}
    \mathcal O_Y^{\dot a I}
    \sim
    D_zY^{\dot a I}\big|_{z=0}. \label{OY}
\end{equation}
Supersymmetry organizes these boundary modes into the $3d$ $\mathcal{N}=4$ flavor-current multiplet \(\mathcal B[0]_1^{(2;0)}\), in the notation of \cite{Cordova:2016emh}.\footnote{This is also commonly denoted $(B,+)_1$ \cite{Liendo:2016ymz}.} Here, the superscripts denote the Dynkin labels of $SU(2)_H \times SU(2)_C$, and the superconformal primary of the BPS multiplet is a scalar of dimension one. This is the moment-map operator, 
\begin{equation}
    \mu_{AB}^I=\mu_{BA}^I,
    \qquad
    \Delta_\mu=1,
    \qquad
    \mu^I\in(\mathbf3,\mathbf1),
\end{equation}
generated microscopically by \eqref{mufromX}. We package its \(SU(2)_H\) indices with a polarization
spinor \(v^A\),
\begin{equation}
    \mathcal O_2^I(x;v)
    \equiv
    \mu_{AB}^I(x)v^A v^B,
    \qquad
    \Delta_{\mathcal O_2}=1,
    \label{eq:moment-map-def}
\end{equation}
where the subscript refers to its \(SU(2)_H\) Dynkin label (equivalently,
its degree in the polarization), rather than its conformal dimension. Explicitly,
$\mu_{AB}^Iv^Av^B=y^a(v)\,\mu^{aI}$, where
$y^a(v)\equiv\frac{i}{\sqrt2}\,v^A(\sigma^a\epsilon)_{AB}v^B$ is a null vector built from the Pauli
matrices $\sigma^a$ and the antisymmetric tensor $\epsilon_{AB}$, normalized so that
$y(v_1)\cdot y(v_2)=(\epsilon_{AB}v_1^Av_2^B)^2$. Then, representation theory identifies the
dimension-two \((\mathbf1,\mathbf3)\) scalar in \eqref{OY} as a superconformal
descendant of the moment map, appearing at the same top level of the multiplet as the dimension-two \((\mathbf1,\mathbf1)\)  conserved current \eqref{JfromF}. The complete multiplet, including its fermionic component $\Psi^I$, is
\begin{equation}
\begin{tikzpicture}[baseline=(current bounding box.center),
  every node/.style={inner sep=3pt},>=stealth]
 \node (mu)  at (0,0)       {$\mu^I$\,:\quad $\Delta=1$,\ \ $j=0$,\ \ $(\mathbf3,\mathbf1)$};
 \node (psi) at (0,-1.5)    {$\Psi^I$\,:\quad $\Delta=\tfrac32$,\ \ $j=\tfrac12$,\ \ $(\mathbf2,\mathbf2)$};
 \node (J)   at (-3.4,-3.0) {$J_i^I$\,:\quad $\Delta=2$,\ \ $j=1$,\ \ $(\mathbf1,\mathbf1)$};
 \node (O)   at (3.4,-3.0)  {$\mathcal O_Y^I$\,:\quad $\Delta=2$,\ \ $j=0$,\ \ $(\mathbf1,\mathbf3)$};
 \draw[->] (mu) -- node[right] {\small $Q$} (psi);
 \draw[->] (psi) -- node[above left=-2pt] {\small $Q$} (J);
 \draw[->] (psi) -- node[above right=-2pt] {\small $Q$} (O);
\end{tikzpicture}
\label{eq:current-multiplet-diagram}
\end{equation}
where each arrow denotes the action of a Poincar\'e supercharge $Q$, and for each component we list its conformal dimension $\Delta$, its spin $j$ and its representation under $SU(2)_H\times SU(2)_C$.

In the equivalent \(\AdS_4\) description, the bulk scalars $X,Y$ are conformally coupled to the curvature, with 
\begin{equation}
    m^2L^2=\Delta(\Delta-3)=-2,
    \qquad
    \Delta_-=1,
    \qquad
    \Delta_+=2 .
\end{equation}
Therefore, under the Weyl map, the Nahm field \(X^a\) gives the
\(\Delta=1\) quantization, while the Dirichlet field \(Y^{\dot a}\)
gives the \(\Delta=2\) quantization.  Equivalently, 
\begin{equation}
\begin{split}
    & X_{\rm flat}^a
    =
    X_{(0)}^a+\cdots
    \quad\Longrightarrow\quad
    X_{\AdS}^a
    \sim z\,X_{(0)}^a+\cdots , \\
    & Y_{\rm flat}^{\dot a}
    =
    z\,Y_{(1)}^{\dot a}+\cdots
    \quad\Longrightarrow\quad
    Y_{\AdS}^{\dot a}
    \sim z^2Y_{(1)}^{\dot a}+\cdots .
\end{split}
\end{equation}
The correlators of the boundary theory are functions of the
complexified coupling 
\begin{equation}
    \tau=\frac{\theta}{2\pi}+\frac{4\pi i}{g^2},
\end{equation}
where $g$ is the conventionally normalized gauge coupling. They are
not invariant under S-duality, which maps the Dirichlet condition to
a different boundary condition, in which the dual gauge fields are
coupled to the three-dimensional theory $T[G]$ \cite{Gaiotto:2008ak}.
The strong-coupling limit of our observable is therefore governed by
the weakly coupled dual bulk theory interacting with $T[G]$.

\paragraph{The displacement multiplet.} Every genuine boundary CFT has a displacement operator because translations
normal to the boundary are broken \cite{McAvity:1993ue,Billo:2016cpy}. In the flat half-space
description, this breaking is encoded in the Ward identity
\begin{equation}
    \partial_\mu T^{\mu z}(x,z)
    =
    \delta(z)\,\mathsf D(x),
    \label{eq:displacement-Ward}
\end{equation}
where $\mathsf D$ is the displacement operator, a scalar of dimension four. For a half-BPS boundary preserving $OSp(4|4)$, the
displacement is a descendant of the short multiplet
$\mathcal B[0]_2^{(4;0)}$
\cite{Cordova:2016emh,Liendo:2016ymz}.
In our conventions, in which the moment map is charged under
$SU(2)_H$, its superconformal
primary is a scalar of dimension two, transforming
with spin two under $SU(2)_H$ and as a singlet
under $SU(2)_C$ and the boundary flavor group $G$.\footnote{The same short multiplet contains operators associated with the broken R-symmetry generators \cite{Cordova:2016emh}.}

\paragraph{The moment-map OPE.}
Our main observable is the four-point function of moment-map operators,
\begin{equation}
\left\langle
\mathcal O_2^{I_1}(x_1;v_1)
\mathcal O_2^{I_2}(x_2;v_2)
\mathcal O_2^{I_3}(x_3;v_3)
\mathcal O_2^{I_4}(x_4;v_4)
\right\rangle .
\label{eq:moment-map-fourpoint}
\end{equation}
Supersymmetry relates it to correlation functions of the other
components of the flavor-current multiplet, including those of the
conserved currents. The boundary operator algebra contains many $OSp(4|4)$ multiplets. For our purposes, we focus on those appearing in the OPE of two moment maps, or equivalently of two flavor-current multiplets. We follow the conventions of \cite{Chang:2019dzt}, with the two $SU(2)$ R-symmetry factors exchanged in order. 

In addition to the superconformal selection rules, Bose symmetry under
exchange of the two external moment-map operators,
\(1\leftrightarrow2\), constrains the quantum numbers of the operator
appearing in their OPE. Since the external operators transform in the
adjoint of the flavor group,
\begin{equation}
    {\rm Adj}\otimes{\rm Adj}
    =
    {\rm Sym}^2({\rm Adj})
    \oplus
    \wedge^2({\rm Adj}) .
\end{equation}
We denote by \(\sigma_{\mathcal R}\) the parity of the corresponding flavor tensor
under \(I_1\leftrightarrow I_2\),
\begin{equation}
    \sigma_{\mathcal R}\equiv
    \begin{cases}
        +1, & \mathcal R\subset {\rm Sym}^2({\rm Adj}),\\
        -1, & \mathcal R\subset \wedge^2({\rm Adj}).
    \end{cases}
\end{equation}
For an exchanged superconformal primary of spacetime spin \(\ell\) and
\(SU(2)_H\) isospin \(j_H\), the spacetime and R-symmetry structures
contribute parities \((-1)^\ell\) and \((-1)^{j_H}\), respectively.
Bose symmetry therefore requires
\begin{equation}
    (-1)^{\ell+j_H}\sigma_{\mathcal R}=1.
\label{eq:moment-map-spin-selection}
\end{equation}
To implement this constraint compactly in the OPE below, we define the allowed flavor sectors at fixed spacetime spin,
\begin{equation} 
\mathsf F_{\ell}^{(+)}
\equiv
\begin{cases}
{\rm Sym}^2({\rm Adj}), & \ell\ {\rm even},\\
\wedge^2({\rm Adj}), & \ell\ {\rm odd},
\end{cases}
\qquad
\mathsf F_{\ell}^{(-)}
\equiv
\begin{cases}
\wedge^2({\rm Adj}), & \ell\ {\rm even},\\
{\rm Sym}^2({\rm Adj}), & \ell\ {\rm odd}.
\end{cases}
\label{eq:flavor-parity-families}
\end{equation}
Then, the moment-map OPE takes the schematic form
\begin{align}
&\bigl(\mathcal B[0]_{1}^{(2;0)},{\rm Adj}\bigr)
\times
\bigl(\mathcal B[0]_{1}^{(2;0)},{\rm Adj}\bigr)
\sim{}
\bigl(\mathcal B[0]_{0}^{(0;0)},\mathbf 1\bigr)
+
\bigl(\mathcal B[0]_{1}^{(2;0)},{\rm Adj}_{A}\bigr)
\notag\\
&\qquad+
\sum_{\mathcal R\subset {\rm Sym}^2({\rm Adj})}
\bigl(\mathcal B[0]_{2}^{(4;0)},\mathcal R\bigr)
+
\sum_{\ell=0}^{\infty}
\sum_{\mathcal R\subset\mathsf F_{\ell}^{(+)}}
\bigl(\mathcal A[2\ell]_{\ell+1}^{(0;0)},\mathcal R\bigr)
+
\sum_{\ell=0}^{\infty}
\sum_{\mathcal R\subset\mathsf F_{\ell}^{(-)}}
\bigl(\mathcal A[2\ell]_{\ell+2}^{(2;0)},\mathcal R\bigr)
\notag\\
&\qquad+
\sum_{\ell=0}^{\infty}
\sum_{\Delta>\ell+1}
\sum_{\mathcal R\subset\mathsf F_{\ell}^{(+)}}
\bigl(\mathcal L[2\ell]_{\Delta}^{(0;0)},\mathcal R\bigr).
\label{eq:moment-map-OPE-schematic}
\end{align}
Some comments are in order. The flavor-current multiplet is exchanged in the antisymmetric adjoint channel. The protected $\mathcal B[0]_2^{(4;0)}$ sector
also contains a distinguished flavor-singlet multiplet whose descendants include the displacement operator. The family $\mathcal A[2\ell]_{\ell+1}^{(0;0)}$ contains
conserved-current multiplets. Its $\ell=0$ member,
$\mathcal A[0]_1^{(0;0)}$, is the three-dimensional stress-tensor
multiplet. These multiplets are allowed by the selection rules, but
they do not appear in our correlator. The three-dimensional
superconformal symmetry is inherited from the four-dimensional
bulk, and there is no conserved boundary stress tensor, nor any
conserved higher-spin current, at any finite value of the bulk coupling.
This is already the case at zero coupling, where the moment map is a
generalized free field of dimension one: the R-symmetry singlets of
spin $\ell$ in its OPE start at dimension $\ell+2$, above the
dimension $\ell+1$ of the superconformal primary of
$\mathcal A[2\ell]_{\ell+1}^{(0;0)}$. The unprotected sector consists of an infinite family of long
multiplets, whose superconformal primaries are R-symmetry
singlets. These will be the main focus of our one-loop analysis.

Finally, the moment-map operators generate a protected one-dimensional topological subsector of the boundary theory through Higgs-branch cohomology
\cite{Chester:2014mea,Beem:2016cbd,Dedushenko:2016jxl,Dedushenko:2020vgd}. This sector provides exact constraints on protected OPE coefficients, complementary to the perturbative data extracted from the four-point function.

%%%%%%%%%%%%%%%%%%%%%%%%%%%%%%%%%%%%%%%%%%%%%%%%%%%%%%%%%%%%%%%%
\section{Tree-level AdS$_4$ amplitude}
%%%%%%%%%%%%%%%%%%%%%%%%%%%%%%%%%%%%%%%%%%%%%%%%%%%%%%%%%%%%%%%%
In this section we derive the tree-level four-point function of the boundary moment-map operators from Dirichlet $4d$ $\mathcal N=4$ SYM on AdS$_4$. We normalize the gauge generators as
\begin{equation}
\operatorname{Tr}(T^I T^J)=\kappa\,\delta^{IJ},
\qquad
[T^I,T^J]=i f^{IJK}T^K. \label{normalizations}
\end{equation}
In the flat-half-space frame, the bosonic part of the Euclidean action is
\begin{equation}
    S=\frac{1}{g_{\rm YM}^2}\int_{z>0}d^4x\,
    {\rm Tr}\!\left[
    \frac14F_{\mu\nu}F^{\mu\nu}
    +\frac12D_\mu\Phi^MD^\mu\Phi^M
    -\frac14[\Phi^M,\Phi^N]^2
    \right],
    \qquad
    \Phi^M=(X^a,Y^{\dot a}),
    \label{eq:SYM-action}
\end{equation}
with $D_\mu=\partial_\mu-i[A_\mu,\,\cdot\,]$ and all fields expanded on the generators, $\Phi^M=\Phi^{MI}T^I$. It is supplemented by the boundary term \eqref{eq:nahm-boundary-functional} discussed below. In components, the action carries an overall factor $\kappa/g_{\mathrm{YM}}^2$, so that the canonically normalized coupling is $g_{\rm YM}^2/\kappa$.\footnote{In components the gauge kinetic term is
$\frac{\kappa}{4g_{\rm YM}^2}F^I_{\mu\nu}F^{I\mu\nu}$. For $\kappa=\frac12$ this is the conventional
$\frac{1}{4g^2}F^a_{\mu\nu}F^{a\mu\nu}$ with $g^2=2g_{\rm YM}^2$.} We strip this common normalization from the intermediate Witten-diagram kernels and restore it in the complete correlator below.

\subsection{Momentum-space representation}
\label{subsec:tree-momentum}
Boundary correlators in the $\AdS_4$ theory admit a perturbative expansion in
momentum-space Witten diagrams. We work in Euclidean Poincar\'e $\AdS_4$ and
set the AdS radius to one. Translation invariance along the boundary allows us
to Fourier transform the boundary coordinates and define the $3d$ momenta $\boldsymbol{k}_i$, with $\sum_i \boldsymbol{k}_i=0$.\footnote{The $3d$ conformal symmetries are inherited from the AdS$_4$ isometries. Their generators are conserved charges constructed from
the bulk stress tensor, and do not require an intrinsic $3d$ stress tensor.} We denote 
\vspace{-0.5em}
\begin{equation}
    k_i\equiv|\boldsymbol{k}_i|,
    \qquad
    k_{ij}\equiv k_i+k_j,
    \qquad
k_{\underline{ij}}\equiv|\boldsymbol{k}_i+\boldsymbol{k}_j|,
    \qquad
    E\equiv\sum_{i=1}^4 k_i.
\end{equation}
The moment-map four-point function also carries adjoint gauge-group indices and $SU(2)_H$ quantum numbers. We organize the former using the three cubic-diagram color factors,
\begin{align}
    \mathtt c_s^{I_1I_2I_3I_4}
    \equiv f^{I_1I_2J}f^{JI_3I_4}, \qquad 
    \mathtt c_t^{I_1I_2I_3I_4}
    \equiv f^{I_1I_4J}f^{JI_2I_3}, \qquad
    \mathtt c_u^{I_1I_2I_3I_4}
    \equiv f^{I_1I_3J}f^{JI_4I_2},
\end{align}
satisfying the Jacobi identity
\begin{equation}
    \mathtt c_s+\mathtt c_t+\mathtt c_u=0.
    \label{eq:color-Jacobi-momentum}
\end{equation}
The \(SU(2)_H\) dependence is instead encoded using the polarization spinors
introduced in \eqref{eq:moment-map-def}. We define the three \(SU(2)_H\)-invariant tensor structures
\begin{align}
    R_s\equiv(v_1\!\cdot v_2)^2(v_3\!\cdot v_4)^2, \qquad
    R_t\equiv(v_1\!\cdot v_4)^2(v_2\!\cdot v_3)^2, \qquad
    R_u\equiv(v_1\!\cdot v_3)^2(v_2\!\cdot v_4)^2,
    \label{eq:Rstructures-momentum}
\end{align}
where $(v_i\cdot v_j)\equiv \epsilon_{AB}v_i^A v_j^B$. The R-symmetry cross-ratio is instead
\begin{equation}
    \alpha\equiv
    \frac{(v_1\!\cdot v_3)(v_2\!\cdot v_4)}
    {(v_1\!\cdot v_2)(v_3\!\cdot v_4)}.
\end{equation}
In particular, notice that
\begin{equation}
    R_u/R_s=\alpha^2,
    \qquad
    R_t/R_s=(1-\alpha)^2.
    \label{eq:Rstructures-alpha}
\end{equation}
We now
collect the bulk ingredients needed to construct the tree-level Witten diagrams. 

\paragraph{External scalar wavefunctions.}
As discussed in section~\ref{sec:setup}, the $X$ scalars have
$m^2L^2=-2$ and the boundary moment-map operator corresponds to their alternate
$\Delta=1$ quantization. Near the AdS boundary, a Fourier mode has the
general asymptotic form
\begin{equation}
    \phi_k(z)=a(k)\,z+b(k)\,z^2+\cdots .
\end{equation}
In the standard quantization, $\Delta=2$, $a(k)$ is
the source and $b(k)$ the response, while in the alternate
quantization, $\Delta=1$, their roles are exchanged
\cite{Klebanov:1999tb,Hartman:2006dy}. Modified Bessel functions solve the full radial equation. Regularity in the AdS interior ($z \to \infty$) then selects
\begin{equation}
    \phi_k(z)\propto z^{3/2}K_{1/2}(kz)
    \propto z e^{-kz}
    =z-kz^2+\cdots ,
\end{equation}
and therefore relates the two near-boundary coefficients by $b=-ka$. With unit source normalization, the bulk-to-boundary wavefunctions are
\begin{equation}
    K_{\Delta=1}(k,z)
    =-\frac{z}{k}e^{-kz}=-\frac{1}{k}K_{\Delta=2}(k,z).
    \label{eq:K1-momentum}
\end{equation}
At tree level, the boundary moment-map four-point function receives bulk
contributions from the gauge-field exchange and the quartic scalar
interaction of the four-dimensional \(\mathcal N=4\) SYM action.

\paragraph{Gauge-field exchange.}
The momentum-space vertex takes the form
\begin{equation}
 V_{XXA}^{\,i}
 =
 f^{I_1I_2J}\,
 \delta^{a_1a_2}\,
 (\boldsymbol k_1-\boldsymbol k_2)^i ,
\end{equation}
where \(i=1,2,3\) labels directions tangent to the AdS boundary, and the gauge-field bulk-to-bulk propagator admits the spectral
representation \cite{Raju:2012zr}
\begin{equation}
 G_{ij}^{IJ}(\boldsymbol q;z,z')
 =
\delta^{IJ}
 \int_0^\infty p\,dp\,
 \frac{(zz')^{1/2}J_{1/2}(pz)J_{1/2}(pz')}
      {q^2+p^2}
 \left(\delta_{ij}+\frac{q_iq_j}{p^2}\right).
\end{equation}
For the $s$-channel, the momentum
flowing through the internal line is $\boldsymbol k_1+\boldsymbol k_2$, with magnitude $k_{\underline{12}}$. Using the external $\Delta=1$ wavefunctions
\eqref{eq:K1-momentum} and performing the radial and spectral
integrals gives the exchange contribution
\begin{equation}
\begin{split}
W_s
={}
-\frac{1}{k_1k_2k_3k_4}
\frac{1}{E\,\mathsf{s}}
\bigg[
(\boldsymbol k_1-\boldsymbol k_2)
\cdot(\boldsymbol k_3-\boldsymbol k_4)
-(k_1^2-k_2^2)(k_3^2-k_4^2)
\frac{E+k_{\underline{12}}}{k_{12}k_{34}k_{\underline{12}}}
\bigg],
\end{split}
\label{eq:Ws-scalar}
\end{equation}
where the Mandelstam-like variable is defined by $\mathsf{s}\equiv(k_{12}+k_{\underline{12}})
      (k_{34}+k_{\underline{34}})$.\footnote{The momentum-space Mandelstam-like variables \cite{Albayrak:2020fyp} satisfy $\mathsf{s}+\mathsf{t}+\mathsf{u}
    =
    E\left(
        E+k_{\underline{12}}
         +k_{\underline{13}}
         +k_{\underline{23}}
    \right)$, which reduces to the usual massless Mandelstam relation in the flat-space limit $E\to0$ \cite{Raju:2012zr}.} The $t$- and $u$-channel diagrams follow by crossing, and the corresponding $R$-symmetry structures are the $R_s$, $R_t$, and $R_u$ introduced in \eqref{eq:Rstructures-momentum}, respectively.

\paragraph{Scalar contact interaction.}
The scalar potential of \(\mathcal N=4\) SYM contains the quartic interaction $f^{IJM}f^{KLM}
    X^{aI}X^{bJ}X^{aK}X^{bL}$. Contracting the external \(SU(2)_H\) indices with the polarization spinors,
the corresponding gauge and \(R\)-symmetry tensor is
\begin{equation}
   \mathcal{T}_4= \mathtt c_s(R_u-R_t)
    +\mathtt c_t(R_s-R_u)
    +\mathtt c_u(R_t-R_s). \label{r4}
\end{equation}
Since the quartic vertex carries no derivatives, the remaining
momentum dependence comes entirely from the radial contact integral, giving
\begin{equation}
    W_c
    =
    -\frac{1}{E\,k_1k_2k_3k_4}.
    \label{eq:Wc-delta1}
\end{equation}

\paragraph{Nahm-induced scalar exchange.}
For the specific D-type theory considered here, the two bulk contributions
above actually do not exhaust the tree-level correlator. The supersymmetric boundary condition obeyed by the \(X\) scalars is non-linear, 
\begin{equation}
    \left(
    D_zX^a-\frac{i}{2}\epsilon^{abc}[X^b,X^c]
    \right)\bigg|_{z=0}=0 .
    \label{eq:Nahm-bc-tree}
\end{equation}
Around the trivial background \(X^a=0\), the boundary condition
linearizes to the ordinary Neumann condition. Together with the free
bulk equation of motion, this selects the Neumann scalar propagator.
In the flat-half-space frame, and stripping the overall coupling
normalization as above, it takes the form
\begin{equation}
 G_N^{aI,bJ}(\boldsymbol q;z,z')
 =
 \delta^{ab}\delta^{IJ}\,
 \frac{1}{2q}
 \left(
 e^{-q|z-z'|}
 +
 e^{-q(z+z')}
 \right).
\end{equation}
The non-linear term in \eqref{eq:Nahm-bc-tree} can be implemented by the variation of the finite boundary
functional 
\begin{equation}
 S_{\rm Nahm}
 =\frac{i}{6g_{\rm YM}^2}\int_{z=0} d^3x\,
 \epsilon^{abc}{\rm Tr}\!\left(X^a[X^b,X^c]\right)
 =-\frac{\kappa}{6g_{\rm YM}^2}\int_{z=0}d^3x\,
 \epsilon^{abc}f^{IJK}X^{aI}X^{bJ}X^{cK}.
 \label{eq:nahm-boundary-functional}
\end{equation}
This is a perturbative
implementation of the supersymmetric boundary condition, rather than an independent boundary interaction.\footnote{A particularly close supersymmetric AdS\(_4\) example is the finite cubic boundary term of \cite{Freedman:2016yue}, which contributes directly to boundary scalar correlators.}

We now evaluate the scalar-exchange contribution generated by two
insertions of the Nahm boundary vertex. Since both vertices are localized
at \(z=0\), the internal \(X\) line is obtained by restricting both
endpoints of the Neumann propagator to the boundary,
\begin{equation}
 G_N^{aI,bJ}(\boldsymbol q;0,0)
 =
 \frac{\delta^{ab}\delta^{IJ}}{q}.
 \label{eq:flat-neumann-boundary-limits}
\end{equation}
The external legs are the corresponding Neumann source wavefunctions,
\(K_N(k,0)=-1/k\), related by the Weyl transformation to the
alternate-quantization wavefunction \eqref{eq:K1-momentum}. Joining the
two boundary vertices with the internal propagator and attaching the four
external legs gives the $s$-channel contribution
\begin{equation}
    B_s=\frac{1}{k_1\,k_2\,k_3\,k_4\,k_{\underline{12}}}.
\end{equation}
The color and \(R\)-symmetry
tensors follow directly from the two cubic vertices: \(B_s\) multiplies
\(\mathtt c_s(R_u-R_t)\), with the \(t\)- and \(u\)-channel structures
obtained by crossing. Notice that this is the same tensorial structure as
the quartic interaction in \eqref{r4}. Their relative coefficient follows
from the common microscopic \(\mathcal N=4\) SYM action and its
supersymmetric boundary completion. We can then combine the scalar sector of the $s$-channel tree-level correlator into
\begin{equation}
    \widetilde B_s
    \equiv B_s+W_c
    =
    \frac{E-k_{\underline{12}}}
{E\,k_1\,k_2\,k_3\,k_4\,k_{\underline{12}}}.
\end{equation}

\paragraph{Complete result.}
To conclude, let us restore the coupling normalization. We normalize the moment map as $\mu^{aI}\equiv(\kappa/g_{\rm YM}^2)\,X^{aI}\big|_{z=0}$. At leading order in weak coupling, its two-point function is
\begin{equation}
 \left\langle\mu^{aI}(\boldsymbol k)
 \mu^{bJ}(-\boldsymbol k)\right\rangle
 =\frac{\kappa}{g_{\rm YM}^2}
 \frac{\delta^{ab}\delta^{IJ}}{k},
 \label{eq:raw-moment-map-two-point}
\end{equation}
where we have suppressed the overall $3d$ momentum-conserving delta function. It is convenient to define momentum-normalized operators
\begin{equation}
 \widehat\mu^{aI}
 \equiv\frac{g_{\rm YM}}{\sqrt\kappa}\,\mu^{aI},
 \qquad
 \langle\widehat\mu^{aI}(\boldsymbol k)
 \widehat\mu^{bJ}(-\boldsymbol k)\rangle
 =
 \frac{\delta^{ab}\delta^{IJ}}{k}.
 \label{eq:moment-map-momentum-normalization}
\end{equation}
Following our definition \eqref{eq:moment-map-def}, the tree-level four-point function can be written as
\begin{equation}
\begin{aligned}
\left\langle
\widehat{\mathcal O}_2^{I_1}
\widehat{\mathcal O}_2^{I_2}
\widehat{\mathcal O}_2^{I_3}
\widehat{\mathcal O}_2^{I_4}
\right\rangle_{\rm tree}
=
\frac{g_{\rm YM}^2}{\kappa}
\Big[
\mathtt c_s
\bigl(
R_s W_s+(R_u-R_t)\widetilde B_s
\bigr)
+\text{crossed}
\Big].
\end{aligned} \label{finalmomentum}
\end{equation}

\subsection{Matching to the supergluon Mellin amplitude}
\label{subsec:tree-mellin}
The moment map belongs to the same short multiplet
as the conserved current, whose AdS realization is
the bulk gauge field. It is therefore useful to place
our observable in the broader context of gauge-theory
correlators in AdS. Tree-level current correlators
in pure Yang--Mills theory were studied in
\cite{Raju:2012zr,Raju:2012zs}, and subsequent developments
introduced compact momentum-space representations,
recursive constructions, and generalized
color--kinematics relations
\cite{Armstrong:2022mfr,Gomez:2026yno,
Armstrong:2020woi,Albayrak:2020fyp}; see also
\cite{Heuveline:2026knw} for the self-dual sector.
On the supersymmetric side, four-point amplitudes
of AdS supergluons were obtained in a broad class
of holographic theories and shown to exhibit related
universal and color--kinematic structures
\cite{Alday:2021odx}. Explicit double-copy relations
have also been obtained in special four-point AdS
settings \cite{Zhou:2021gnu}.\footnote{KLT relations between AdS open-
and closed-string building blocks have recently
been established in \cite{Alday:2025bjp,Nocchi:2026yms}.}
In these holographic constructions, however,
the gauge fields typically arise from flavor-brane
or compactification sectors of a higher-dimensional
gravitational theory. The problem considered here
is different: the gauge field is that of four-dimensional
$\mathcal N=4$ SYM itself, propagating on rigid $\AdS_4$.

We now introduce the tree-level
$d=3,\,p=2$ supergluon Mellin amplitude of
\cite{Alday:2021odx}, while establishing the relation
to the microscopic momentum-space result derived above.
These supergluons are scalar superprimaries of short
superconformal multiplets whose descendants have spin
at most one. We denote their Kaluza--Klein label by $p$,
to avoid confusion with the momentum magnitudes $k_i$
introduced above.

The lowest multiplet, $p=2$, is the flavor-current
multiplet. In the $3d$ $\mathcal N=4$ description, its
bosonic components include the dimension-one scalar
superprimary $s_2^I$, the dimension-two vector
$A_{2,\mu}^I$, and the dimension-two scalar $t_2^I$,
transforming respectively in $(\mathbf3,\mathbf1)$,
$(\mathbf1,\mathbf1)$, and $(\mathbf1,\mathbf3)$ under the R-symmetry 
$SU(2)_a\times SU(2)_b$.\footnote{
The explicit $d=3$ flavor-brane model of
\cite{Alday:2021odx} is flavored ABJM and preserves
$\mathcal N=3$ supersymmetry. Its vector multiplet
spectrum, however, admits the $\mathcal N=4$
organization used here. For the $p=2$ sector, this gives the same bosonic exchanged content.
}
Here $I$ is an adjoint flavor index. We identify the
two R-symmetry factors with those preserved by our
boundary conditions, $SU(2)_H \times SU(2)_C$. At the level of the AdS fields,
\begin{equation}
    s_2^I
    \longleftrightarrow
    X^{a,I}\quad(\Delta=1),
    \qquad
    A_{2,\mu}^I
    \longleftrightarrow
    A_\mu^I,
    \qquad
    t_2^I
    \longleftrightarrow
    Y^{\dot a,I}\quad(\Delta=2).
\end{equation}
For the four-point function of the $p=2$
superprimaries, the exchanged current multiplet
contributes through the scalar superprimary $s_2$
and the vector descendant $A_2$, while $t_2$
transforms non-trivially under $SU(2)_C$ and is
therefore not exchanged in the $s_2\times s_2$ OPE.

The tree-level answer of \cite{Alday:2021odx} is fixed
by superconformal symmetry, crossing, and factorization.
In this sense, it is universal. Our microscopic
calculation provides a direct realization of this
structure in $\mathcal N=4$ SYM. Moreover, in the holographic supergluon constructions, the
leading large-$N$ dynamics is governed by the
self-interactions of the vector multiplet, while
graviton exchange and other contributions from
the gravitational sector are subleading. In our
setup, gravity is instead non-dynamical from
the outset.

Let us now make the comparison explicit by introducing
the Mellin representation, which will also provide
the framework for the one-loop analysis. For four
external operators of dimension one, we write
\begin{equation}
\begin{split}
\left\langle
    \prod_{i=1}^4 \mathcal O_2^{I_i}(x_i;v_i)\right\rangle
=
\left(
    \frac{(v_1\!\cdot v_2)(v_3\!\cdot v_4)}
         {x_{12}x_{34}}
\right)^2
\mathcal G^{I_1I_2I_3I_4}(U,V;\alpha),
\end{split}
\end{equation}
with the conformal cross-ratios
\begin{equation}
    U\equiv\frac{x_{12}^2x_{34}^2}{x_{13}^2x_{24}^2},
    \qquad
    V\equiv\frac{x_{14}^2x_{23}^2}{x_{13}^2x_{24}^2}.
\end{equation}
The connected part of the dynamical function admits the Mellin representation
\cite{Mack:2009mi,Fitzpatrick:2011ia}
\begin{equation}
\mathcal G^{I_1I_2I_3I_4}(U,V;\alpha)
=
\int_{-i\infty}^{i\infty}
\frac{ds\,dt}{(4\pi i)^2}\,
U^{s/2}V^{t/2-1}\mathcal K(s,t)
\mathcal M^{I_1I_2I_3I_4}(s,t;\alpha),
\label{eq:mellin-representation}
\end{equation}
where $s+t+u=4$, and the corresponding Mellin measure reads
\begin{equation}
    \mathcal K(s,t)
    \equiv
    \Gamma\left(\frac{2-s}{2}\right)^2
    \Gamma\left(\frac{2-t}{2}\right)^2
    \Gamma\left(\frac{2-u}{2}\right)^2.
    \label{eq:mellin-kernel}
\end{equation}
The tree-level supergluon amplitude takes the color
decomposition
\begin{equation}
    \mathcal M^{I_1I_2I_3I_4}(s,t;\alpha)
    = (C_{222})^2 \left(
    \mathtt c_s M_s
    +\mathtt c_t M_t
    +\mathtt c_u M_u \right),
    \label{eq:mellin-color-decomposition}
\end{equation}
where $C_{222}$ is the OPE
coefficient of the unit-normalized flavor-current multiplet, and 
\begin{equation}
M_s(s,t;\alpha)
=
\big[(1-\alpha)s+t-2\big]
\left[
\frac{1}{\pi^2s}
-
\frac{1}{\pi^{5/2}s}
\frac{\Gamma\left(\frac12-\frac{s}{2}\right)}
     {\Gamma\left(1-\frac{s}{2}\right)}
\right],
\label{Ms}
\end{equation}
while $M_t$ and $M_u$ are obtained by crossing. In the conventions of appendix A of \cite{Alday:2021odx}, we can read the contributions of the vector and scalar exchanges, $\mathcal{M}_{\Delta,\ell}$, whose relative coefficient is fixed by superconformal symmetry:
\begin{equation}
M_s(s,t;\alpha)
=
-\frac1{16}\mathcal M^{(s)}_{2,1}(s,t)
+
(2\alpha-1)
\left[
\frac12\mathcal M^{(s)}_{1,0}(s)
-\frac1{2\pi^2}
\right].
\label{eq:Ms-superexchange-decomposition}
\end{equation}
The comparison with our microscopic result
\eqref{finalmomentum} is now immediate.
The three contributions are identified as\footnote{
We factored out the common overall factor
$1/(16\pi^4)$ associated with the required
Fourier transform.
}
\begin{equation}
\begin{aligned}
W_s
&\longleftrightarrow
-\frac1{16}\mathcal M^{(s)}_{2,1}(s,t),
\qquad
B_s
\longleftrightarrow
\frac12\mathcal M^{(s)}_{1,0}(s),
\qquad
W_c
\longleftrightarrow
-\frac1{2\pi^2}.
\end{aligned}
\label{eq:boxed-momentum-mellin-dictionary}
\end{equation}
Thus the bulk gauge-field exchange supplies the
vector component $A_2$ of the exchanged current
multiplet, while the boundary scalar exchange
induced by the non-linear Nahm condition supplies
precisely its scalar component $s_2$.
The quartic scalar interaction provides the
corresponding contact contribution.
Together, these three terms reproduce
\eqref{eq:Ms-superexchange-decomposition}, with the
relative R-symmetry factor
$(R_u-R_t)/R_s=2\alpha-1$ following from our
momentum-space result \eqref{finalmomentum}.

Finally, converting from our momentum normalization
\eqref{eq:moment-map-momentum-normalization} to the
unit position-space normalization,
$\mathcal O_2=\sqrt{2}\pi\,\widehat{\mathcal O}_2$,
the tree-level matching fixes the leading
weak-coupling normalization,
\begin{equation}
    (C_{222})^2
    =
    \frac{g_{\rm YM}^2}{4\kappa}
    +O(g_{\rm YM}^4).
    \label{eq:C222-gym}
\end{equation}

%%%%%%%%%%%%%%%%%%%%%%%%%%%%%%%%%%%%%%%%%%%%%%%%%%%%%%%%%%%%%%%%
\section{Tree-level CFT data and one-loop strategy}
\label{sec:one-loop-strategy}
%%%%%%%%%%%%%%%%%%%%%%%%%%%%%%%%%%%%%%%%%%%%%%%%%%%%%%%%%%%%%%%%

At tree level, the four-point function is determined entirely by the
external current multiplet and its allowed cubic and quartic
interactions. At one loop, by contrast, the answer is sensitive to
the complete spectrum of states that can propagate as intermediate
states. We expand the four-point correlator at weak coupling as
\begin{equation}
    \mathcal G
    =
    \mathcal G_{\rm GFF}
    +\varepsilon\,\mathcal G_{\rm tree}
    +\varepsilon^2\,\mathcal G_{\rm 1-loop}
    +\cdots ,
    \label{eq:perturbative-correlator-loop}
\end{equation}
where $\varepsilon\equiv (C_{222})^2=8/C_J$, and \(C_J\) is the flavor central charge in the
conventions of \cite{Alday:2021odx}.\footnote{The expansion in
$\varepsilon$ differs from that in the microscopic
Yang--Mills coupling beyond leading order, as is clear from \eqref{eq:C222-gym}.} The leading term is the disconnected generalized-free-field (GFF)
correlator, which contains the double-trace families of the
dimension-one moment-map operator,
\begin{equation}
    [\mathcal O_2\mathcal O_2]_{n,\ell},
    \qquad
    \Delta^{(0)}_{n,\ell}
    =2\Delta_{\mathcal O_2}+2n+\ell
    =2+2n+\ell.
    \label{doubletrace}
\end{equation}
Interactions perturb these data starting at order $\varepsilon$.
For a double-trace family of spin $\ell$, the dimension and
squared OPE coefficient admit the expansions
\begin{align}
    \Delta_{n,\ell}
    &=
    \Delta^{(0)}_{n,\ell}
    +\varepsilon\gamma^{(1)}_{n,\ell}
    +\varepsilon^2\gamma^{(2)}_{n,\ell}
    +\cdots ,
    \label{eq:dimension-expansion-loop}
    \\
    a_{n,\ell}
    &=
    A_{n,\ell}
    +\varepsilon a^{(1)}_{n,\ell}
    +\varepsilon^2 a^{(2)}_{n,\ell}
    +\cdots .
    \label{eq:ope-expansion-loop}
\end{align}
Here $A_{n,\ell}$ denotes the GFF coefficient fixed by the disconnected correlator, while $\gamma^{(k)}_{n,\ell}$ are the interaction-induced anomalous dimensions of the exchanged long multiplets. The dimensions of protected short multiplets, by contrast, are fixed by supersymmetry.

These anomalous dimensions generate logarithmic terms in the
position-space correlator. To see this, recall that a bosonic
conformal block behaves in the OPE limit $U\to0$ at fixed $V$ as
\begin{equation}
    G_{\Delta,\ell}(U,V)
    =
    U^{\frac{\Delta-\ell}{2}}
    \left[g_{\Delta,\ell}(V)+O(U)\right].
\end{equation}
Substituting the perturbative expansion of the dimension into
the leading power of $U$ gives
\begin{equation}
U^{\frac{\Delta_{n,\ell}-\ell}{2}}
=
U^{\frac{\Delta^{(0)}_{n,\ell}-\ell}{2}}
\bigg[
1
+\varepsilon\frac{\gamma^{(1)}_{n,\ell}}{2}\log U
+\varepsilon^2
\left(
\frac{\gamma^{(2)}_{n,\ell}}{2}\log U
+
\frac{\left(\gamma^{(1)}_{n,\ell}\right)^2}{8}
\log^2 U
\right)
+\cdots
\bigg].
\label{eq:origin-of-logs}
\end{equation}
The tree-level anomalous dimensions $\gamma^{(1)}_{n,\ell}$ generate both the
single logarithm at order $\varepsilon$ and the double logarithm
at order $\varepsilon^2$. In particular, the corresponding
coefficients for an individual exchanged family are
\begin{equation}
\mathcal G_{\rm tree}
\supset
\frac12 A_{n,\ell}\gamma^{(1)}_{n,\ell}\log U,
\qquad
\mathcal G_{\rm 1-loop}
\supset
\frac18 A_{n,\ell}
\left(\gamma^{(1)}_{n,\ell}\right)^2
\log^2 U ,
\label{eq:leading-logs}
\end{equation}
where the appropriate unperturbed conformal-block
dependence is understood, and flavor labels are suppressed
for the moment. At one loop, the correlator therefore has the schematic expansion
\begin{equation}
\mathcal G_{\rm 1-loop}(U,V)
=
\mathcal F_2(U,V)\log^2 U
+\mathcal F_1(U,V)\log U
+\mathcal F_0(U,V).
    \label{eq:oneloop-schematic-logs}
\end{equation}
The crucial observation is that the coefficient $\mathcal F_2(U,V)$ is determined entirely by the tree-level anomalous dimensions and the leading GFF OPE coefficients. By contrast, $\mathcal F_1(U,V)$ and $\mathcal F_0(U,V)$ also contain genuinely new one-loop CFT data, including the second-order anomalous dimensions and corrections to the OPE coefficients. To isolate the information contained in $\mathcal F_2(U,V)$, we can take the double-discontinuity of the one-loop expression:\footnote{Equivalently, one may work in a crossed OPE channel, for example with the double-discontinuity in $V$.}
\begin{equation}
    {\rm dDisc}_U\,\mathcal G_{\rm 1-loop}(U,V)
    =
    4\pi^2 \mathcal F_2(U,V).
\end{equation}
Indeed, the unperturbed double-trace operators contribute with integer powers of $U$, and the double-discontinuity annihilates analytic and single-logarithmic terms. 

This is the basis of the AdS unitarity method
\cite{Aharony:2016dwx,Meltzer:2019nbs}.
In flat space, unitarity relates the discontinuity of
a loop amplitude to intermediate physical states
going on shell. The corresponding unitarity cut is
computed by replacing the cut propagators with
on-shell states and gluing lower-order amplitudes
\cite{Cutkosky:1960sp,Bern:1994zx}.
Although AdS has no conventional $S$-matrix,
the analogous information is encoded in the
double-discontinuity of boundary correlators.
At one loop, the two-particle cut is obtained by
gluing tree-level boundary correlators and summing
over the intermediate double-trace states
\eqref{doubletrace}. This is where the microscopic spectrum of the theory becomes essential. In the $\AdS_{d+1}\times S^3$
supergluon constructions, four external $p=2$
operators can couple across the cut to higher
Kaluza--Klein multiplets. The one-loop $\langle 2222 \rangle$
correlator therefore requires tree-level amplitudes
of the form $\langle22pp\rangle$ for all allowed $p$,
and operators belonging to different KK levels can
mix. In contrast, our theory has no internal compact
space and hence no KK tower. Under the identification
established at tree level, the two-particle cut is
obtained by gluing the $p=2$ sector of \cite{Alday:2021odx} to itself.

Moreover, there is no mixing problem to solve. At order
$\varepsilon^0$, the states that contribute to the cut are
two-particle states of the flavor-current multiplet, which
transform in the tensor product of two copies of
$\mathcal B[0]_1^{(2;0)}$. In this tensor product, each long
multiplet $\mathcal L[2\ell]_\Delta^{(0;0)}$ in a given flavor
representation appears at most once. Indeed, such a multiplet
contains exactly one conformal primary with $SU(2)_H$ spin two and
$SU(2)_C$ spin zero, of dimension $\Delta+2$ and spin $\ell$: this is
the top component discussed below, see
\eqref{eq:alpha2-long-top-block}. In the two-particle space,
$SU(2)_H$ spin two can only be reached with two moment maps, since
all other components of the multiplet carry $SU(2)_H$ spin at most
one half, and there is exactly one such generalized-free
double-trace primary for each dimension, spin and flavor symmetry.
 The double-trace long multiplets are
therefore non-degenerate, and the one-loop double logarithm is
controlled by $A_{n,\ell}\bigl(\gamma^{(1)}_{n,\ell}\bigr)^2$ for each
individual multiplet, with no average over degenerate operators.
As usual in the AdS unitarity method \cite{Aharony:2016dwx},
states with more than two particles do not contribute to the
double-discontinuity at this order.

The double-discontinuity alone does not determine
the remaining single-logarithmic and analytic terms.
Together with crossing symmetry and analyticity,
however, the cut data can be used to reconstruct
the non-local part of the correlator, up to the
familiar low-spin or contact-term ambiguities.
In position space, this can be implemented through
the Lorentzian inversion formula
\cite{Caron-Huot:2017vep} or conformal dispersion
relations \cite{Carmi:2019cub}. In the next section, we will instead avoid performing this reconstruction directly in position space and translate the double-discontinuity data into
Mellin space, where the problem becomes a much simpler algebraic reconstruction of the one-loop amplitude.

\paragraph{The highest R-symmetry channel.}
The four-point function of moment-map operators is naturally
decomposed into $OSp(4|4)$ superconformal blocks rather than
ordinary conformal blocks. Each superconformal block is a
fixed linear combination of three-dimensional bosonic
conformal blocks $G_{\Delta,\ell}(U,V)$, multiplied by
$SU(2)_H$ R-symmetry harmonics $\mathcal Y_{2j}(\alpha)$,
with relative coefficients determined by superconformal
symmetry \cite{Chang:2019dzt}. Since the dependence on the
R-symmetry cross-ratio is a polynomial of degree two,
it is particularly convenient to focus on the highest
R-symmetry channel, namely the coefficient of $\alpha^2$.
Equivalently, in the harmonic basis this isolates the
highest R-symmetry harmonic
$\mathcal Y_4(\alpha)=\alpha^2/2-\alpha/2+1/12$,
in the conventions of \cite{Alday:2021odx}.

The unprotected long multiplets allowed by the moment-map
OPE have R-symmetry-singlet superconformal primaries
\eqref{eq:moment-map-OPE-schematic}. Although their full
superconformal blocks contain several bosonic conformal
blocks, the highest R-symmetry component contains only one:
a bosonic conformal primary at the second bosonic level
of the supermultiplet, with dimension $\Delta+2$ and
the same spacetime spin $\ell$. Therefore, the projection of the superconformal block is
\begin{equation}
    \left[\alpha^2\right]
    \mathfrak G_{\mathcal L[2\ell]_{\Delta}^{(0;0)}}
    (U,V;\alpha)
    =
    q_{\Delta,\ell}\,
    G_{\Delta+2,\ell}(U,V),
    \label{eq:alpha2-long-top-block}
\end{equation}
where $q_{\Delta,\ell}$ includes the fixed superconformal-block
prefactor and the normalization of the R-symmetry harmonic.
For the double-trace long operators appearing perturbatively
in our correlator \eqref{doubletrace}, the selected bosonic
block has dimension $4+2n+\ell$. We therefore introduce
the shorthand
\begin{equation}
    G_{n,\ell}(U,V)
    \equiv
    G_{4+2n+\ell,\ell}(U,V).
    \label{eq:bosonic-top-block-definition}
\end{equation}
This simplification applies independently in each allowed
flavor representation. After projecting onto a fixed flavor
channel and stripping the corresponding group-theory factor,
the tree-level $\log U$ term can be expanded directly in
the bosonic top-block basis,
\begin{equation}
    \left.
    [\alpha^2]\mathcal G_{\rm tree}(U,V)
    \right|_{\log U}
    =
    \sum_{n,\ell}
    \omega_{n,\ell}\,
    G_{n,\ell}(U,V).
    \label{eq:alpha2-tree-log-block-expansion-schematic}
\end{equation}
The coefficients $\omega_{n,\ell}$ are the universal
color-stripped coefficients of these bosonic top blocks.
After restoring the flavor projection, they are related
to the physical CFT data by
$\omega_{n,\ell}\propto A_{n,\ell}\gamma^{(1)}_{n,\ell}$,
as discussed below.\footnote{
The fixed prefactor $q_{\Delta,\ell}$ in
\eqref{eq:alpha2-long-top-block} is already included
in the definition of $\omega_{n,\ell}$, and the same
bosonic normalization must be used for the GFF
coefficients $A_{n,\ell}$ entering the one-loop
unitarity relation.
} To determine these coefficients, we first rewrite the tree-level
amplitude in position space and extract its $\log U$ term. Then, expanding at small $U$ and around $V=1$, we match to the corresponding lightcone expansion of the $3d$ bosonic blocks $G_{n,\ell}$, determining
$\omega_{n,\ell}$ recursively. 

\subsection{Position-space representation}
\label{subsec:dbar-representation}
In many familiar holographic examples, exchange Witten diagrams truncate
to finite sums of contact diagrams, or $D$-functions
\cite{DHoker:1999mqo,Rastelli:2017udc,Rastelli:2017ymc}.
Although the usual truncation condition is not satisfied in the present
$\AdS_4$ problem, the
tree-level Mellin amplitude still admits a finite representation in terms of
normalized $\bar D$-functions, similarly to other $3d$ holographic correlators \cite{Alday:2021ymb}. The required weights include
half-integer and vanishing values, which are defined by analytic
continuation of their Mellin--Barnes representation. Our conventions and
the corresponding Mellin-to-position-space dictionary are collected in
appendix~\ref{app:Mellin-to-position}. The $\alpha^2$ projection of the Mellin amplitude \eqref{Ms} is
\begin{align}
[\alpha^2]M_s&=0, \nonumber \\
[\alpha^2]M_t
&=
-\frac{2(s-2)}{\pi^3}
\left[
\frac{\pi}{2t}
+
\frac{\sqrt{\pi}\,
\Gamma\!\left(\frac{1-t}{2}\right)}
{t^2\Gamma\!\left(-\frac t2\right)}
\right],
\nonumber\\
[\alpha^2]M_u
&=
\frac{2(s-2)}{\pi^3}
\left[
\frac{\pi}{2u}
+
\frac{\sqrt{\pi}\,
\Gamma\!\left(\frac{1-u}{2}\right)}
{u^2\Gamma\!\left(-\frac u2\right)}
\right],
\end{align}
and the corresponding position-space correlator takes the form
\begin{equation}
[\alpha^2]\,\G_{\rm tree}(U,V)
=
U^2\left[
\mathtt c_t\,T_t(U,V)
+\mathtt c_u\,T_u(U,V)
\right],
\label{eq:alpha2-tree-dbar-exact}
\end{equation}
with the two
channel functions
\begin{equation}
\begin{split}
T_u(U,V)
&\equiv
\frac{1}{\pi^3}
\Big(
\pi\,\Dbar_{2,1,1,0}(U,V)
-\sqrt{\pi}\,\Dbar_{\frac32,1,\frac12,0}(U,V)
\Big), \\
T_t(U,V)
&\equiv
-\frac{1}{\pi^3}
\Big(
\pi V^{-1}\Dbar_{2,1,0,1}(U,V)
-\sqrt{\pi}\,V^{-1/2}\Dbar_{1,\frac32,\frac12,0}(U,V)
\Big).
\end{split}
\label{eq:Tt-dbar}
\end{equation}
Here, we have used the color Jacobi identity \eqref{eq:color-Jacobi-momentum} and chosen $(\mathtt c_t,\mathtt c_u)$ as independent color factors.

\paragraph{Logarithmic term.}
To extract the tree-level CFT data entering the one-loop unitarity
construction, we only need the coefficient of $\log U$ in
\eqref{eq:alpha2-tree-dbar-exact}. It can be obtained directly from the
Mellin--Barnes representation of the analytically continued
$\bar D$-functions, giving
\begin{equation}
\left.
\Dbar_{r_1r_2r_3r_4}(U,V)
\right|_{\log U}
=
\sum_{m,n\geq0}
d_{m,n}(r_i)\,
U^m(1-V)^n,
\label{eq:dbar-logU-expansion-main}
\end{equation}
where the explicit coefficients $d_{m,n}(r_i)$ are given in
appendix~\ref{app:Mellin-to-position}.

\subsection{Tree-level data}

The two functions $T_t$ and $T_u$ introduced above are convenient for the
explicit $\bar D$-function representation, since they multiply the two
independent color factors $\mathtt c_t$ and $\mathtt c_u$.  However, before performing
the conformal-block decomposition, it is useful to restore a
Jacobi-symmetric three-channel representation,
\begin{equation}
    \frac{[\alpha^2]\mathcal G_{\rm tree}}
    {U^2}
    =
    \mathtt c_s H_s+\mathtt c_t H_t+\mathtt c_u H_u .
\end{equation}
The functions $H_x$ are defined only up to a common shift
$H_x\to H_x+\chi$, which drops out by the color Jacobi identity \eqref{eq:color-Jacobi-momentum}. We fix this redundancy by imposing
\begin{equation}
    H_s+H_t+H_u=0 .
\end{equation}
With this choice, $T_t=H_t-H_s$ and $T_u=H_u-H_s$. 

The coefficient of $\log U$ is decomposed in the basis of
three-dimensional bosonic top blocks \eqref{eq:bosonic-top-block-definition}. We define the universal dynamical coefficient $\omega_{n,\ell}$ through
the Jacobi-symmetric channel decomposition,
\begin{equation}
    U^2H_x(U,V)\big|_{\log U}
    =
    \sum_{n,\ell\geq0}
    q_x(\ell)\,
    \omega_{n,\ell}\,
    G_{n,\ell}(U,V),
    \qquad x=s,t,u ,
    \label{eq:Hx-omega-decomposition}
\end{equation}
where
\begin{equation}
    q_s(\ell)=\frac{1-(-1)^\ell}{2},  \quad q_t(\ell)=-\left(1+\frac{(-1)^\ell}{2}\right), \quad 
    q_u(\ell)=\frac12+(-1)^\ell .
\label{eq:channel-parity-vectors}
\end{equation}
Notice that $q_s(\ell)+q_t(\ell)+q_u(\ell)=0$, consistently with the
Jacobi-symmetric choice above. All explicit channel-dependent spin-parity normalization is carried by \(q_x(\ell)\); the closed formula for \(\omega_{n,\ell}\) contains no additional piecewise even/odd prefactor. Translating to the two functions used in the explicit $\bar D$ representation, we obtain
\begin{align}
    U^2T_t(U,V)\big|_{\log U}
    &=
    -\frac32
    \sum_{n,\ell\geq0}
    \omega_{n,\ell}\,
    G_{n,\ell}(U,V),
    \label{eq:Tt-omega}\\
    U^2T_u(U,V)\big|_{\log U}
    &=
    \frac32
    \sum_{n,\ell\geq0}
    (-1)^\ell\,
    \omega_{n,\ell}\,
    G_{n,\ell}(U,V).
    \label{eq:Tu-omega}
\end{align}
The recursive block extraction can be organized using the small-$U$ lightcone expansion of the three-dimensional scalar blocks, which gives
\begin{equation}
\begin{aligned}
    \omega_{n,\ell}
    ={}
    \frac{2^{4-2n}}{3\pi}\,
    \frac{(2n+1)!}{(n!)^2}\,
    \frac{(2\ell+1)(2\ell+2n+3)}{\ell+1}
    \frac{
        \Gamma(\ell+n+2)\Gamma(\ell+2n+2)
    }{
        \Gamma(\ell+n+\frac52)
        \Gamma(\ell+2n+\frac52)
    }\,
    \mathcal Q_{n,\ell},
\end{aligned}
\label{eq:omega-closed-form}
\end{equation}
where $\mathcal Q_{n,\ell}$ is a terminating hypergeometric function,
\begin{equation}
    \mathcal Q_{n,\ell}
    \equiv
    {}_4F_3\!\left(
    \begin{matrix}
    1,\ 1,\ -n,\ -n-\frac12\\[2pt]
    \frac32,\ \ell+2,\ -\ell-2n-1
    \end{matrix};1\right).
\end{equation}

\subsection{Generalized-free data}
For four identical external scalars of dimension one, the GFF
double-trace family has
\begin{equation}
    \Delta^{(0)}_{m,\ell}
    =
    2+2m+\ell,
    \qquad
    m\in\mathbb Z_{\geq0}.
\end{equation}
In the $12\to34$ channel, the complete disconnected $\alpha^2$ component is
\begin{equation}
    [\alpha^2]\mathcal G_{\rm GFF}^{I_1I_2I_3I_4}(U,V)
    =
    U\,\delta^{I_1I_3}\delta^{I_2I_4}
    +\frac{U}{V}\,\delta^{I_1I_4}\delta^{I_2I_3}.
    \label{eq:gff-full-alpha2}
\end{equation}
We define the coefficients $A^{\rm Wick}_{m,\ell}$ of the single $U/V$ Wick
contraction by
\begin{equation}
    \frac{U}{V}
    =
    \sum_{m,\ell\geq0}
    A^{\rm Wick}_{m,\ell}\,
    G_{2+2m+\ell,\ell}(U,V),
    \label{eq:gff-single-channel-decomposition}
\end{equation}
while exchanging points $3$ and $4$ gives the other contraction, with a parity factor $(-1)^\ell$. In the normalization used for the $3d$ blocks, for $m\geq1$ the single-channel coefficients are\footnote{The endpoint $m=0$ must be treated separately, with $A^{\rm Wick}_{0,\ell}=16\pi(\ell!)^2/\Gamma(\ell+\frac12)^2$, but this does not enter the long top-block sector considered here, for which $m=n+1\geq1$.}

\begin{equation}
A^{\rm Wick}_{m,\ell}=
\frac{16(2\ell+1)\sqrt{\pi}\,
\Gamma(m+\frac12)\Gamma(m+\ell+1)\Gamma(2m+\ell+1)}
{\Gamma(m+1)\Gamma(m+\ell+\frac32)\Gamma(2m+\ell+\frac12)} .
\end{equation}
The top blocks relevant to the long sector of the $\alpha^2$ component have
dimension $4+2n+\ell$, implying that $m=n+1$.  We therefore define the
single-channel coefficient 
\begin{equation}
    A^{\rm ch}_{n,\ell}
  = A^{\rm Wick}_{n+1,\ell}.
\end{equation}

\subsection{Flavor projection and physical anomalous dimensions}
The coefficient $\omega_{n,\ell}$ extracted above is a universal
color-stripped dynamical quantity. To obtain physical CFT data, we must
resolve the flavor structure in the $12\to34$ channel. The
Jacobi-symmetric decomposition \eqref{eq:Hx-omega-decomposition} already
determines the color operator multiplying a spin-$\ell$ block,
\begin{equation}
    \mathbb C_\ell
    \equiv
    \sum_{x=s,t,u}q_x(\ell)\mathtt c_x
    =
    \begin{cases}
    \dfrac32(\mathtt c_u-\mathtt c_t),
        &\ell\ {\rm even},\\[6pt]
    \dfrac32\mathtt c_s,
        &\ell\ {\rm odd}. \label{cases}
    \end{cases}
\end{equation}
We therefore only need to diagonalize $\mathbb C_\ell$ on the irreducible
components $\mathcal R\subset{\rm Adj}\otimes{\rm Adj}$,
\begin{equation}
    \mathbb C_\ell P_{\mathcal R}
    =
    \lambda_{\mathcal R,\ell}P_{\mathcal R},
\end{equation}
where $P_{\mathcal R}$ is the projector onto $\mathcal R$. The color tensors naturally act as operators on
${\rm Adj}\otimes{\rm Adj}$. With the conventions of
\eqref{normalizations}, one has, for instance
$\mathtt c_u=T_1\!\cdot T_2$, so their action on a fixed irreducible
representation is determined by the corresponding quadratic Casimirs.
Using also the Bose-symmetry selection rule $\sigma_{\mathcal R}=(-1)^\ell$, we obtain
\begin{equation}
\lambda_{\mathcal R,\ell}
=
\begin{cases}
\dfrac32\left(C_{\mathcal R}-2C_A\right),
& \ell\ {\rm even},\qquad
\mathcal R\subset{\rm Sym}^2({\rm Adj}),\\[8pt]
\dfrac32 C_A,
& \ell\ {\rm odd},\qquad \, \,
\mathcal R={\rm Adj}_A,
\end{cases}
\label{eq:lambda-flavor-casimir}
\end{equation}
where $C_A$ and $C_{\mathcal R}$ denote the quadratic Casimirs
in the adjoint representation and in $\mathcal R$, respectively,
with $f^{IKL}f^{JKL}=C_A\delta^{IJ}$. For odd spin, all other flavor representations have vanishing
eigenvalue. Therefore,  
\begin{equation}
    \omega^{\mathcal R}_{n,\ell}
    =
    \lambda_{\mathcal R,\ell}\,
    \omega_{n,\ell}.
    \label{eq:omega-flavor-factorization}
\end{equation}
The same flavor projection must be applied to the generalized-free
correlator. The two non-trivial Wick contractions give the flavor
operator $(-1)^\ell\mathbf 1+\Pi$, where $\Pi$ exchanges the
two adjoint indices. Using the Bose-symmetry selection rule,
we obtain
\begin{equation}
    A^{\mathcal R}_{n,\ell}
    =
    2(-1)^\ell A^{\rm ch}_{n,\ell},
    \label{eq:gff-flavor-independence}
\end{equation}
which is the signed coefficient of the
projected top block. To extract the anomalous dimensions, recall that the bosonic
top block selected by the $\alpha^2$ projection is a descendant
of a long double-trace superprimary with
\begin{equation}
    \Delta^{\rm sp}_{n,\ell,\mathcal R}
    =
    2+2n+\ell
    +
    \varepsilon\,\gamma^{(1)}_{n,\ell,\mathcal R}
    +O(\varepsilon^2).
\end{equation}
Its dimension is shifted by two units relative to the
superprimary, so both acquire the same anomalous dimension.
The coefficient of the tree-level logarithm is therefore
\begin{equation}
    \omega^{\mathcal R}_{n,\ell}
    =
    \frac12
    A^{\mathcal R}_{n,\ell}
    \gamma^{(1)}_{n,\ell,\mathcal R}.
\end{equation}
Combining the flavor projections of the logarithmic and generalized-free coefficients, we obtain
\begin{equation}
    \gamma^{(1)}_{n,\ell,\mathcal R}
    =
    (-1)^\ell\lambda_{\mathcal R,\ell}
    \frac{\omega_{n,\ell}}
         {A^{\rm ch}_{n,\ell}}.
    \label{eq:physical-anomalous-dimension-channel}
\end{equation}
As a normalization check, at the lowest level we find
$\gamma^{(1)}_{0,0,\mathcal R}=\lambda_{\mathcal R,0}/\pi^2$,
corresponding to a long double-trace superprimary of
unperturbed dimension two, whose bosonic top component
has dimension four.

%%%%%%%%%%%%%%%%%%%%%%%%%%%%%%%%%%%%%%%%%%%%%%%%%%%%%%%
\section{One-loop AdS$_4$ amplitude}
\label{sec:mellin-reconstruction}
%%%%%%%%%%%%%%%%%%%%%%%%%%%%%%%%%%%%%%%%%%%%%%%%%%%%%%%

\subsection{Algorithm for the one-loop coefficients} 
We are now ready to use the tree-level CFT data to
fix the non-local part of the one-loop amplitude, via the AdS unitarity method. 

In the higher-dimensional examples where this construction was first
implemented, the superconformal Ward identity can be solved in terms of a
single reduced correlator, with the full R-symmetry dependence generated
by differential operators in position space, or equivalently by
finite-difference operators in Mellin space \cite{Rastelli:2016nze,Rastelli:2017udc,Rastelli:2017ymc}. In three dimensions, the standard reduced position-space representation involves non-local operators \cite{Dolan:2004mu}. We therefore construct the one-loop double-discontinuity directly from the full $\alpha^2$ moment-map correlator, and only afterwards reorganize the result in terms of a reduced Mellin amplitude. This step uses the recent reduced Mellin-space construction of
\cite{Virally:2025nnl}, adapted to three-dimensional theories with eight
supercharges in \cite{Chester:2025jxg}.\footnote{For a position-space analysis of reduced superconformal
blocks in theories with eight supercharges, including three dimensions,
see \cite{Woolley:2026anb}.}

The one-loop input is the $\log^2U$ coefficient in the highest
R-symmetry channel of the full correlator. For a fixed flavor irreducible representation $\mathcal R$, the unitarity
relation gives
\begin{equation}
    \left.
    [\alpha^2]\mathcal G_{\mathcal R}^{\rm 1-loop}(U,V)
    \right|_{\log^2 U}
    =
    \frac12
    \sum_{n,\ell}
    \frac{\left(\omega^{\mathcal R}_{n,\ell}\right)^2}
         {A^{\mathcal R}_{n,\ell}}
    G_{n,\ell}(U,V).
    \label{eq:one-loop-log2-omega}
\end{equation}
As in the tree-level analysis, we now separate the universal kinematic
spin sums from their flavor dressing. We define
\begin{align}
    F_{\rm even, \, odd}(U,V)
    &\equiv
    \frac14 
    \sum_{n,\ell\geq0} \frac{1\pm(-1)^\ell}{2}\,
    \frac{\omega_{n,\ell}^{\,2}}
         {A^{\rm ch}_{n,\ell}}\,
    G_{n,\ell}(U,V).
    \label{eq:F-definition}
\end{align}
These are the two universal functions that we need to reconstruct. The flavor dependence is restored by multiplying each spin-$\ell$ block
by $(-1)^\ell\lambda_{\mathcal R,\ell}^2$. Thus, an
allowed even-spin channel is proportional to $F_{\rm even}$, whereas an
allowed odd-spin channel is proportional to $-F_{\rm odd}$. This provides the position-space input for the AdS
unitarity reconstruction.

\paragraph{Summary of the strategy.}
To expose the analytic structure of the one-loop double-discontinuity,
we expand the universal functions at small $U$,
\begin{equation}
F_{\rm even, \,odd}(U,V)
=
\sum_{p\geq2}U^p h_p^\pm(V).
\label{eq:Fpm-U-slices-conceptual}
\end{equation}
At fixed $p$, only finitely many double-trace levels contribute, while
the spin sum remains infinite. The highest logarithmic singularity in the small-$V$ expansion of the functions $h_p^\pm(V)$ is proportional to $\log^2V$. Since $F_{\rm even, \, odd}$ determine the coefficient of
$\log^2U$ in the one-loop correlator, these terms correspond to
$\log^2U\,\log^2V$ contributions. 

Their Mellin-space interpretation is
particularly simple. The Mellin measure has double poles at the
double-trace locations \eqref{eq:mellin-kernel}, while an additional simple pole of the Mellin
amplitude at the same location produces a triple pole and hence a
squared logarithm. Simultaneous $\log^2U$ and $\log^2V$ terms therefore
signal simultaneous simple poles in the full Mellin amplitude. Motivated by the remarkably simple analytic structure found for one-loop reduced Mellin amplitudes of AdS supergluons (see, for example, \cite{Alday:2021ajh,Huang:2023ppy}), and the recent Mellin reduced formalism proposed in \cite{Chester:2025jxg}, we ask whether the present full-correlator data admit an analogous reduced representation. We therefore make the minimal reduced ansatz\footnote{In our conventions, the first reduced pole at $m=1$ is mapped to the first physical long sector contribution beginning at order $U^2$.}
\begin{equation}
B(s,t)
=
\sum_{m,n\geq1}
\frac{c_{mn}}
{(s-2m)(t-2n)}.
\label{eq:conceptual-reduced-mellin-ansatz}
\end{equation}
The even- and odd-spin reduced amplitudes are obtained from the appropriate
crossed sum and difference of this kernel. They are then mapped to the full
$\alpha^2$ component by the Mellin-space Ward/difference operator of
\cite{Chester:2025jxg}.

The coefficients $c_{mn}$ are first fixed by requiring the Ward-lifted
amplitude to reproduce the highest crossed-channel logarithms,
$\log^2U\,\log^2V$, of the full correlator. Since the Ward operator shifts
the Mellin variables only by finite amounts, a given $c_{mn}$ contributes
to a finite set of neighboring $U$-slices. The completeness of the simultaneous-pole ansatz is then tested by comparing the reconstructed amplitude with the full \(\log^2U\) data, including the terms subleading in \(\log V\). 

The conceptual reconstruction can be summarized as
\begin{equation}
\omega_{n,\ell}
\;\longrightarrow\;
\frac{1}{4}\frac{\omega_{n,\ell}^2}{A^{\rm ch}_{n,\ell}}
\;\longrightarrow\;
F_{\rm even},\,F_{\rm odd}
\;\longrightarrow\;
h_p^\pm(V)
\;\longrightarrow\;
\left[\log^2 V\right]h_p^\pm(V)
\;\longrightarrow\;
c_{mn}.
\label{eq:reconstruction-chain}
\end{equation}
The first steps are statements about CFT data and the one-loop double-discontinuity, followed by the reconstruction of the non-local amplitude. We now report the explicit results.

\paragraph{Step 1: $U$-slices.} The starting point is the small-$U$ expansion \eqref{eq:Fpm-U-slices-conceptual}. We define $\widehat h_p^\pm(V) \equiv \pi^4 h_p^\pm(V)$. The spin-summed slices take the finite rational--polylogarithmic form
\begin{equation}
\begin{split}
    \widehat h_p^\pm(V)
    =
    \frac{1}{(1-V)^{2p-2}}
    \bigg[
    R_{p,0}^\pm(V)
    +R_{p,1}^\pm(V)\log V
    +R_{p,2}^\pm(V)\log^2V
    +
    R_{p,3}^\pm(V)\Li_2(1-V)
    \bigg],
\end{split}
\label{eq:slice-ansatz}
\end{equation}
where the $R_{p,i}^\pm(V)$ are polynomials. 
For example, for the first two slices we find
\begin{align}
    \widehat h_2^+(V)
    &=
    \frac{\log^2V}{36(1-V)^2},
    \label{eq:h2-even}
    \\
    \widehat h_2^-(V)
    &=
    \frac{
    \log V\bigl(4+\log V\bigr)
    +4\Li_2(1-V)}
    {36(1-V)^2},
    \label{eq:h2-odd}
\end{align}
and
\begin{align}
    \widehat h_3^+(V)
    &=
    \frac{
    5(1-V)\log V
    +(2+V)\log^2V
    +2(1-V)\Li_2(1-V)}
    {36(1-V)^4},
    \label{eq:h3-even}
    \\
    \widehat h_3^-(V)
    &=
    \frac{
    10(1-V)
    +11(1+V)\log V
    +(2+V)\log^2V
    +6(1+V)\Li_2(1-V)}
    {36(1-V)^4}.
    \label{eq:h3-odd}
\end{align}

\paragraph{Step 2: The $\log^2V$ coefficient.}
Despite their different finite parts, the two parity sectors have the
same highest crossed-channel logarithm. We find the general form
\begin{equation}
    [\log^2V]\,\widehat h_p^+(V)
    =
    [\log^2V]\,\widehat h_p^-(V)
    \equiv
    \widehat L_p(V)
    =
    \frac{P_p(V)}
    {36(1-V)^{2p-2}},
    \label{eq:Lp}
\end{equation}
where
\begin{equation}
    P_p(V)
    =
    \sum_{k=0}^{p-2}
    \binom{p-1}{k+1}
    \binom{p-2}{k}
    V^k =
(p-1)\,{}_2F_1(2-p,2-p;2;V).
    \label{eq:Pp}
\end{equation}

\paragraph{Step 3: Triangular recursion for the Mellin coefficients.}
The last step is to determine the coefficients of the reduced simultaneous-pole
ansatz from the highest crossed-channel logarithms. Since the
$\log^2U\,\log^2V$ terms probe simultaneous poles in the $s$ and $t$
channels, it is sufficient at this stage to consider the common $st$ kernel
$B(s,t)$. Its contribution to the full $\alpha^2$ Mellin amplitude is
obtained by the Ward lift \cite{Chester:2025jxg},
\begin{align}
    \mathcal D_2 B(s,t)
    ={}&
    (2-s)^2(4-s)\,B(s-4,t)
    +(2-s)(3-s)(2-t)\,B(s-2,t-2)
    \nonumber\\
    &+
    (3-s)(s-2)(2-s-t)\,B(s-2,t).
    \label{eq:D2}
\end{align}
To make the matching explicit, suppose that near a simultaneous
double-trace location the full $\alpha^2$ Mellin amplitude behaves as
\begin{equation}
    [\alpha^2]\mathcal M^{\rm full}(s,t)
    \sim
    \frac{R_{K,L}}
    {(s-2K)(t-2L)} .
    \label{eq:full-simultaneous-pole}
\end{equation}
The Mellin measure \eqref{eq:mellin-kernel} has double poles at the same locations, so the full
integrand has triple poles in both $s$ and $t$. Their inverse Mellin
transform therefore produces
\begin{equation}
    \rho_{K,L}\,R_{K,L}\,
    U^K V^{L-1}\log^2U\,\log^2V,
    \qquad
    \rho_{K,L}
    =
    \frac{\Gamma(K+L-1)^2}
    {16\,[(K-1)!]^2[(L-1)!]^2},
    \label{eq:rho-KL}
\end{equation}
where $\rho_{K,L}$ follows from the Mellin measure.

On the CFT side, the $\log^2U\,\log^2V$ part of the full
$\alpha^2$ correlator is encoded in the functions $\widehat L_p(V)$ \eqref{eq:Lp}. Expanding
\begin{equation}
    \widehat L_p(V)
    =
    \sum_{r\geq0}\widehat L_{p,r}V^r,
    \label{eq:Lpr-definition}
\end{equation}
the coefficient $\widehat L_{p,r}$ multiplies
$U^pV^r\log^2U\,\log^2V$ in the full correlator. The Mellin and CFT descriptions are therefore related by
\begin{equation}
    \widehat L_{K,L-1}
    =
    \rho_{K,L}\,\widehat R_{K,L},
    \label{eq:L-R-matching}
\end{equation}
where we have rescaled the coefficients by a common factor of \(\pi^4\), denoting the resulting quantities by \(\widehat c_{mn}\) and \(\widehat R_{K,L}\), respectively. Combining the full simultaneous-pole coefficient generated by
\eqref{eq:D2} with \eqref{eq:L-R-matching} gives
\begin{align}
    \widehat L_{m+1,n-1}
    ={}&
    \rho_{m+1,n}
    \bigg[
    -8m^2(m-1)\,\widehat c_{m-1,n}
    -4m(2m-1)(n-1)\,\widehat c_{m,n-1}
    \nonumber\\
    &\hspace{1.6cm}
    +4m(2m-1)(m+n)\,\widehat c_{m,n}
    \bigg],
    \label{eq:triangular-recursion}
\end{align}
with
\begin{equation}
    \rho_{m+1,n}
    =
    \frac{\Gamma(m+n)^2}
    {16(m!)^2((n-1)!)^2},
    \qquad
    \widehat c_{0,n}
    =
    \widehat c_{m,0}
    =
    0.
\end{equation}
Since the first two terms in \eqref{eq:triangular-recursion} involve coefficients with lower indices, while the last contains the new coefficient $\widehat c_{m,n}$, the system is triangular and determines the residue
matrix recursively. The resulting coefficients admit the compact symmetric expression
\begin{equation}
    \widehat c_{mn}
    =
    \frac{1}{9} 
    \left[
    \frac{
    \sqrt{\pi}\,
    \Gamma(m)\Gamma(n)
    \Gamma\!\left(m+n+\frac12\right)}
    {
    \Gamma\!\left(m+\frac12\right)
    \Gamma\!\left(n+\frac12\right)
    \Gamma(m+n+1)}
    -
    \frac{1}{mn}
    \right].
    \label{eq:cmn-symmetric}
\end{equation}
The manifest symmetry $\widehat c_{mn}=\widehat c_{nm}$ is a highly non-trivial consistency check of the ansatz. It is not imposed by the triangular recursion itself: the recursion determines entries row by row
from the $U$-slice data. The fact that the resulting matrix is symmetric is therefore strong evidence
that the full $\alpha^2$ data really come from a crossing-symmetric reduced Mellin object. Since the two
parity sectors have the same highest-logarithmic data
\eqref{eq:Lp}, the same residue matrix $\widehat c_{mn}$ governs both.

Although the
coefficients $\widehat c_{mn}$ were fixed using only the
$\log^2U\log^2V$ terms, after the Ward lift and inverse Mellin transform
the reconstructed amplitudes reproduce the complete functions
$F_{\rm even}(U,V)$ and $F_{\rm odd}(U,V)$, including the
single-logarithmic and regular terms in $V$. We have checked the
reconstruction in exact arithmetic for both parity sectors: for
$p=2,\ldots,10$, the inverse Mellin transform of the Ward-lifted kernels
$\mathcal D_2\bigl[B(s,t)\pm B(s,\widetilde u)\bigr]$, with
$\widetilde u=-s-t$, agrees with the independently computed CFT block
sum through order $(1-V)^{12}$, including the full slice rather than
only its $\log^2V$ coefficient. The closed position-space expressions
for $p=2,\ldots,7$ also agree identically with the inverse Mellin
result. These checks support the conclusion that no additional reduced
single-pole contribution is required.

\subsection{Flavor projection and complete one-loop amplitude}
\label{subsec:complete-one-loop}
We now restore the flavor structure and crossing symmetry and assemble the
complete reduced one-loop amplitude. The Mellin variables of the full and
reduced correlators obey different constraints. For the full correlator,
$s+t+u=4$, while the modified Mellin transform defining the reduced
amplitude is naturally parametrized by
\begin{equation}
    \widetilde u\equiv u-4,
    \qquad
    s+t+\widetilde u=0 ,
    \label{eq:reduced-mellin-constraint}
\end{equation}
in terms of which the crossing properties of the reduced amplitude take their simplest form \cite{Chester:2025jxg}. Using the common kernel $B(s,t)$ reconstructed above, we introduce the crossed parity
combinations entering the $s$-channel cut,
\begin{equation}
    f_{\rm even}
    \equiv
    B(s,t)+B(s,\widetilde u),
    \qquad
    f_{\rm odd}
    \equiv
    B(s,t)-B(s,\widetilde u).
    \label{eq:parity-reduced-kernels}
\end{equation}
Recall from \eqref{cases} that the tree-level flavor operator
$\mathbb C_\ell$ takes different forms for even and odd spins.
The one-loop unitarity relation involves the squares of the
corresponding flavor eigenvalues. Combining these with the
even- and odd-spin reduced kernels, and using the Jacobi identity, the $s$-channel cut can be reorganized in terms of color-box tensors:
\begin{equation}
    \mathbb C_{\rm even}^{\,2}f_{\rm even}
    -
    \mathbb C_{\rm odd}^{\,2}f_{\rm odd}
    =
    9\left[
    \mathtt d_{st}B(s,t)
    +
    \mathtt d_{su}B(s,\widetilde u)
    \right],
    \label{eq:s-channel-cut-color-boxes}
\end{equation}
where
\begin{equation}
    \mathtt d_{st}^{I_1I_2I_3I_4}
    \equiv
    f^{JI_1K}f^{KI_2L}f^{LI_3M}f^{MI_4J},
    \label{dst}
\end{equation}
along with the crossed expressions. The complete non-local reduced
one-loop amplitude is 
\begin{equation}
    \widetilde{\mathcal M}_{\rm cut}^{\rm 1-loop}(s,t)
    =
    9
    \left[
    \mathtt d_{st}B(s,t)
    +
    \mathtt d_{su}B(s,\widetilde u)
    +
    \mathtt d_{tu}B(t,\widetilde u)
    \right],
    \label{eq:complete-reduced-one-loop-amplitude}
\end{equation}
where the kernel $B$ is given by
\eqref{eq:conceptual-reduced-mellin-ansatz}, with coefficients
\eqref{eq:cmn-symmetric}. This is the same color-box organization found in
holographic studies.

As a final remark, notice that the double-discontinuity determines the non-local part of the reduced
amplitude but is insensitive to regular Mellin polynomials. In
appendix~\ref{app:local-ambiguities} we argue that such polynomial
local completions are excluded: requiring that the large-radius limit
of the correlator reproduces the flat-space one-loop amplitude, any
polynomial term would produce a power-law growth in $L$ that the
amplitude does not have. Protected OPE data provide
an independent check at finite AdS radius. We therefore take
\begin{equation}
    \widetilde{\mathcal M}^{\rm 1-loop}_{\rm tot}
    =
    \widetilde{\mathcal M}^{\rm 1-loop}_{\rm cut}.
\end{equation}
The full R-symmetry-dependent amplitude is obtained by the Ward lift,
\begin{equation}
    \mathcal M^{\rm 1-loop}_{\rm full}(s,t;\alpha)
    =
    \sum_{J=0}^{2}
    \alpha^J\,
    \mathcal D_J
    \widetilde{\mathcal M}^{\rm 1-loop}_{\rm tot}(s,t),
    \label{eq:full-ward-lift}
\end{equation}
where the finite-difference operators $\mathcal D_J$ implement the superconformal Ward identities. The highest R-symmetry operator $\mathcal D_2$ was given in \eqref{eq:D2}, while the remaining two read \cite[eq.~(B.11)]{Chester:2025jxg}
\begin{align}
\mathcal D_0={}&
(s+t)(s+t-2)^2 P_{0,0}
+(2-s)(s+t-1)(s+t-2)P_{-2,0}
\nonumber\\
&+
(2-t)(s+t-1)(s+t-2)P_{0,-2},
\label{eq:D0} 
\\[1mm]
\mathcal D_1={}&
-(s+t)(s+t-2)^2 P_{0,0}
+(t-2)(s+2t-4)(s+t-2)P_{0,-2}
\nonumber\\
&+
(s-2)(t+2)(s+t-2)P_{-2,0}
+(s-2)(t-2)(s-t)P_{-2,-2}
\nonumber\\
&+
(s-2)^2(s-4)P_{-4,0}
+(4-t)(t-2)^2P_{0,-4},
\label{eq:D01}
\end{align}
where $P_{a,b}$ denotes the shift operator, $P_{a,b}f(s,t)\equiv f(s+a,t+b)$.

Notice that the simplicity of the reduced amplitude does not imply the
absence of single poles in the full Mellin amplitude. The finite-difference
Ward lift can generate single-pole terms and Mellin-variable-dependent
residues from the simultaneous-pole kernel.

%%%%%%%%%%%%%%%%%%%%%%%%%%%%%%%%%%%%%%%%%%%%%%%%%%%%%%%%%%%%%%%%%%%
\section{The flat-space limit}
\label{sec:flat-limit}

We now study the large-radius limit of the reconstructed one-loop amplitude. The external moment-map operators select bulk scalar states belonging to the same four-dimensional $\mathcal N=4$ SYM multiplet as the gluons. Their effective
AdS mass satisfies $m^2=-2/L^2$, so these states become massless in the flat-space limit. We denote the Mellin variables by $s,t,u$, while
$S,T,U$ are the four-dimensional Mandelstam invariants,
\begin{equation}
S=-(p_1+p_2)^2,\qquad
T=-(p_1+p_4)^2,\qquad
U=-(p_1+p_3)^2,
\qquad S+T+U=0.
\end{equation}
In this section, $U$ always denotes the Mandelstam invariant, not the
conformal cross-ratio of the previous sections, which does not appear
here. All external momenta are incoming and obey $p_i^2=0$.
We use a mostly-plus Lorentzian signature. 

The prescription originally proposed in \cite{Penedones:2010ue} extracts a local flat-space scattering amplitude from the large-radius limit of the Mellin representation of an AdS boundary correlator,
\begin{equation}
 \displaystyle
 \mathcal A_4(S,T;\alpha)
 \equiv8\pi^2\lim_{L\to\infty}
 \mathcal I_\beta\!\left[
   \mathcal M\!\left(
     \frac{L^2S}{2\beta},\frac{L^2T}{2\beta};\alpha
   \right)
 \right],
 \label{eq:flat-transform-normalized}
\end{equation}
where we have stripped the momentum-conserving delta function and the
external polarization factor $R_s$, and 
\begin{equation}
 \qquad
 \mathcal I_\beta[F]
 \equiv\sqrt\pi\int_{\mathcal C}\frac{d\beta}{2\pi i}
       e^\beta\beta^{-1/2}F(\beta).
 \label{eq:flat-beta-transform}
\end{equation}
Following \cite{Fitzpatrick:2011hu}, the contour $\mathcal C$ is
oriented upward and chosen to lie to the right of the poles inherited from
the Mellin amplitude and of the branch cut generated by $\beta^{-1/2}$.
We use the principal branch, $-\pi<\arg\beta<\pi$, with the cut along the
negative real axis, and the same branch for $\log\beta$ below. Introducing 
\begin{equation}
    \lambda\equiv\frac{L^2}{2\beta},
\end{equation}
for fixed flat-space kinematics, the Mellin amplitude is evaluated at
\begin{equation}
 s=\lambda S,\qquad t=\lambda T,\qquad
 u=4+\lambda U.
 \label{eq:flat-scaling}
\end{equation}
As the radius of AdS goes to infinity, $ \lambda \to \infty$ and the transform probes the simultaneous large-Mellin regime at fixed ratios $s/t=S/T$. However, in the present massless one-loop problem, the strict limit in \eqref{eq:flat-transform-normalized} is infrared divergent. We therefore
define the auxiliary large-radius asymptotic transform
\begin{equation}
 \mathcal A_{4,L}(S,T;\alpha)
 \equiv8\pi^2\mathcal I_\beta\!\left[
   \mathcal M\!\left(
     \frac{L^2S}{2\beta},\frac{L^2T}{2\beta};\alpha
   \right)
 \right]_{\rm asymp}.
\end{equation}
We first take the large-$L$ asymptotic expansion of the integrand at fixed $\beta$ and fixed $S,T,U$, and then compute the inverse Laplace moments. At large but finite $L$, the result is finite, and in this sense the AdS radius regulates the infrared. We stress, however, that $L$ is not a parameter of the boundary correlator: the theory is conformal, and the Mellin amplitude $\mathcal M(s,t;\alpha)$ does not depend on $L$. The radius enters only through the scaling \eqref{eq:flat-scaling}, which converts Mellin variables into flat-space momenta, so that the logarithms of $L$ found below are logarithms of large Mellin variables. The order in which the Mellin sums, the Ward lift, and this asymptotic expansion must be performed is important and will be specified below. 

We first derive the large-Mellin asymptotics in a region that does not cross the Mellin poles, and then analytically continue the result to physical Lorentzian kinematics.

\paragraph{Reduced versus full Mellin amplitude.}
Recall the one-loop reduced kernel
\begin{equation}
 B(s,t)
 =\sum_{m,n\geq1}\frac{c_{mn}}
 {(s-2m)(t-2n)},
 \qquad
 c_{mn}=\frac{1}{9\pi^4}\left(a_{mn}-\frac1{mn}\right),
 \label{eq:flat-reduced-kernel}
\end{equation}
where we have defined
\begin{equation}
 a_{mn}\equiv
 \frac{\sqrt\pi\,\Gamma(m)\Gamma(n)
       \Gamma(m+n+\frac12)}
 {\Gamma(m+\frac12)\Gamma(n+\frac12)\Gamma(m+n+1)}.
 \label{eq:flat-a-coefficients}
\end{equation}
The discrete AdS poles have a direct flat-space interpretation.
At fixed $\beta$, the scaling $s=\lambda S$ maps the pole
$s=2m$ to $S_m=2m/\lambda$. The spacing between neighboring
poles is therefore
\begin{equation}
    S_{m+1}-S_m=\frac{2}{\lambda}
    \xrightarrow[\lambda\to\infty]{}0,
    \label{eq:flat-pole-condensation}
\end{equation}
and similarly in the other channels. As the curvature is removed,
the discrete AdS poles become dense in $S$. Conversely, probing a
fixed nonzero value of $S$ as $\lambda \to \infty$ requires $m\sim\lambda S/2$, so that
the relevant pole levels become parametrically large. For
simultaneous poles, the corresponding scaling region has
$m,n=O(\lambda)$. This motivates a continuum approximation to
the large-level pole sums, whose resummation is expected to
reproduce the logarithmic non-analyticities and branch cuts
characteristic of flat-space loop amplitudes.

There is, however, an important subtlety: the leading continuum
approximation to the reduced amplitude does not necessarily
capture the full flat-space limit. The outcome also depends on
the pole residues and, crucially, on the action of the
superconformal Ward identities. The flat-space scaling \eqref{eq:flat-scaling} is formulated in
terms of the full Mellin variables, whereas the reduced amplitude
introduced in the preceding section uses a shifted third
variable, $\widetilde u=u-4$. Although this constant shift is
subleading at large Mellin variables, the relation between the
reduced and full amplitudes involves a non-trivial finite-difference Ward
operator, whose action must be taken into account.

To see why this matters, consider the asymptotic behavior of
the reduced coefficients. At simultaneously large $m,n$ with
fixed ratio, we find
\begin{equation}
a_{mn}\sim\frac{\sqrt\pi}{\sqrt{mn(m+n)}}.
\label{eq:flat-a-density}
\end{equation}
Replacing the double sum by its continuum approximation then
gives a leading contribution to the reduced Mellin amplitude
that is homogeneous of degree $-3/2$, as in the five-dimensional
massless box of \cite{Alday:2021ajh}.

In many familiar holographic examples, the superconformal Ward
difference operator reduces, in the flat-space limit, to
multiplication by a universal kinematic polynomial
\cite{Alday:2019nin}. Here, instead, its leading large-Mellin
action retains a differential operator. Defining
$\mathcal E\equiv s\partial_s+t\partial_t$, we find
\begin{equation}
\mathcal D_0+\alpha\mathcal D_1+\alpha^2\mathcal D_2
\sim
(\widetilde u+\alpha s)^2(3+2\mathcal E).
\label{eq:flat-leading-full-ward}
\end{equation}
In particular, for a homogeneous function $f_{-3/2}$,
\begin{equation}
(3+2\mathcal E)f_{-3/2}=0.
\label{eq:flat-annihilation}
\end{equation}
Thus, the leading continuum contribution to the reduced amplitude
is annihilated by the leading Ward operator in every
R-symmetry component. The naive leading large-Mellin behavior
of the reduced amplitude is therefore insufficient to determine
the physical four-dimensional result. Subleading contributions, including those
associated with the $1/(mn)$ term and the endpoints of the
pole sums, cannot be discarded a priori. To retain their
combined effect, we must first apply the exact Ward lift and
resum the resulting full Mellin amplitude, and only then take
the large-Mellin limit.

\paragraph{Exact resummation of the full amplitude.}
Let us define the \(\alpha^2\)-component of the elementary Ward-lifted \(st\) block
\begin{equation}
 \widehat{\mathcal W}_{st}(s,t)
 \equiv\pi^4\mathcal D_2B(s,t).
 \label{eq:flat-hatted-block}
\end{equation}
We find the exact resummation
\begin{equation}
 \widehat{\mathcal W}_{st}(s,t)
 =(s-2)\left[
  \frac19 H\!\left(1-\frac{s}{2}\right)
             \frac{H(-t/2)}{t}
  +\tau(t)
 \right],
 \label{eq:flat-exact-W}
\end{equation}
where $H(z)=\psi(z+1)+\gamma_E$ is the analytic continuation of the harmonic numbers, \(\psi(z)=\Gamma'(z)/\Gamma(z)\) is the digamma function, and \(\gamma_E\) is the Euler–Mascheroni constant. The one-variable completion $\tau(t)$ admits the absolutely convergent pole representation
\begin{equation}
 \tau(t)=\frac19\sum_{n=1}^{\infty}\frac{b_n}{t-2n},
 \qquad
 b_n=\frac1{n^2}\left(\frac{4^n}{\binom{2n}{n}}-1\right).
 \label{eq:flat-tau-poles}
\end{equation}
Equivalently, it can be written in closed hypergeometric
form as 
\begin{align}
 \tau(t)={}
 -\frac{2}{9t^2}H\!\left(-\frac t2\right)
 +\frac{\pi^2}{27t}
 +\frac{2\pi^{3/2}}{9t^2}
       \frac{\Gamma(\frac12-\frac t2)}{\Gamma(-\frac t2)}
 -\frac{4}{9(t-1)t}
 {}_3F_2\!\left(
 \begin{matrix}\frac12,\frac12,1\\[1mm]
 \frac32,\frac32-\frac t2\end{matrix};1\right).
 \label{eq:flat-tau-hypergeometric}
\end{align}
The product of harmonic functions in \eqref{eq:flat-exact-W} carries the simultaneous poles, while
$\tau$ contains the simple-pole terms generated by the same Ward lift. It is
therefore fixed by the reduced kernel and is not an independent addition. The pole representation implies \(t\,\tau(t)\to\pi^2/27\) for large negative \(t\). Together with the large-argument expansion of the harmonic functions, this gives the flat-space scaling
\begin{equation}
 \widehat{\mathcal W}_{st}(\lambda S,\lambda T)
 =\frac19\frac ST
   \left[\mathcal L_S\mathcal L_T+\frac{\pi^2}{3}\right]
   +o(1),
 \label{eq:flat-W-asymptotics}
\end{equation}
where we have defined
\begin{equation}
 \mathcal L_X(\beta,L)
 \equiv\log\!\left(\frac{L^2(-X-i0)}{4}\right)-\log\beta+\gamma_E,
 \qquad X\in\{S,T,U\}.
 \label{eq:flat-beta-log}
\end{equation}
The other two R-symmetry components do not require a separate
resummation. Writing $\widehat{\mathcal W}_J\equiv\pi^4\mathcal D_JB$, so that
$\widehat{\mathcal W}_2=\widehat{\mathcal W}_{st}$, the symmetry
$B(s,t)=B(t,s)$ and the explicit form of the Ward operators imply the
exact identities
\begin{equation}
\begin{split}
\widehat{\mathcal W}_0(s,t)
={}&(s+t-2)\left[
\frac{\widehat{\mathcal W}_2(s+2,t)}{s}
+\frac{\widehat{\mathcal W}_2(t+2,s)}{t}\right], \\
\widehat{\mathcal W}_1(s,t)
={}&\widehat{\mathcal W}_2(t,s)-\widehat{\mathcal W}_2(s,t)
-\widehat{\mathcal W}_0(s,t).
\label{eq:W0W1-identity}
\end{split}
\end{equation}
Together with \eqref{eq:flat-exact-W}, they determine the other two
components exactly, including their single-pole completions. In the
flat-space scaling, \eqref{eq:flat-W-asymptotics} then gives
\begin{equation}
\pi^4
\left.
\left[
\left(
\mathcal D_0
+\alpha\,\mathcal D_1
+\alpha^2\,\mathcal D_2
\right)
B(s,t)
\right]
\right|_{s=\lambda S,\;t=\lambda T}
=
\frac{K(S,T;\alpha)}{9ST}
\left[
\mathcal L_S\mathcal L_T
+\frac{\pi^2}{3}
\right]
+o(1),
\end{equation}
where $K(S,T;\alpha)=(U+\alpha S)^2$. The crossed blocks have this same overall scalar kinematic factor, with the pair \(ST\) in the denominator and logarithms replaced by \(SU\) and \(TU\), respectively. The equation above makes the four-dimensional box structure manifest.
Although the $\alpha^2$-component in
\eqref{eq:flat-W-asymptotics} scales as $S/T$, the Ward lift supplies
the scalar kinematic factor, with $[\alpha^2]K=S^2$. After stripping
this factor, the result is proportional to $1/(ST)$, as required for
the four-dimensional massless scalar box.

\subsection{The AdS radius as an infrared regulator}
\label{subsec:flat-AdS-regulator}
Applying the flat-space transform to the resummed Mellin block \eqref{eq:flat-W-asymptotics} fixes the dimensionless AdS-regulated box to
\begin{equation}
 \displaystyle
 \mathcal I^{\rm AdS}_{XY,L}
 =\Lambda_X(L)\Lambda_Y(L)-\frac{\pi^2}{6}
 \qquad
 (XY)=(ST),(SU),(TU),
 \label{eq:flat-AdS-box}
\end{equation}
where we have defined 
\begin{equation}
 \Lambda_X(L)\equiv\log\,\bigl(L^2(-X-i0)\bigr)+2\gamma_E.
 \label{eq:flat-Lambda}
\end{equation}
At fixed kinematics, the box diverges logarithmically as $L\to\infty$: the radius $L$ plays the role of an infrared regulator. The constants that accompany the logarithms, namely the $2\gamma_E$ in \eqref{eq:flat-Lambda} and the $-\pi^2/6$ in \eqref{eq:flat-AdS-box}, are tied to the prescription \eqref{eq:flat-transform-normalized}--\eqref{eq:flat-beta-transform} that converts Mellin variables into flat-space momenta. They should be regarded as part of the definition of the regulator, in the same way as the scheme-dependent constants of dimensional regularization. The regulator-independent information is the structure of the divergent terms and the kinematic dependence of the finite remainder. Restoring the flavor dressing and coupling normalization gives the one-loop regularized amplitude
\begin{equation}
 \displaystyle
 \mathcal A^{\rm 1-loop}_{4,L}
 =\frac{8(C_{222})^4}{\pi^2}\,
 K(S,T;\alpha)
 \sum_{(XY)}\frac{\mathtt d_{XY}}{XY}
 \mathcal I^{\rm AdS}_{XY,L}+o(1).
 \label{eq:flat-full-result}
\end{equation}
Here $\mathtt d_{ST}\equiv\mathtt d_{st}$, defined in \eqref{dst}, and similarly in the crossed
channels. This has the structure of the one-loop four-point amplitude
of $\mathcal N=4$ SYM in flat space \cite{Green:1982sw}: a universal
kinematic factor multiplying a sum of scalar box integrals, one for
each pair of channels, dressed by the color factors $\mathtt d_{XY}$.

\paragraph{AdS minimal-subtraction scheme.} Now, to isolate the dependence on the geometric infrared regulator from the
kinematic logarithms, let us introduce an arbitrary reference scale $\mu$.\footnote{The reference scale \(\mu\) is auxiliary and does not introduce a physical scale into the conformal boundary theory. Its dependence cancels in the complete AdS-regulated amplitude, while the infrared-subtracted finite remainder is scheme dependent.} We define
\begin{equation}
 \eta\equiv\log(L^2\mu^2),\qquad
 \ell_X\equiv\log\!\left(\frac{-X-i0}{\mu^2}\right).
 \label{eq:flat-geometric-log}
\end{equation}
The quantity $\eta$ contains all the dependence on the AdS radius. In particular, $\eta\to+\infty$ when $L\to\infty$, namely when the AdS infrared regulator
is removed.  On the other hand, $\ell_X$ contains only the dependence on the
flat-space kinematics. Since $\Lambda_X(L)=\eta+\ell_X+2\gamma_E$, equation \eqref{eq:flat-AdS-box} becomes
\begin{align}
\mathcal I^{\rm AdS}_{XY,L}
={}\eta^2+\eta(\ell_X+\ell_Y+4\gamma_E)+(\ell_X+2\gamma_E)(\ell_Y+2\gamma_E)
-\frac{\pi^2}{6}.
\label{eq:flat-geometric-expansion}
\end{align}
We define a natural geometric minimal-subtraction scheme by removing the
terms that diverge with $\eta$,
\begin{equation}
 \mathcal S^{\rm AdS}_{XY}
 \equiv\eta^2+\eta(\ell_X+\ell_Y+4\gamma_E),
 \label{eq:flat-AdS-MS-subtraction}
\end{equation}
which leaves the finite remainder
\begin{equation}
 \mathcal H^{\rm AdS}_{XY}(\mu)
 =(\ell_X+2\gamma_E)(\ell_Y+2\gamma_E)-\frac{\pi^2}{6}.
 \label{eq:flat-AdS-MS-remainder}
\end{equation}

\paragraph{Universal infrared structure.}
The divergent terms \eqref{eq:flat-AdS-MS-subtraction} have the color
structure required by infrared factorization. The color-box tensors obey
\begin{equation}
 \mathtt d_{st}-\mathtt d_{tu}=\frac{C_A}{2}\,\mathtt c_t,
 \qquad
 \mathtt d_{su}-\mathtt d_{tu}=-\frac{C_A}{2}\,\mathtt c_u,
 \label{eq:box-differences}
\end{equation}
where we recall that $C_A$ is the quadratic Casimir of the adjoint representation. Moreover, the flat-space limit of the
tree-level amplitude \eqref{eq:mellin-color-decomposition} is
\begin{equation}
 \mathcal A_4^{\rm tree}
 =-8(C_{222})^2\,K(S,T;\alpha)
 \left(\frac{\mathtt c_t}{ST}-\frac{\mathtt c_u}{SU}\right).
 \label{eq:flat-tree}
\end{equation}
Using $S+T+U=0$, the divergent part of \eqref{eq:flat-full-result} can
then be written as
\begin{equation}
 \mathcal A^{\rm 1-loop}_{4,L}\Big|_{\rm div}
 =\frac{(C_{222})^2}{\pi^2}
 \sum_{j=2}^{4}\mathbf T_1\!\cdot\mathbf T_j
 \left[\frac12\eta^2+\eta\left(\ell_{1j}+2\gamma_E\right)\right]
 \mathcal A_4^{\rm tree},
 \label{eq:flat-dipole}
\end{equation}
where $\ell_{12}\equiv\ell_S$, $\ell_{14}\equiv\ell_T$,
$\ell_{13}\equiv\ell_U$, and
$\mathbf T_i\cdot\mathbf T_j\equiv\sum_J\mathbf T_i^J\mathbf T_j^J$, with
$\mathbf T_i^J$ the adjoint generator $(\mathbf T^J)_{IK}=-if^{JIK}$
acting on the color index of the $i$-th particle. Color conservation
gives 
\begin{equation}
\sum_{j=2}^{4}\mathbf T_1\!\cdot\mathbf T_j=-C_A, 
\end{equation}
so that the double
logarithm is proportional to the tree-level amplitude,
\begin{equation}
 \mathcal A^{\rm 1-loop}_{4,L}\Big|_{\eta^2}
 =-\frac{(C_{222})^2\,C_A}{2\pi^2}\,\eta^2\,\mathcal A_4^{\rm tree},
 \label{eq:flat-double-log}
\end{equation}
while the single logarithms have the color-dipole form familiar from
the infrared structure of gauge-theory amplitudes
\cite{Catani:1998bh,Becher:2009qa,Gardi:2009qi}. This holds for any gauge group
and does not rely on the planar limit.

\paragraph{Comparison with dimensional regularization.}
Translating the AdS infrared regularization to a standard scheme such as
dimensional regularization is not automatic.
In general, there is no universal identification of the geometric
logarithm with a Laurent pole in $\epsilon_{\rm IR}$. We now present an explicit comparison between two infrared
subtraction schemes.

Let $D=4-2\epsilon_{\rm IR}$. Factoring out the standard
kinematic and one-loop prefactor, we define the dimensionless
box function $\mathcal I^{\rm DR}_{XY}$ by
\begin{equation}
 I_{\rm box}^{\rm DR}(X,Y)
 \equiv\frac{\mathcal I^{\rm DR}_{XY}}{8\pi^2XY}.
\end{equation}
In the dimensionally regulated normalization of \cite{Bern:1992em,Duplancic:2000sk}, the massless scalar box reads\footnote{We include the normalization factor
$r_\Gamma(\epsilon_{\rm IR})
\equiv e^{\gamma_E\epsilon_{\rm IR}}
\frac{\Gamma(1+\epsilon_{\rm IR})
\Gamma^2(1-\epsilon_{\rm IR})}
{\Gamma(1-2\epsilon_{\rm IR})}
=1-\frac{\pi^2}{12}\epsilon_{\rm IR}^2
+O(\epsilon_{\rm IR}^3)$. The quadratic term multiplies the leading double infrared pole and contributes to the finite part of the box.}
\begin{equation}
 \mathcal I^{\rm DR}_{XY}
 =\frac{2}{\epsilon_{\rm IR}^2}
 -\frac{\ell_X+\ell_Y}{\epsilon_{\rm IR}}
 +\ell_X\ell_Y-\frac{2\pi^2}{3}
 +O(\epsilon_{\rm IR}).
 \label{eq:flat-DR-box-expanded}
\end{equation}
Dimensional minimal subtraction is defined
by removing only the Laurent poles. The resulting finite remainder is
\begin{equation}
 \mathcal H^{\rm DR}_{XY}(\mu)
 =\ell_X\ell_Y-\frac{2\pi^2}{3}.
 \label{eq:flat-DR-MS}
\end{equation}
To compare with the finite remainder in the AdS scheme, we allow for a finite conversion
between the reference scales of the two regulators. We write
\begin{equation}
 \ell_X^{\rm AdS}
 \equiv\log\!\left(\frac{-X-i0}{\mu_{\rm AdS}^2}\right),
 \qquad
 \ell_X^{\rm DR}
 \equiv\log\!\left(\frac{-X-i0}{\mu_{\rm DR}^2}\right).
\end{equation}
We choose $ \mu_{\rm DR}=e^{-\gamma_E}\mu_{\rm AdS}$. This aligns the kinematic logarithms, while clearly the remaining constant depends on the infrared subtraction scheme,
\begin{equation}
    \mathcal{H}_{XY}^{{\rm AdS}}(\mu_{{\rm AdS}})=\mathcal{H}_{XY}^{{\rm DR}}(e^{-\gamma_E}\mu_{{\rm AdS}})+\frac{\pi^2}{2}.
\end{equation}

\paragraph{Comparison with the Higgs regulator.}
It is also instructive to compare with the Higgs regulator of
\cite{Alday:2009zm}, in which planar color-ordered amplitudes are
regulated by masses $m$ on the Coulomb branch. In the conventions of
that reference, the one-loop four-point amplitude is governed by the
box function
\begin{equation}
 I^{(1)}(s,t;m^2)
 =2\log\frac{m^2}{s}\log\frac{m^2}{t}-\pi^2+O(m^2),
 \label{eq:flat-Higgs-box}
\end{equation}
written in the Euclidean region, where the logarithms are real. This
has the same bilinear form as \eqref{eq:flat-AdS-box}. Identifying
their $s,t$, which are positive in the Euclidean region, with
$-X,-Y$, we find
\begin{equation}
 \mathcal I^{\rm AdS}_{XY,L}
 =\frac12\,I^{(1)}\Big|_{m^2=e^{-2\gamma_E}/L^2}+\frac{\pi^2}{3}+o(1),
 \label{eq:flat-AdS-vs-Higgs}
\end{equation}
so that the AdS radius plays the role of an inverse mass,
$m\leftrightarrow e^{-\gamma_E}/L$, and the two regulators differ by a
finite scheme constant. The overall normalization agrees as well. 

In the planar limit, each color-ordered amplitude receives a single
box, whose coefficient relative to the tree-level amplitude is
given by \eqref{eq:flat-double-log}. For $SU(N)$, using $C_A=2\kappa N$ and
\eqref{eq:C222-gym}, this becomes, to leading order
in the coupling,
\begin{equation}
-\frac{(C_{222})^2C_A}{2\pi^2}
=-\frac{g_{\rm YM}^2N}{4\pi^2}=-a, \qquad a\equiv\frac{g^2N}{8\pi^2},
\qquad g^2=2g_{\rm YM}^2,
\end{equation}
where $g$ is the conventionally normalized gauge coupling. Given
\eqref{eq:flat-AdS-vs-Higgs}, this reproduces the coefficient of
$I^{(1)}$ in the planar amplitude $M_4=1-\frac a2I^{(1)}+\dots$ of
\cite{Alday:2009zm}. There is, however, an important difference of
scope. The Higgs regulator renders finite only planar color-ordered
amplitudes: beyond the planar limit, the massless fields of the
unbroken gauge group run in loops and their infrared divergences
remain. The AdS-regulated result \eqref{eq:flat-full-result}, by
contrast, holds for the complete color structure.

%%%%%%%%%%%%%%%%%%%%%%%%%%%%%%%%%%%%%%%%%%%%%%%%%%%%%%%%%%%%%%%%
\section{Discussion}
\label{sec:discussion}
%%%%%%%%%%%%%%%%%%%%%%%%%%%%%%%%%%%%%%%%%%%%%%%%%%%%%%%%%%%%%%%%
We have studied the four-point function of boundary moment-map
operators in four-dimensional $\mathcal N=4$ SYM on fixed
$\AdS_4$ with Dirichlet boundary conditions. Starting from the
microscopic theory, we have reproduced the $d=3, \, p=2$ tree-level supergluon amplitude of \cite{Alday:2021odx} and then used it to reconstruct the non-local one-loop
correlator via AdS unitarity. Remarkably, the double-discontinuity
computed from the full tree-level correlator admits a simple
reduced Mellin representation involving only simultaneous poles,
with manifestly symmetric coefficients. The agreement with the complete one-loop double-discontinuity
and the flat-space limit provides non-trivial checks of this
structure. A further independent check at finite AdS radius
comes from the matching of protected OPE coefficients, as
discussed in appendix~\ref{app:local-ambiguities}.

The main conceptual lesson of our calculation is an explicit
realization of the AdS radius as an infrared regulator. The boundary correlator
itself is infrared safe and independent of $L$; the radius enters only
when the Mellin variables are converted into flat-space momenta. In
the resulting large-radius limit, the discrete Mellin poles condense
into the branch cuts of the four-dimensional massless box, and the
infrared divergences of the flat-space amplitude reappear as
logarithms of $L$, with the regulated one-loop answer containing both
$\log^2 L$ and $\log L$ terms. Their color structure is the universal
one dictated by infrared factorization, for any gauge group. The
comparison with dimensional regularization and with the Higgs
regulator makes this correspondence quantitative, including the
finite differences between the infrared subtraction schemes. Our
result thus provides a concrete example of how the infrared
singularities of a massless gauge theory are regulated by placing the
theory on AdS and recovered as $L\to\infty$.

Beyond perturbation theory, the main virtue of this observable is that
nothing in its definition refers to weak coupling or to the planar
limit. For a given gauge group, the moment-map correlator is a single
function of the cross-ratios and of the complexified coupling $\tau$,
and the results of this paper describe its expansion around one corner
of the conformal manifold. Other corners can be reached by different
means. S-duality maps the Dirichlet condition to the coupling to
$T[G]$ \cite{Gaiotto:2008ak}, so that the strong-coupling regime is
described by a weakly coupled bulk theory interacting with a strongly
coupled boundary theory. At large $N$ and large 't~Hooft coupling,
$\mathcal N=4$ SYM on a half-space, or equivalently on $\AdS_4$
\cite{Aharony:2010ay}, with half-BPS boundary conditions is dual to
type IIB supergravity on the $\AdS_4\times S^2\times S^2\times\Sigma$
solutions of \cite{DHoker:2007a,DHoker:2007b}, as identified in
\cite{Aharony:2011}. Computing the moment-map four-point function in
the solution dual to the Dirichlet condition would give the
strong-coupling counterpart of the correlator studied here. In
between, the correlator is constrained by crossing symmetry and
unitarity, together with protected data that can be computed exactly.
The one-loop anomalous dimensions and OPE-coefficient corrections,
which can be extracted from our result, will provide, with the
tree-level data and the protected OPE coefficients, weak-coupling
benchmarks for the numerical conformal bootstrap of this correlator at
finite coupling, to which we will return in
\cite{BoundaryProtectedOPE:inprep}.

Within perturbation theory, a natural next step is to extend this
analysis to higher loops.
In flat space, infrared singularities of gauge-theory amplitudes
obey universal factorization and exponentiation properties.
It would be interesting to understand how these structures
are realized in AdS, and whether higher-loop infrared
divergences can be systematically organized in terms of
powers of $\log L$. In particular, one may ask how the
usual infrared anomalous dimensions and the associated
scheme dependence emerge in this geometric regularization.
This perspective may also provide
a useful starting point for studying infrared physics
beyond weak coupling. In particular, it is natural to conjecture that the double logarithms
of \eqref{eq:flat-dipole} exponentiate, as they do in flat space
\cite{Bern:2005iz,Becher:2009qa}. At finite coupling, the boundary
correlator would then be Sudakov suppressed at large Mellin variables
and fixed angle, with a color structure that reduces at one loop to
the dipole form found here. The exponent would be governed by a
function of the coupling whose leading term is fixed by our one-loop
result. In dimensional regularization, this function is the cusp
anomalous dimension
\cite{Korchemsky:1987wg,Beisert:2006ez,Henn:2019swt}, but it is not
regulator independent: on the Coulomb branch it interpolates between
the cusp and the octagon anomalous dimensions, which differ beyond
one loop \cite{Caron-Huot:2021usw,Belitsky:2022,Alday:2026}.
Which function arises on $\AdS_4$ is an open question. We have not
tested this beyond one loop.

It would also be interesting to obtain a closed position-space
representation of the complete one-loop correlator. Although the
individual $U$-slices of the double-discontinuity involve only ordinary
logarithms and dilogarithms, an integral representation of the resummed
answer reveals a richer analytic structure. As shown in
appendix~\ref{app:elliptic}, the inverse Mellin transform of the
one-variable completion $\tau(t)$ in \eqref{eq:flat-tau-poles} involves an algebraic kernel defining
a genus-one curve at generic cross-ratios. After a change of variables,
this curve coincides with the one appearing in the one-loop Witten
diagram of a conformally coupled scalar with $\Delta=1$ in $\AdS_4$
\cite{Heckelbacher:2022fbx,Stawinski:2023}. Whether an irreducible
elliptic contribution survives in the complete crossing-symmetric
correlator, and whether the answer can be expressed in terms of
elliptic multiple polylogarithms, remains to be understood.

Another direction is to investigate the one-loop
correlator directly in momentum space, providing an
independent approach to its analytic structure and
flat-space limit. Supersymmetry could also be used
to reconstruct correlators of the conserved-current
descendants of the moment map, allowing a more direct
comparison with gluon amplitudes
\cite{Bala:2026bdx}. It would furthermore be interesting
to extend the analysis to other supersymmetric boundary
conditions, possibly involving additional three-dimensional
degrees of freedom, and ultimately beyond supersymmetry.
In particular, generalizing the construction to pure
Yang--Mills theory on AdS would help identify which
features of the one-loop answer follow from the
geometry and which rely on supersymmetric constraints.

The relation to flat-space scattering also raises
the question of whether additional symmetries of
$\mathcal N=4$ SYM amplitudes survive, in a generalized
form, at finite AdS radius. Planar flat-space amplitudes
possess dual conformal and Yangian symmetries, closely
related to their integrable structure \cite{Drummond:2010km,Henn:2011xk}.
Related Yangian structures have also been identified
for certain Witten diagrams \cite{Rigatos:2022eos}.
It would be interesting to investigate whether a
deformed or twisted realization of these symmetries
acts on the full boundary correlator, with the
effects of the AdS curvature and boundary conditions
encoded in the corresponding Ward identities.
At strong coupling, planar flat-space amplitudes
admit a description in terms of minimal surfaces
in AdS, where integrability plays a central role \cite{Alday:2007hr,Alday:2010vh}.
Although this holographic construction is distinct
from gauge theory on rigid $\AdS_4$, understanding
whether any of these structures have a counterpart
in the present setting is an intriguing open problem.

We close by returning to the question posed at the beginning. The
scattering amplitudes of $\mathcal N=4$ SYM have been computed to
remarkably high orders and have revealed a wealth of hidden structure,
yet they are infrared divergent, they are defined only order by order
in perturbation theory, and their most striking properties are confined
to the planar limit. The boundary correlator studied here encodes the
amplitude in its behavior at large Mellin variables, but it is a
well-defined function of the coupling for any gauge group. It obeys
crossing symmetry and unitarity, it is constrained by superconformal
symmetry and by exact results from localization, and it can be
approached from weak coupling, from its S-dual description and, at
large $N$, from supergravity. It is natural to hope that this makes it
one of the few observables of an interacting four-dimensional gauge
theory that can be determined at finite coupling. The one-loop
analysis presented here is a first step in that direction.

\bigskip

\section*{Acknowledgments}

We are grateful to Cl\'ement Virally for helpful discussions
on his reduced Mellin-space construction
of the superconformal Ward identities. We also thank Lorenzo Di Pietro for
stimulating conversations,
and Matteo Sacchi for insightful discussions on related topics. We acknowledge the use of AI agents for assistance with some technical calculations and some editing of the draft. The work of LR is supported in part by NSF grant PHY-2513893 and by
the Simons Foundation grant 681267 (Simons Investigator Award).

%%%%%%%%%%%%%%%%%%%%%%%%%%%%%%%%%%%%%%%%%%%%%%%%%%%%%%%%%%%%%%%%
\appendix
%%%%%%%%%%%%%%%%%%%%%%%%%%%%%%%%%%%%%%%%%%%%%%%%%%%%%%%%%%%%%%%%

\section{Three-dimensional conformal blocks}
\label{app:blocks}
Generic three-dimensional scalar blocks do not admit
the finite holomorphic factorization familiar in even
dimensions. It is therefore useful to introduce a
systematic lightcone expansion at small $U$ with fixed
$W\equiv1-V$:
\begin{equation}
  G_{\Delta,\ell}(U,V)
  =\sum_{q=0}^{\infty}
  U^{\frac{\Delta-\ell}{2}+q}
  g^{(q)}_{\Delta,\ell}(W).
\end{equation}
We normalize the blocks such that, in the radial OPE limit,
\begin{equation}
G_{\Delta,\ell}(U,V)
\sim
\frac{U^{\Delta/2}}{4^\Delta}
P_\ell\left(
\frac{1+U-V}{2\sqrt U}
\right),
\end{equation}
with the argument of the Legendre polynomial held fixed.
In this convention, the leading lightcone block is
\begin{equation}
  g^{(0)}_{\Delta,\ell}(W)
  =\frac{(\frac12)_\ell}{4^\Delta\ell!}
  W^\ell\,{}_2F_1\!\left(\frac{\Delta+\ell}{2},\frac{\Delta+\ell}{2};\Delta+\ell;W\right).
\end{equation}
The first subleading lightcone correction can be written as
\begin{align}
  g^{(1)}_{\Delta,\ell}(W)=\frac{(\frac12)_\ell}{4^\Delta\ell!}
  \bigg[&a_{10}(\Delta,\ell)
  W^\ell\,{}_2F_1\!\left(\frac{\Delta+\ell}{2}+1,\frac{\Delta+\ell}{2}+1;\Delta+\ell+2;W\right)\nonumber\\
  &+a_{12}(\ell)
  W^{\ell-2}\,{}_2F_1\!\left(\frac{\Delta+\ell}{2}-1,\frac{\Delta+\ell}{2}-1;\Delta+\ell-2;W\right)\bigg],
\end{align}
where
\begin{equation}
  a_{10}(\Delta,\ell)=
  \frac{(\Delta-1)_2\left(\frac{\Delta+\ell}{2}\right)^4}
  {(\Delta+\ell-1)_2(\Delta+\ell)_2(\Delta-\frac12)},
\end{equation}
while
\begin{equation}
  a_{12}(\ell)=
  \begin{cases}
  -\dfrac{2\ell(\ell-1)}{2\ell-1},& \ell\ge2,\\[0.8em]
  0,& \ell=0,1.
  \end{cases}
\end{equation}
Higher lightcone orders can be generated recursively from
the three-dimensional quadratic Casimir equation or from
the closed-form coefficients of \cite{Li:2019cwm}.
For the blocks relevant to our calculation, the contribution
at order $U^{2+p}$ receives terms with $n+q=p$.
Thus, after subtracting the contributions of lower-twist
families, the leading lightcone blocks determine the
coefficients $\omega_{p,\ell}$ recursively. This is the
procedure used to match the position-space $\Dbar$
representation to the conformal block decomposition.

\section{Mellin to position space}
\label{app:Mellin-to-position}

In this appendix we collect the Mellin--Barnes dictionary used in
section~\ref{subsec:dbar-representation} to rewrite the tree-level Mellin
amplitude in terms of normalized $\bar D$-functions and to extract its
$\log U$ contribution. 

For generic positive weights $r_i$, a scalar contact Witten diagram can be written as 
\begin{equation}
  D_{r_1r_2r_3r_4}(x_i)
  \equiv
  \int_{\AdS_{d+1}}
  \frac{\dd z_0\,\dd^d z}{z_0^{d+1}}
  \prod_{i=1}^4
  \left(
    \frac{z_0}{z_0^2+(\vec z-\vec x_i)^2}
  \right)^{r_i}.
\end{equation}
After extracting the standard conformal prefactor, the normalized function $\bar D_{r_1r_2r_3r_4}$ is defined by
\begin{equation}
\frac{2 \pi^{-d/2}\prod_{i=1}^4\Gamma(r_i)}
{\Gamma\!\left(\frac12\sum_i r_i-\frac d2\right)}
D_{r_1r_2r_3r_4}(x_i)
\equiv
\frac{
  (x_{14}^2)^{\frac12\sum_i r_i-r_1-r_4}
  (x_{34}^2)^{\frac12\sum_i r_i-r_3-r_4}
}{
  (x_{13}^2)^{\frac12\sum_i r_i-r_4}
  (x_{24}^2)^{r_2}
}
\Dbar_{r_1r_2r_3r_4}(U,V).
\label{eq:appendix-dbar-definition}
\end{equation}
With this normalization, the reduced function $\Dbar$ is independent of
the spacetime dimension $d$. It admits the Mellin--Barnes representation
\begin{align}
  \Dbar_{r_1r_2r_3r_4}(U,V)
  ={}&
  \int\frac{\dd \sigma\,\dd \tau}{(4\pi i)^2}
  U^{\sigma/2}V^{\tau/2}\,
  \mathscr D_{r_i}(\sigma,\tau),
  \label{eq:appendix-dbar-mellin-barnes}\\
  \mathscr D_{r_i}(\sigma,\tau)
  ={}&
  \Gamma\!\left(-\frac{\sigma}{2}\right)
  \Gamma\!\left(-\frac{\tau}{2}\right)
  \Gamma\!\left(r_2+\frac{\sigma+\tau}{2}\right)
  \Gamma\!\left(
  \frac{-r_1-r_2+r_3+r_4-\sigma}{2}
  \right)
  \nonumber\\
  &\times
  \Gamma\!\left(
  \frac{r_1-r_2-r_3+r_4-\tau}{2}
  \right)
  \Gamma\!\left(
  \frac{r_1+r_2+r_3-r_4+\sigma+\tau}{2}
  \right).
  \nonumber
\end{align}
For the non-positive and half-integer weights encountered below, \eqref{eq:appendix-dbar-definition} should not be interpreted as a
literal convergent contact integral. Instead, we define the corresponding
$\bar D$-functions by analytic continuation in the parameters $r_i$ of
the Mellin--Barnes representation
\eqref{eq:appendix-dbar-mellin-barnes}.

Our four-point function is written using the Mellin convention
\begin{equation}
\mathcal G(U,V)
=
\int\frac{\dd s\,\dd t}{(4\pi i)^2}\,
U^{s/2}V^{t/2-1}
\mathcal K(s,t)\,
\mathcal M(s,t),
\end{equation}
where
\begin{equation}
\mathcal K(s,t)
\equiv
\Gamma\!\left(\frac{2-s}{2}\right)^2
\Gamma\!\left(\frac{2-t}{2}\right)^2
\Gamma\!\left(\frac{2-u}{2}\right)^2,
\qquad
s+t+u=4.
\label{eq:appendix-mellin-kernel}
\end{equation}
Consider now a position-space term
\begin{equation}
U^aV^b\,
\Dbar_{r_1r_2r_3r_4}(U,V).
\end{equation}
Using \eqref{eq:appendix-dbar-mellin-barnes}, it takes the form
\begin{equation}
U^aV^b\Dbar_{r_1r_2r_3r_4}(U,V)
=
\int\frac{\dd\sigma\,\dd\tau}{(4\pi i)^2}\,
U^{\sigma/2+a}
V^{\tau/2+b}\,
\mathscr D_{r_i}(\sigma,\tau).
\end{equation}
Matching the powers of the cross-ratios with
$U^{s/2}V^{t/2-1}$ gives
\begin{equation}
\sigma=s-2a,
\qquad
\tau=t-2-2b.
\label{eq:appendix-variable-shift}
\end{equation}
The Mellin amplitude represented by this term is therefore
\begin{equation}
U^aV^b\Dbar_{r_1r_2r_3r_4}(U,V)
\quad\longleftrightarrow\quad
\frac{
\mathscr D_{r_i}(s-2a,t-2-2b)
}{
\mathcal K(s,t)
},
\label{eq:dbar-practical-dictionary}
\end{equation}
with $u=4-s-t$ understood throughout. This formula is particularly useful
because ratios of the Gamma functions in $\mathscr D_{r_i}$ against the
universal external kernel $\mathcal K(s,t)$ frequently collapse to the
rational and Gamma-function structures appearing directly in the
tree-level Mellin amplitude.

Applying this dictionary to the tree-level Mellin amplitude gives a finite
$\bar D$-function representation for each of its three R-symmetry
components. Suppressing the common arguments $(U,V)$ of the
$\bar D$-functions, we find
\begin{align}
[\alpha^0]\,\G_{\rm tree}
={}&
\frac{1}{\pi^3}
\Bigg[
-\mathtt c_s
\Big(
\pi U\,\Dbar_{2,1,1,0}
-\sqrt{\pi}\,U^{1/2}\Dbar_{\frac32,\frac12,1,0}
\Big) +\mathtt c_t
\Big(
\pi \frac{U}{V}\Dbar_{1,1,0,2}
-\sqrt{\pi}\,\frac{U}{V}\Dbar_{\frac12,1,0,\frac32}
\Big)
\Bigg], \nonumber\\[6pt]
[\alpha]\,\G_{\rm tree}
={}&
\frac{1}{\pi^3}
\Bigg\{
\mathtt c_s
\Bigg[
\pi U
\Big(
\Dbar_{2,1,0,1}
+\Dbar_{2,1,1,0}
\Big)
-\sqrt{\pi}\,U^{1/2}
\Big(
\Dbar_{\frac32,\frac12,0,1}
+\Dbar_{\frac32,\frac12,1,0}
\Big)
\Bigg]
\nonumber\\
&\quad
+\mathtt c_t
\Bigg[
\pi \frac{U^2}{V}\Dbar_{2,1,0,1}
-\sqrt{\pi}\,U^2V^{-1/2}\Dbar_{1,\frac32,\frac12,0}
-\pi U V^{-1}\Dbar_{1,1,0,2}
+\sqrt{\pi}\,\frac{U}{V}\Dbar_{\frac12,1,0,\frac32}
\Bigg]
\nonumber\\
&\quad
-\mathtt c_u
\Bigg[
\pi U^2\Dbar_{2,1,1,0}
-\sqrt{\pi}\,U^2\Dbar_{\frac32,1,\frac12,0}
+\pi U\Dbar_{2,0,1,1}
-\sqrt{\pi}\,U\Dbar_{\frac32,0,\frac12,1}
\Bigg]
\Bigg\},
\nonumber\\[6pt]
[\alpha^2]\,\G_{\rm tree}
={}&
\frac{U^2}{\pi^3}
\Bigg[
\mathtt c_u
\Big(
\pi\,\Dbar_{2,1,1,0}
-\sqrt{\pi}\,\Dbar_{\frac32,1,\frac12,0}
\Big) -\mathtt c_t
\Big(
\pi V^{-1}\Dbar_{2,1,0,1}
-\sqrt{\pi}\,V^{-1/2}\Dbar_{1,\frac32,\frac12,0}
\Big)
\Bigg].
\end{align}
The highest R-symmetry component in the last line is the representation used in the main text to extract the logarithmic term required for the
tree-level CFT data.

\paragraph{Extraction of the $\log U$ term.} The coefficient of $\log U$ can
be obtained directly from the Mellin--Barnes representation, without
requiring a closed form for the complete $\bar D$-functions. Following \cite{Dolan:2000ut,Alday:2021ymb}, 
\begin{equation}
\left.
\Dbar_{r_1r_2r_3r_4}(U,V)
\right|_{\log U}
=
\sum_{m,n\geq0}
d_{m,n}(r_i)\,
U^m(1-V)^n,
\label{eq:dbar-logU-expansion}
\end{equation}
where the coefficients obtained from the Mellin--Barnes residues are
\begin{align}
d_{m,n}(r_i)
={}&
-\exp\!\left[
\frac{i\pi}{2}
(r_1+r_2-r_3-r_4)
\right]
\frac{
\Gamma(r_1+m)
\Gamma(r_2+m+n)
}{
\Gamma(m+1)
\Gamma(n+1)
\Gamma(r_1+r_2+2m+n)
}
\nonumber\\
&\times
\frac{
\Gamma\!\left(
\frac{r_1+r_2-r_3+r_4}{2}+m
\right)
\Gamma\!\left(
\frac{r_1+r_2+r_3-r_4}{2}+m+n
\right)
}{
\Gamma\!\left(
\frac{r_1+r_2-r_3-r_4+2}{2}+m
\right)
}.
\label{eq:dbar-logU-coefficients}
\end{align}
As above, this expression is understood by analytic continuation in the
weights whenever one or more of the $r_i$ are non-positive or
half-integer. For the four combinations in
\eqref{eq:Tt-dbar}, the continuation gives real
coefficients in the branch convention used throughout the paper.

\section{Local ambiguities}
\label{app:local-ambiguities}

The simultaneous-pole reconstruction provides strong evidence that the complete one-loop \(\log^2 U\) coefficient is captured without additional reduced single-pole terms. We now examine possible local contact-term ambiguities at
separated boundary points. Finite-derivative contact interactions are
represented by polynomial full Mellin amplitudes \cite{Alday:2021odx}.
In the reduced representation used for the cut, we first consider a
polynomial completion (with flavor indices understood),
\begin{equation}
\widetilde{\M}_{\rm tot}^{\rm 1-loop}
=\widetilde{\M}_{\rm cut}^{\rm 1-loop}+P_{\rm loc}(s,t).
\label{eq:local-reduced-completion}
\end{equation}
We do not assume that every polynomial full Ward solution admits a
polynomial reduced primitive.\footnote{Completeness of the reduced Mellin representation has not been established in general; see section~6 of \cite{Virally:2025nnl}. Our exclusion of polynomial full Mellin contact terms below does not rely on this assumption.}

\paragraph{Visibility under the Ward lift.}
The full amplitude is reconstructed from the reduced one by the
superconformal Ward operators \cite{Chester:2025jxg},
\begin{equation}
\M_{\rm full}(s,t;\alpha)
=\sum_{J=0}^{2}\alpha^J\mathcal D_J\widetilde{\M}(s,t).
\label{eq:full-Ward-lift}
\end{equation}
Let $P_q$ be the nonzero homogeneous part of highest degree $q\geq0$ of
$P_{\rm loc}$, so that
\[
P_{\rm loc}(\lambda s,\lambda t)
=\lambda^q P_q(s,t)+O(\lambda^{q-1}).
\]
Expanding the finite shifts in the Ward operators at large Mellin
variables, the nominally leading cubic terms cancel. Defining
$\mathcal E=s\partial_s+t\partial_t$, the first non-vanishing contribution is
\begin{equation}
\sum_{J=0}^{2}\alpha^J\mathcal D_J
\sim(\widetilde u+\alpha s)^2(3+2\mathcal E),
\qquad \widetilde u=-s-t.
\end{equation}
In particular,
\begin{equation}
\bigl[\mathcal D_2P_{\rm loc}\bigr]_{q+2}
=(3+2q)s^2P_q(s,t),
\label{eq:D2-leading-polynomial}
\end{equation}
which cannot vanish for a nonzero $P_q$. Thus a reduced polynomial of
degree $q$ produces a full Mellin polynomial of degree $q+2$, and the
Ward lift is injective on reduced polynomials, already in the $\alpha^2$
component. The first cases, $q=0,1$, have the derivative counting of
supersymmetric $F^4$ and $D^2F^4$ contact interactions, respectively.

\paragraph{Exclusion from the flat-space limit.}
To exclude contact terms without assuming that all polynomial Ward
solutions have reduced polynomial primitives, consider directly an
arbitrary nonzero polynomial addition to the full Mellin amplitude,
with highest homogeneous part $Q_d(s,t;\alpha)$ of degree $d\geq0$.
Any polynomial decomposes into homogeneous parts in $s,t$, and its
highest-degree part controls the leading large-$L$ behavior.
In the large-radius scaling of section~\ref{sec:flat-limit},
\[
s=\frac{L^2S}{2\beta},\qquad
 t=\frac{L^2T}{2\beta},
\]
its leading contribution under the normalized Penedones transform
\eqref{eq:flat-transform-normalized} is
\begin{equation}
\delta\mathcal A_{4,L}^{\rm poly}
=\frac{8\pi^2}{(2d-1)!!}\,L^{2d}
Q_d(S,T;\alpha)+O(L^{2d-2}),
\label{eq:local-full-contact-growth}
\end{equation}
with $(-1)!!=1$. Here $Q_d$ is evaluated at generic fixed
four-dimensional kinematics. For a reduced polynomial,
\eqref{eq:D2-leading-polynomial} gives $d=q+2$ and
\[
Q_{q+2}(S,T;\alpha)
=(3+2q)K(S,T;\alpha)P_q(S,T),
\qquad K(S,T;\alpha)=\left((\alpha-1)S-T\right)^2,
\]
which is nonzero for $P_q\neq0$ at generic kinematics.

For $d\geq1$, \eqref{eq:local-full-contact-growth} produces a power
divergence in $L$. By contrast, the massless one-loop four-point
amplitude of four-dimensional $\mathcal N=4$ SYM is cut-constructible,
built from scalar boxes, with no independent rational remainder
\cite{Green:1982sw,Bern:1994zx,Britto:2004nc}. Its infrared singularities are
logarithmic, in agreement with our reconstructed cut in
section~\ref{sec:flat-limit}:
\[
\mathcal A_{4,L}^{\rm cut}=O(\log^2 L).
\]
Thus a nonzero $Q_d$ with $d\geq1$ would introduce an uncancelled
power divergence absent from the microscopic amplitude. A constant
full polynomial ($d=0$) would instead give an independent rational
flat-space term, which is likewise absent; such a polynomial is also
forbidden by the full Ward identities together with crossing symmetry,
as found in appendix~D of \cite{Alday:2021odx}. UV finiteness \cite{Mandelstam:1982cb,Brink:1982wv} by itself
would not exclude finite local terms.

Nor can the coefficient of a polynomial addition be made to vanish
by an inverse power of $L$. Since the theory is Weyl-equivalent to
$\mathcal N=4$ SYM on the flat half-space, with fixed conformal
boundary conditions, the normalized boundary correlator is independent
of $L$. Its dimensionless Mellin coefficients therefore cannot acquire
inverse powers of $L$. In particular, independent higher-derivative
curvature-suppressed contact terms would require an additional
dimensionful coupling, absent from the microscopic theory. This does
not exclude curvature couplings fixed by the conformal completion of
the action, which are already part of the theory. Flat-space matching
therefore excludes every nonzero full polynomial contact term.
Combined with the injectivity of the Ward lift, this also requires
$P_{\rm loc}=0$.

We stress the status of this argument. Because $L$ enters only
through the scaling of the Mellin variables, the absence of power
divergences in $L$ is the same statement as a bound on the growth of
the full Mellin amplitude at large $s,t$ with fixed ratio. This bound
is not an independent assumption: it follows from requiring that the
large-radius limit of the correlator reproduce the flat-space
one-loop amplitude, as in section~\ref{sec:flat-limit}. That
requirement is a physical input. Crossing symmetry and the
double-discontinuity alone do not fix local terms.

\paragraph{Independent protected OPE check.}

An independent check at finite AdS radius comes from the
protected $\mathcal B[0]_2^{(4;0)}$ multiplets in the
moment-map OPE. Their squared OPE coefficients can be
computed independently using localization of the
mass-deformed hemisphere partition function and the
associated protected twisted-trace algebra
\cite{Dedushenko:2020vgd}, as we show explicitly in \cite{BoundaryProtectedOPE:inprep}. For $SU(N\geq4)$, these multiplets appear in the
symmetric flavor representations
\begin{equation}
    {\rm Sym}^2({\rm Adj})
=\mathbf 1\oplus{\rm Adj}_S\oplus \mathcal R_-\oplus \mathcal R_+,
\end{equation}
where $\mathcal R_-$ and $\mathcal R_+$ denote the two additional
irreducible representations, characterized by the Casimirs 
\begin{equation}
    C_2(\mathcal R_\pm)=4(N\pm1), \qquad C_A=2N,
\end{equation}
in our
normalization $\kappa=1$. In the following, we denote by $\mathcal G^{(2)}_{\mathcal R}$ and
$\widetilde{\M}^{(2)}_{\mathcal R}$ the
order-$\varepsilon^2$ coefficients of the
correlator and reduced Mellin amplitude,
respectively, in the flavor representation
$\mathcal R$, with the overall factor
$\varepsilon^2$ extracted.

The protected multiplet $\mathcal B[0]_2^{(4;0)}$
has a scalar superprimary of dimension $\Delta=2$
and $SU(2)_H$ spin two. In our normalization, its superconformal block starts
at $V=1$ as 
\begin{equation}
    U(1-6\alpha+6\alpha^2)
=12U\mathcal Y_4(\alpha),
\end{equation}
in terms of the
R-symmetry harmonic introduced in
section~\ref{sec:one-loop-strategy}. The leading $U$ term
is therefore extracted from the $s=2$ pole of
the Mellin measure. Near this pole, the Ward lift gives
\[
\begin{aligned}
\mathcal D_2\widetilde{\M}^{(2)}_{\mathcal R}
=(s-2)\big[
(t-2)\widetilde{\M}^{(2)}_{\mathcal R}(0,t-2)-t\widetilde{\M}^{(2)}_{\mathcal R}(0,t)
\big]+O((s-2)^2).
\end{aligned}
\]
The factor $(s-2)$ cancels one of the double
poles of the Mellin measure, leaving a term
proportional to $U$, without $\log U$.
At $V=1$, the remaining Gamma-function weight
is periodic under $t\to t+2$, allowing the
contour integral to be reduced to the residue
at $t=0$. For the contact polynomials and
the reconstructed cut kernel, this gives
\[
[U\alpha^2]\mathcal G^{(2)}_{\mathcal R}(U,1)
=4\widetilde{\M}^{(2)}_{\mathcal R}(0,0).
\]
Comparing with the coefficient $6$ of
$U\alpha^2$ in the protected superblock,
we obtain
\begin{equation}
a_{\mathcal B,\mathcal R}^{(2)}
=\frac23\widetilde{\M}^{(2)}_{\mathcal R}(0,0),
\label{eq:protected-origin-functional}
\end{equation}
where $a_{\mathcal B,\mathcal R}^{(2)}$ denotes the
order-$\varepsilon^2$ correction to the squared OPE
coefficient of the protected multiplet
$\mathcal B\equiv\mathcal B[0]_2^{(4;0)}$
in the flavor representation $\mathcal R$. For the reconstructed cut, $B(0,0)=1/1620$,
and the flavor projections give
\begin{equation}
\left(
a_{\mathcal B,\mathbf1}^{(2)},
a_{\mathcal B,{\rm Adj}_S}^{(2)},
a_{\mathcal B,\mathcal R_-}^{(2)},
a_{\mathcal B,\mathcal R_+}^{(2)}
\right)_{\rm cut}
=
\left(
\frac{N^2}{27},\frac{N^2}{135},
\frac{6-N}{135},\frac{6+N}{135}
\right),
\label{eq:local-protected-values}
\end{equation}
in agreement with the independent protected-sector
calculation presented in \cite{BoundaryProtectedOPE:inprep}. An additional reduced contact polynomial
$P_{\mathcal R}(s,t)$ would shift these OPE
coefficients by $\frac23 P_{\mathcal R}(0,0)$.
The exact agreement therefore requires
$P_{\mathcal R}(0,0)=0$ in every flavor channel.

To see what this constrains, a crossing-symmetric
constant contact for $SU(N\geq4)$ has the form
\[
P_0=aT_\delta+bT_{\rm tr},
\]
where
\[
\begin{aligned}
T_\delta^{IJKL}
&=\delta^{IJ}\delta^{KL}
 +\delta^{IK}\delta^{JL}
 +\delta^{IL}\delta^{JK},\\
T_{\rm tr}^{IJKL}
&=\operatorname{Tr}_{\rm f}
(t^{(I}t^Jt^Kt^{L)}),
\end{aligned}
\]
with averaged symmetrization and
$\operatorname{Tr}_{\rm f}(t^It^J)=\delta^{IJ}$.
Its projections onto $\mathcal R_-$ and $\mathcal R_+$ are
\[
P_{0,\mathcal R_-}=2a-\frac b3,\qquad
P_{0,\mathcal R_+}=2a+\frac b3.
\]
Requiring both to vanish fixes $a=b=0$.
For $SU(2)$ and $SU(3)$, the single independent
constant structure is likewise fixed by
the protected data.

To summarize, within polynomial Mellin contact ambiguities, the flat-space limit of
the microscopic four-dimensional theory excludes any nonzero local
completion: the Ward lift turns a degree-$q$ reduced polynomial into
a term whose Penedones transform grows as $L^{2q+4}$, whereas the
massless one-loop amplitude has only logarithmic infrared divergences.
The exact protected OPE matching independently eliminates the constant
contact sector directly at finite AdS radius. We therefore conclude,
for the polynomial local completions considered here, that
\begin{equation}
\widetilde{\M}_{\rm tot}^{\rm 1-loop}
=\widetilde{\M}_{\rm cut}^{\rm 1-loop}.
\label{eq:no-local-completion}
\end{equation}

\section{An elliptic curve in the position-space representation}
\label{app:elliptic}

We derive an integral representation for the contribution of $\tau(t)$
to the inverse Mellin transform of the Ward-lifted $st$ block
\eqref{eq:flat-exact-W}. Throughout this appendix, $U,V$ denote the
conformal cross-ratios. We work with
$\widehat{\mathcal W}_{st}=\pi^4\mathcal D_2B$, suppressing the overall
color and coupling factors.

The pole representation \eqref{eq:flat-tau-poles} of $\tau$ can be
converted into an integral using the generating function
\begin{equation}
g(x)\equiv\sum_{n=1}^{\infty}\frac{x^n}{n^2}
\left(\frac{4^n}{\binom{2n}{n}}-1\right)
=2\arcsin^2\!\sqrt{x}-\Li_2(x).
\label{eq:elliptic-g}
\end{equation}
Indeed, integrating each term gives
\begin{equation}
\tau(t)=-\frac{1}{18}\int_0^1\frac{\dd x}{x}\,x^{-t/2}\,g(x),
\qquad {\rm Re}\,t<2,
\label{eq:elliptic-tau-integral}
\end{equation}
with other values obtained by analytic continuation.

Let $\Phi(U,V)\equiv\Dbar_{1111}(U,V)$ denote the conformal scalar box.
In our Mellin convention \eqref{eq:mellin-representation}, the inverse transform of the Gamma-function measure
alone is $U\,\Phi(U,V)$. Inserting \eqref{eq:elliptic-tau-integral} therefore replaces \(V\) by \(V/x\) in this kernel and introduces an additional factor of \(1/x\), due to the power \(V^{t/2-1}\), while multiplication by
$s-2$ acts as $2U\partial_U-2$. The contribution of $(s-2)\tau(t)$ to
the position-space transform of $\widehat{\mathcal W}_{st}$ is
consequently
\begin{align}
\mathcal G^{st}_\tau(U,V)
={}&-\frac{1}{18}\int_0^1\frac{\dd x}{x}\,g(x)\,
(2U\partial_U-2)\left[\frac{U}{x}\,
\Phi\!\left(U,\frac{V}{x}\right)\right]
\nonumber\\
={}&-\frac{U^2}{9}\,\partial_U\int_0^1\frac{\dd x}{x^2}\,g(x)\,
\Phi\!\left(U,\frac{V}{x}\right).
\label{eq:elliptic-G-tau}
\end{align}
The algebraic structure follows directly from the scalar-box kernel. A
useful representation is
\begin{equation}
\Phi(U,V)=\int_0^\infty\dd\xi\,
\frac{\log\left[\dfrac{(\xi+V)(1+\xi)}{U\xi}\right]}
{\xi^2+(1+V-U)\xi+V}\,.
\label{eq:elliptic-box}
\end{equation}
The quadratic denominator has discriminant
\begin{equation}
\lambda(1,U,V)\equiv(1-U-V)^2-4UV.
\label{eq:elliptic-kallen}
\end{equation}
In \eqref{eq:elliptic-G-tau}, the substitution
\begin{equation}
x=\frac{r^2}{1+r^2},\qquad 0<r<\infty,
\end{equation}
gives $\arcsin\sqrt{x}=\arctan r$ and
\begin{equation}
\lambda\!\left(1,U,\frac{V}{x}\right)=\frac{P_4(r)}{r^4},
\qquad
P_4(r)=\lambda(1,U,V)\,r^4+2V(V-1-U)\,r^2+V^2.
\label{eq:elliptic-P4}
\end{equation}
The resulting algebraic kernel thus involves the curve $y^2=P_4(r)$.
Its discriminant is
\begin{equation}
{\rm Discr}\,P_4=4096\,U^2V^6\,\lambda(1,U,V),
\end{equation}
which is nonzero at generic cross-ratios, so the smooth
compactification of this curve has genus one.

There is a direct relation to the curve found in
\cite{Stawinski:2023}. Denoting the monic quartic in eq.~(4.2) of that
reference by $P_4^{\rm St}(\sigma;x,y,z)$, one finds
\begin{equation}
\frac{P_4(i\sigma)}{\lambda(1,U,V)}
=P_4^{\rm St}(\sigma;V,U,1).
\label{eq:elliptic-match}
\end{equation}
Thus the two representations involve the same complex elliptic curve,
after a rescaling of the vertical coordinate. This identifies the
elliptic geometry of the kernel in \eqref{eq:elliptic-G-tau}; it does
not establish that an irreducible elliptic contribution survives in the
complete crossing-symmetric correlator.

\bibliographystyle{JHEP}
\bibliography{references_draft}

@article{Aharony:2010ay,
    author = "Aharony, Ofer and Marolf, Donald and Rangamani, Mukund",
    title = "{Conformal field theories in anti-de Sitter space}",
    eprint = "1011.6144",
    archivePrefix = "arXiv",
    primaryClass = "hep-th",
    reportNumber = "DCPT-10-37, WIS-16-10-NOV-DPPA",
    doi = "10.1007/JHEP02(2011)041",
    journal = "JHEP",
    volume = "02",
    pages = "041",
    year = "2011"
}

@article{Gaiotto:2008sa,
    author = "Gaiotto, Davide and Witten, Edward",
    title = "{Supersymmetric Boundary Conditions in N=4 Super Yang-Mills Theory}",
    eprint = "0804.2902",
    archivePrefix = "arXiv",
    primaryClass = "hep-th",
    doi = "10.1007/s10955-009-9687-3",
    journal = "J. Statist. Phys.",
    volume = "135",
    pages = "789--855",
    year = "2009"
}

@article{Gaiotto:2008ak,
    author = "Gaiotto, Davide and Witten, Edward",
    title = "{S-Duality of Boundary Conditions In N=4 Super Yang-Mills Theory}",
    eprint = "0807.3720",
    archivePrefix = "arXiv",
    primaryClass = "hep-th",
    doi = "10.4310/ATMP.2009.v13.n3.a5",
    journal = "Adv. Theor. Math. Phys.",
    volume = "13",
    number = "3",
    pages = "721--896",
    year = "2009"
}

@article{Dedushenko:2020vgd,
    author = "Dedushenko, Mykola and Gaiotto, Davide",
    title = "{Algebras, traces, and boundary correlators in $ \mathcal{N} $ = 4 SYM}",
    eprint = "2009.11197",
    archivePrefix = "arXiv",
    primaryClass = "hep-th",
    doi = "10.1007/JHEP12(2021)050",
    journal = "JHEP",
    volume = "12",
    pages = "050",
    year = "2021"
}

@article{Chester:2014mea,
    author = "Chester, Shai M. and Lee, Jaehoon and Pufu, Silviu S. and Yacoby, Ran",
    title = "{Exact Correlators of BPS Operators from the 3d Superconformal Bootstrap}",
    eprint = "1412.0334",
    archivePrefix = "arXiv",
    primaryClass = "hep-th",
    reportNumber = "PUPT-2476, MIT-CTP-4614",
    doi = "10.1007/JHEP03(2015)130",
    journal = "JHEP",
    volume = "03",
    pages = "130",
    year = "2015"
}

@article{Dedushenko:2016jxl,
    author = "Dedushenko, Mykola and Pufu, Silviu S. and Yacoby, Ran",
    title = "{A one-dimensional theory for Higgs branch operators}",
    eprint = "1610.00740",
    archivePrefix = "arXiv",
    primaryClass = "hep-th",
    reportNumber = "CALT-TH-2016-029, PUPT-2509",
    doi = "10.1007/JHEP03(2018)138",
    journal = "JHEP",
    volume = "03",
    pages = "138",
    year = "2018"
}

@article{Cordova:2016emh,
    author = "Cordova, Clay and Dumitrescu, Thomas T. and Intriligator, Kenneth",
    title = "{Multiplets of Superconformal Symmetry in Diverse Dimensions}",
    eprint = "1612.00809",
    archivePrefix = "arXiv",
    primaryClass = "hep-th",
    doi = "10.1007/JHEP03(2019)163",
    journal = "JHEP",
    volume = "03",
    pages = "163",
    year = "2019"
}

@article{Liendo:2016ymz,
    author = "Liendo, Pedro and Meneghelli, Carlo",
    title = "{Bootstrap equations for $ \mathcal{N} $ = 4 SYM with defects}",
    eprint = "1608.05126",
    archivePrefix = "arXiv",
    primaryClass = "hep-th",
    doi = "10.1007/JHEP01(2017)122",
    journal = "JHEP",
    volume = "01",
    pages = "122",
    year = "2017"
}

@article{McAvity:1993ue,
    author = "McAvity, D. M. and Osborn, H.",
    title = "{Energy momentum tensor in conformal field theories near a boundary}",
    eprint = "hep-th/9302068",
    archivePrefix = "arXiv",
    reportNumber = "DAMTP-93-01",
    doi = "10.1016/0550-3213(93)90005-A",
    journal = "Nucl. Phys. B",
    volume = "406",
    pages = "655--680",
    year = "1993"
}

@article{Billo:2016cpy,
    author = "Bill{\`o}, Marco and Gon{\c{c}}alves, Vasco and Lauria, Edoardo and Meineri, Marco",
    title = "{Defects in conformal field theory}",
    eprint = "1601.02883",
    archivePrefix = "arXiv",
    primaryClass = "hep-th",
    doi = "10.1007/JHEP04(2016)091",
    journal = "JHEP",
    volume = "04",
    pages = "091",
    year = "2016"
}

@article{Raju:2012zr,
    author = "Raju, Suvrat",
    title = "{New Recursion Relations and a Flat Space Limit for AdS/CFT Correlators}",
    eprint = "1201.6449",
    archivePrefix = "arXiv",
    primaryClass = "hep-th",
    reportNumber = "HRI-ST-1201",
    doi = "10.1103/PhysRevD.85.126009",
    journal = "Phys. Rev. D",
    volume = "85",
    pages = "126009",
    year = "2012"
}

@article{Gomez:2026yno,
    author = "Gomez, Humberto and Lipinski Jusinskas, Renann and Lipstein, Arthur and Lopez-Arcos, Cristhiam",
    title = "{On the amplitude expansion of gluon correlators in $\textrm{AdS}_4$}",
    eprint = "2606.23776",
    archivePrefix = "arXiv",
    primaryClass = "hep-th",
    month = "6",
    year = "2026"
}

@article{Hartman:2006dy,
    author = "Hartman, Thomas and Rastelli, Leonardo",
    title = "{Double-trace deformations, mixed boundary conditions and functional determinants in AdS/CFT}",
    eprint = "hep-th/0602106",
    archivePrefix = "arXiv",
    reportNumber = "PUPT-2192, YITP-SB-06-04",
    doi = "10.1088/1126-6708/2008/01/019",
    journal = "JHEP",
    volume = "01",
    pages = "019",
    year = "2008"
}

@article{Alday:2021odx,
    author = "Alday, Luis F. and Behan, Connor and Ferrero, Pietro and Zhou, Xinan",
    title = "{Gluon Scattering in AdS from CFT}",
    eprint = "2103.15830",
    archivePrefix = "arXiv",
    primaryClass = "hep-th",
    doi = "10.1007/JHEP06(2021)020",
    journal = "JHEP",
    volume = "06",
    pages = "020",
    year = "2021"
}

@article{Freedman:2016yue,
    author = "Freedman, Daniel Z. and Pilch, Krzysztof and Pufu, Silviu S. and Warner, Nicholas P.",
    title = "{Boundary Terms and Three-Point Functions: An AdS/CFT Puzzle Resolved}",
    eprint = "1611.01888",
    archivePrefix = "arXiv",
    primaryClass = "hep-th",
    reportNumber = "MIT-CTP-4851, PUPT-2510",
    doi = "10.1007/JHEP06(2017)053",
    journal = "JHEP",
    volume = "06",
    pages = "053",
    year = "2017"
}

@article{Chang:2019dzt,
    author = "Chang, Chi-Ming and Fluder, Martin and Lin, Ying-Hsuan and Shao, Shu-Heng and Wang, Yifan",
    title = "{3d N=4 Bootstrap and Mirror Symmetry}",
    eprint = "1910.03600",
    archivePrefix = "arXiv",
    primaryClass = "hep-th",
    reportNumber = "IPMU19-0116, CALT-TH 2019-024, PUPT-2596",
    doi = "10.21468/SciPostPhys.10.4.097",
    journal = "SciPost Phys.",
    volume = "10",
    number = "4",
    pages = "097",
    year = "2021"
}

@article{Aharony:2016dwx,
    author = "Aharony, Ofer and Alday, Luis F. and Bissi, Agnese and Perlmutter, Eric",
    title = "{Loops in AdS from Conformal Field Theory}",
    eprint = "1612.03891",
    archivePrefix = "arXiv",
    primaryClass = "hep-th",
    doi = "10.1007/JHEP07(2017)036",
    journal = "JHEP",
    volume = "07",
    pages = "036",
    year = "2017"
}

@article{Meltzer:2019nbs,
    author = "Meltzer, David and Perlmutter, Eric and Sivaramakrishnan, Allic",
    title = "{Unitarity Methods in AdS/CFT}",
    eprint = "1912.09521",
    archivePrefix = "arXiv",
    primaryClass = "hep-th",
    reportNumber = "CALT-TH-2019-052",
    doi = "10.1007/JHEP03(2020)061",
    journal = "JHEP",
    volume = "03",
    pages = "061",
    year = "2020"
}

@article{Alday:2021ajh,
    author = "Alday, Luis F. and Bissi, Agnese and Zhou, Xinan",
    title = "{One-loop gluon amplitudes in AdS}",
    eprint = "2110.09861",
    archivePrefix = "arXiv",
    primaryClass = "hep-th",
    reportNumber = "UUITP-51/21",
    doi = "10.1007/JHEP02(2022)105",
    journal = "JHEP",
    volume = "02",
    pages = "105",
    year = "2022"
}

@article{DHoker:1999mqo,
    author = "D'Hoker, Eric and Freedman, Daniel Z. and Rastelli, Leonardo",
    title = "{AdS / CFT four point functions: How to succeed at z integrals without really trying}",
    eprint = "hep-th/9905049",
    archivePrefix = "arXiv",
    reportNumber = "MIT-CTP-2857, UCLA-99-TEP-16",
    doi = "10.1016/S0550-3213(99)00526-X",
    journal = "Nucl. Phys. B",
    volume = "562",
    pages = "395--411",
    year = "1999"
}

@article{Alday:2021ymb,
    author = "Alday, Luis F. and Chester, Shai M. and Raj, Himanshu",
    title = "{ABJM at strong coupling from M-theory, localization, and Lorentzian inversion}",
    eprint = "2107.10274",
    archivePrefix = "arXiv",
    primaryClass = "hep-th",
    doi = "10.1007/JHEP02(2022)005",
    journal = "JHEP",
    volume = "02",
    pages = "005",
    year = "2022"
}

@article{Li:2019cwm,
    author = "Li, Wenliang",
    title = "{Lightcone expansions of conformal blocks in closed form}",
    eprint = "1912.01168",
    archivePrefix = "arXiv",
    primaryClass = "hep-th",
    doi = "10.1007/JHEP06(2020)105",
    journal = "JHEP",
    volume = "06",
    pages = "105",
    year = "2020"
}

@article{Virally:2025nnl,
    author = "Virally, Cl{\'e}ment",
    title = "{Solving superconformal Ward identities in Mellin space}",
    eprint = "2503.09703",
    archivePrefix = "arXiv",
    primaryClass = "hep-th",
    doi = "10.1007/JHEP06(2025)236",
    journal = "JHEP",
    volume = "06",
    pages = "236",
    year = "2025"
}

@article{Chester:2025jxg,
    author = "Chester, Shai M. and Mouland, Rishi and van Muiden, Jesse and Virally, Cl{\'e}ment",
    title = "{Extremal couplings and gluon scattering in M-theory}",
    eprint = "2512.04057",
    archivePrefix = "arXiv",
    primaryClass = "hep-th",
    doi = "10.1007/JHEP06(2026)090",
    journal = "JHEP",
    volume = "06",
    pages = "090",
    year = "2026"
}

@misc{BoundaryProtectedOPE:inprep,
  author = {Chen, Siqi and Nocchi, Maria and Rastelli, Leonardo and Sacchi, Matteo},
  howpublished = {\textit{In preparation}}
}

@article{Penedones:2010ue,
  author = {Penedones, Joao},
  title = {Writing {CFT} correlation functions as {AdS} scattering amplitudes},
  year = {2010},
  eprint = {1011.1485},
  archivePrefix = {arXiv},
  primaryClass = {hep-th}
}

@article{Bern:2005iz,
  author = {Bern, Zvi and Dixon, Lance J. and Smirnov, Vladimir A.},
  title = {Iteration of Planar Amplitudes in Maximally Supersymmetric {Yang--Mills} Theory at Three Loops and Beyond},
  year = {2005},
  eprint = {hep-th/0505205},
  archivePrefix = {arXiv}
}

@article{Fitzpatrick:2011hu,
    author = "Fitzpatrick, A. Liam and Kaplan, Jared",
    title = "{Analyticity and the Holographic S-Matrix}",
    eprint = "1111.6972",
    archivePrefix = "arXiv",
    primaryClass = "hep-th",
    reportNumber = "SLAC-PUB-14841",
    doi = "10.1007/JHEP10(2012)127",
    journal = "JHEP",
    volume = "10",
    pages = "127",
    year = "2012"
}

@article{Duplancic:2000sk,
    author = "Duplancic, G. and Nizic, B.",
    title = "{Dimensionally regulated one loop box scalar integrals with massless internal lines}",
    eprint = "hep-ph/0006249",
    archivePrefix = "arXiv",
    reportNumber = "IRB-TH-8-00",
    doi = "10.1007/s100520100675",
    journal = "Eur. Phys. J. C",
    volume = "20",
    pages = "357--370",
    year = "2001"
}

@article{Bern:1992em,
    author = "Bern, Zvi and Dixon, Lance J. and Kosower, David A.",
    title = "{Dimensionally regulated one loop integrals}",
    eprint = "hep-ph/9212308",
    archivePrefix = "arXiv",
    reportNumber = "SLAC-PUB-6001, CERN-TH-6756-92, UCLA-92-42",
    doi = "10.1016/0370-2693(93)90400-C",
    journal = "Phys. Lett. B",
    volume = "302",
    pages = "299--308",
    year = "1993",
    note = "[Erratum: Phys.Lett.B 318, 649 (1993)]"
}

@article{Mack:2009mi,
    author = "Mack, Gerhard",
    title = "{D-independent representation of Conformal Field Theories in D dimensions via transformation to auxiliary Dual Resonance Models. Scalar amplitudes}",
    eprint = "0907.2407",
    archivePrefix = "arXiv",
    primaryClass = "hep-th",
    month = "7",
    year = "2009"
}

@article{Fitzpatrick:2011ia,
    author = "Fitzpatrick, A. Liam and Kaplan, Jared and Penedones, Joao and Raju, Suvrat and van Rees, Balt C.",
    title = "{A Natural Language for AdS/CFT Correlators}",
    eprint = "1107.1499",
    archivePrefix = "arXiv",
    primaryClass = "hep-th",
    reportNumber = "SLAC-PUB-14506, HRI-ST-1107",
    doi = "10.1007/JHEP11(2011)095",
    journal = "JHEP",
    volume = "11",
    pages = "095",
    year = "2011"
}

@article{Britto:2004nc,
    author = "Britto, Ruth and Cachazo, Freddy and Feng, Bo",
    title = "{Generalized unitarity and one-loop amplitudes in N=4 super-Yang-Mills}",
    eprint = "hep-th/0412103",
    archivePrefix = "arXiv",
    doi = "10.1016/j.nuclphysb.2005.07.014",
    journal = "Nucl. Phys. B",
    volume = "725",
    pages = "275--305",
    year = "2005"
}

@article{Li:2021snj,
    author = "Li, Yue-Zhou",
    title = "{Notes on flat-space limit of AdS/CFT}",
    eprint = "2106.04606",
    archivePrefix = "arXiv",
    primaryClass = "hep-th",
    doi = "10.1007/JHEP09(2021)027",
    journal = "JHEP",
    volume = "09",
    pages = "027",
    year = "2021"
}

@article{Callan:1989em,
    author = "Callan, Jr., Curtis G. and Wilczek, Frank",
    title = "{INFRARED BEHAVIOR AT NEGATIVE CURVATURE}",
    reportNumber = "IASSNS-HEP-90-4, PUPT-1168",
    doi = "10.1016/0550-3213(90)90451-I",
    journal = "Nucl. Phys. B",
    volume = "340",
    pages = "366--386",
    year = "1990"
}

@article{Aharony:2012jf,
    author = "Aharony, Ofer and Berkooz, Micha and Tong, David and Yankielowicz, Shimon",
    title = "{Confinement in Anti-de Sitter Space}",
    eprint = "1210.5195",
    archivePrefix = "arXiv",
    primaryClass = "hep-th",
    reportNumber = "WIS-14-12-AUG-DPPA",
    doi = "10.1007/JHEP02(2013)076",
    journal = "JHEP",
    volume = "02",
    pages = "076",
    year = "2013"
}

@article{Copetti:2023sya,
    author = "Copetti, Christian and Di Pietro, Lorenzo and Ji, Ziming and Komatsu, Shota",
    title = "{Taming Mass Gaps with Anti{\textendash}de Sitter Space}",
    eprint = "2312.09277",
    archivePrefix = "arXiv",
    primaryClass = "hep-th",
    doi = "10.1103/PhysRevLett.133.081601",
    journal = "Phys. Rev. Lett.",
    volume = "133",
    number = "8",
    pages = "081601",
    year = "2024"
}

@article{Ciccone:2024guw,
    author = "Ciccone, Riccardo and De Cesare, Fabiana and Di Pietro, Lorenzo and Serone, Marco",
    title = "{Exploring confinement in Anti-de Sitter space}",
    eprint = "2407.06268",
    archivePrefix = "arXiv",
    primaryClass = "hep-th",
    doi = "10.1007/JHEP12(2024)218",
    journal = "JHEP",
    volume = "12",
    pages = "218",
    year = "2024",
    note = "[Erratum: JHEP 06, 037 (2025)]"
}

@article{Mandelstam:1982cb,
    author = "Mandelstam, Stanley",
    title = "{Light Cone Superspace and the Ultraviolet Finiteness of the N=4 Model}",
    reportNumber = "UCB-PTH-82-15",
    doi = "10.1016/0550-3213(83)90179-7",
    journal = "Nucl. Phys. B",
    volume = "213",
    pages = "149--168",
    year = "1983"
}

@article{Brink:1982wv,
    author = "Brink, Lars and Lindgren, Olof and Nilsson, Bengt E. W.",
    title = "{The Ultraviolet Finiteness of the N=4 Yang-Mills Theory}",
    reportNumber = "UTTG-1-82",
    doi = "10.1016/0370-2693(83)91210-8",
    journal = "Phys. Lett. B",
    volume = "123",
    pages = "323--328",
    year = "1983"
}

@article{Heuveline:2026knw,
    author = "Heuveline, Simon and Ruzziconi, Romain and Sheta, Ahmed and Strominger, Andrew",
    title = "{Self-Dual Yang-Mills in AdS$_4$}",
    eprint = "2608.27731",
    archivePrefix = "arXiv",
    primaryClass = "hep-th",
    month = "8",
    year = "2026"
}

@article{Liendo:2012hy,
    author = "Liendo, Pedro and Rastelli, Leonardo and van Rees, Balt C.",
    title = "{The Bootstrap Program for Boundary CFT$_d$}",
    eprint = "1210.4258",
    archivePrefix = "arXiv",
    primaryClass = "hep-th",
    reportNumber = "YITP-SB-12-37",
    doi = "10.1007/JHEP07(2013)113",
    journal = "JHEP",
    volume = "07",
    pages = "113",
    year = "2013"
}

@article{Catani:1998bh,
    author = "Catani, Stefano",
    title = "{The Singular behavior of QCD amplitudes at two loop order}",
    eprint = "hep-ph/9802439",
    archivePrefix = "arXiv",
    reportNumber = "CERN-TH-98-42, LPTHE-ORSAY-97-57",
    doi = "10.1016/S0370-2693(98)00332-3",
    journal = "Phys. Lett. B",
    volume = "427",
    pages = "161--171",
    year = "1998"
}

@article{Becher:2009qa,
    author = "Becher, Thomas and Neubert, Matthias",
    title = "{On the Structure of Infrared Singularities of Gauge-Theory Amplitudes}",
    eprint = "0903.1126",
    archivePrefix = "arXiv",
    primaryClass = "hep-ph",
    reportNumber = "FERMILAB-PUB-09-061-T, MZ-TH-09-05",
    doi = "10.1088/1126-6708/2009/06/081",
    journal = "JHEP",
    volume = "06",
    pages = "081",
    year = "2009",
    note = "[Erratum: JHEP 11, 024 (2013)]"
}

@article{DiPietro:2025ozw,
    author = "Di Pietro, Lorenzo and Kousvos, Stefanos R. and Meineri, Marco and Piazza, Alessandro and Serone, Marco and Vichi, Alessandro",
    title = "{A bootstrap study of confinement in AdS}",
    eprint = "2512.00150",
    archivePrefix = "arXiv",
    primaryClass = "hep-th",
    doi = "10.1007/JHEP08(2026)057",
    journal = "JHEP",
    volume = "08",
    pages = "057",
    year = "2026"
}

@article{Armstrong:2020woi,
    author = "Armstrong, Connor and Lipstein, Arthur E. and Mei, Jiajie",
    title = "{Color/kinematics duality in AdS$_{4}$}",
    eprint = "2012.02059",
    archivePrefix = "arXiv",
    primaryClass = "hep-th",
    doi = "10.1007/JHEP02(2021)194",
    journal = "JHEP",
    volume = "02",
    pages = "194",
    year = "2021"
}

@article{Albayrak:2020fyp,
    author = "Albayrak, Soner and Kharel, Savan and Meltzer, David",
    title = "{On duality of color and kinematics in (A)dS momentum space}",
    eprint = "2012.10460",
    archivePrefix = "arXiv",
    primaryClass = "hep-th",
    doi = "10.1007/JHEP03(2021)249",
    journal = "JHEP",
    volume = "03",
    pages = "249",
    year = "2021"
}

@article{Zhou:2021gnu,
    author = "Zhou, Xinan",
    title = "{Double Copy Relation in AdS Space}",
    eprint = "2106.07651",
    archivePrefix = "arXiv",
    primaryClass = "hep-th",
    doi = "10.1103/PhysRevLett.127.141601",
    journal = "Phys. Rev. Lett.",
    volume = "127",
    number = "14",
    pages = "141601",
    year = "2021"
}

@article{Armstrong:2022mfr,
    author = "Armstrong, Connor and Gomez, Humberto and Lipinski Jusinskas, Renann and Lipstein, Arthur and Mei, Jiajie",
    title = "{New recursion relations for tree-level correlators in anti{\textendash}de Sitter spacetime}",
    eprint = "2209.02709",
    archivePrefix = "arXiv",
    primaryClass = "hep-th",
    doi = "10.1103/PhysRevD.106.L121701",
    journal = "Phys. Rev. D",
    volume = "106",
    number = "12",
    pages = "L121701",
    year = "2022"
}

@article{Raju:2012zs,
    author = "Raju, Suvrat",
    title = "{Four Point Functions of the Stress Tensor and Conserved Currents in AdS$_4$/CFT$_3$}",
    eprint = "1201.6452",
    archivePrefix = "arXiv",
    primaryClass = "hep-th",
    reportNumber = "HRI-ST-1202",
    doi = "10.1103/PhysRevD.85.126008",
    journal = "Phys. Rev. D",
    volume = "85",
    pages = "126008",
    year = "2012"
}

@article{Alday:2025bjp,
    author = {Alday, Luis F. and Nocchi, Maria and Sangar{\'e}, Aur{\'e}lie Str{\"o}mholm},
    title = "{Stringy KLT Relations on AdS}",
    eprint = "2504.19973",
    archivePrefix = "arXiv",
    primaryClass = "hep-th",
    doi = "10.1007/JHEP02(2026)124",
    journal = "JHEP",
    volume = "02",
    pages = "124",
    year = "2026"
}

@article{Nocchi:2026yms,
    author = {Nocchi, Maria and Pitombo, Rodrigo Schmidt and Str{\"o}mholm Sangar{\'e}, Aur{\'e}lie and Tao, Yi-Xiao},
    title = "{All-multiplicity monodromy and KLT relations for AdS string integrals}",
    eprint = "2606.09769",
    archivePrefix = "arXiv",
    primaryClass = "hep-th",
    month = "6",
    year = "2026"
}

@article{Bala:2026bdx,
    author = "Bala, Aswini and Jain, Sachin and S., Dhruva K. and Rao, Adithya A.",
    title = "{Super-Grassmannians for $\mathcal{N}=2$ to $4$ SCFT$_3$: From AdS$_4$ Correlators to $\mathcal{N}=4$ SYM scattering Amplitudes}",
    eprint = "2604.07503",
    archivePrefix = "arXiv",
    primaryClass = "hep-th",
    month = "4",
    year = "2026"
}

@article{Alday:2007hr,
    author = "Alday, Luis F. and Maldacena, Juan Martin",
    title = "{Gluon scattering amplitudes at strong coupling}",
    eprint = "0705.0303",
    archivePrefix = "arXiv",
    primaryClass = "hep-th",
    reportNumber = "SPIN-07-16, ITP-UU-07-24",
    doi = "10.1088/1126-6708/2007/06/064",
    journal = "JHEP",
    volume = "06",
    pages = "064",
    year = "2007"
}

@article{Rastelli:2017ymc,
    author = "Rastelli, Leonardo and Zhou, Xinan",
    title = "{Holographic Four-Point Functions in the (2, 0) Theory}",
    eprint = "1712.02788",
    archivePrefix = "arXiv",
    primaryClass = "hep-th",
    doi = "10.1007/JHEP06(2018)087",
    journal = "JHEP",
    volume = "06",
    pages = "087",
    year = "2018"
}

@article{Rastelli:2017udc,
    author = "Rastelli, Leonardo and Zhou, Xinan",
    title = "{How to Succeed at Holographic Correlators Without Really Trying}",
    eprint = "1710.05923",
    archivePrefix = "arXiv",
    primaryClass = "hep-th",
    reportNumber = "YITP-SB-2017-44",
    doi = "10.1007/JHEP04(2018)014",
    journal = "JHEP",
    volume = "04",
    pages = "014",
    year = "2018"
}

@article{Dolan:2000ut,
    author = "Dolan, F. A. and Osborn, H.",
    title = "{Conformal four point functions and the operator product expansion}",
    eprint = "hep-th/0011040",
    archivePrefix = "arXiv",
    reportNumber = "DAMTP-2000-125",
    doi = "10.1016/S0550-3213(01)00013-X",
    journal = "Nucl. Phys. B",
    volume = "599",
    pages = "459--496",
    year = "2001"
}

@article{Huang:2023ppy,
    author = "Huang, Zhongjie and Wang, Bo and Yuan, Ellis Ye and Zhou, Xinan",
    title = "{Simplicity of AdS super Yang-Mills at one loop}",
    eprint = "2309.14413",
    archivePrefix = "arXiv",
    primaryClass = "hep-th",
    doi = "10.1007/JHEP01(2024)190",
    journal = "JHEP",
    volume = "01",
    pages = "190",
    year = "2024"
}

@article{Alday:2019nin,
    author = "Alday, Luis F. and Zhou, Xinan",
    title = "{Simplicity of AdS Supergravity at One Loop}",
    eprint = "1912.02663",
    archivePrefix = "arXiv",
    primaryClass = "hep-th",
    reportNumber = "PUPT-2604",
    doi = "10.1007/JHEP09(2020)008",
    journal = "JHEP",
    volume = "09",
    pages = "008",
    year = "2020"
}

@article{tHooft:1972tcz,
    author = "'t Hooft, Gerard and Veltman, M. J. G.",
    title = "{Regularization and Renormalization of Gauge Fields}",
    doi = "10.1016/0550-3213(72)90279-9",
    journal = "Nucl. Phys. B",
    volume = "44",
    pages = "189--213",
    year = "1972"
}

@article{Alday:2009zm,
    author = "Alday, Luis F. and Henn, Johannes M. and Plefka, Jan and Schuster, Theodor",
    title = "{Scattering into the fifth dimension of N=4 super Yang-Mills}",
    eprint = "0908.0684",
    archivePrefix = "arXiv",
    primaryClass = "hep-th",
    reportNumber = "HU-EP-09-35",
    doi = "10.1007/JHEP01(2010)077",
    journal = "JHEP",
    volume = "01",
    pages = "077",
    year = "2010"
}

@article{Caron-Huot:2017vep,
    author = "Caron-Huot, Simon",
    title = "{Analyticity in Spin in Conformal Theories}",
    eprint = "1703.00278",
    archivePrefix = "arXiv",
    primaryClass = "hep-th",
    doi = "10.1007/JHEP09(2017)078",
    journal = "JHEP",
    volume = "09",
    pages = "078",
    year = "2017"
}

@article{Carmi:2019cub,
    author = "Carmi, Dean and Caron-Huot, Simon",
    title = "{A Conformal Dispersion Relation: Correlations from Absorption}",
    eprint = "1910.12123",
    archivePrefix = "arXiv",
    primaryClass = "hep-th",
    doi = "10.1007/JHEP09(2020)009",
    journal = "JHEP",
    volume = "09",
    pages = "009",
    year = "2020"
}

@article{Cutkosky:1960sp,
    author = "Cutkosky, R. E.",
    title = "{Singularities and discontinuities of Feynman amplitudes}",
    doi = "10.1063/1.1703676",
    journal = "J. Math. Phys.",
    volume = "1",
    pages = "429--433",
    year = "1960"
}

@article{Bern:1994zx,
    author = "Bern, Zvi and Dixon, Lance J. and Dunbar, David C. and Kosower, David A.",
    title = "{One loop n point gauge theory amplitudes, unitarity and collinear limits}",
    eprint = "hep-ph/9403226",
    archivePrefix = "arXiv",
    reportNumber = "SLAC-PUB-6415, SACLAY-SPH-T-94-20, UCLA-TEP-94-4, SWAT-94-17",
    doi = "10.1016/0550-3213(94)90179-1",
    journal = "Nucl. Phys. B",
    volume = "425",
    pages = "217--260",
    year = "1994"
}

@article{Heckelbacher:2022fbx,
    author = "Heckelbacher, Till and Sachs, Ivo and Skvortsov, Evgeny and Vanhove, Pierre",
    title = "{Analytical evaluation of AdS$_{4}$ Witten diagrams as flat space multi-loop Feynman integrals}",
    eprint = "2201.09626",
    archivePrefix = "arXiv",
    primaryClass = "hep-th",
    reportNumber = "IPhT-t22/001, LMU-ASC 03/22",
    doi = "10.1007/JHEP08(2022)052",
    journal = "JHEP",
    volume = "08",
    pages = "052",
    year = "2022"
}

@article{Drummond:2010km,
    author = "Drummond, J. M.",
    title = "{Review of AdS/CFT Integrability, Chapter V.2: Dual Superconformal Symmetry}",
    eprint = "1012.4002",
    archivePrefix = "arXiv",
    primaryClass = "hep-th",
    reportNumber = "CERN-PH-TH-2010-307",
    doi = "10.1007/s11005-011-0519-4",
    journal = "Lett. Math. Phys.",
    volume = "99",
    pages = "481--505",
    year = "2012"
}

@article{Henn:2011xk,
    author = "Henn, Johannes M.",
    title = "{Dual conformal symmetry at loop level: massive regularization}",
    eprint = "1103.1016",
    archivePrefix = "arXiv",
    primaryClass = "hep-th",
    reportNumber = "HU-EP-11-01",
    doi = "10.1088/1751-8113/44/45/454011",
    journal = "J. Phys. A",
    volume = "44",
    pages = "454011",
    year = "2011"
}

@article{Rigatos:2022eos,
    author = "Rigatos, Konstantinos C. and Zhou, Xinan",
    title = "{Yangian Symmetry in Holographic Correlators}",
    eprint = "2206.07924",
    archivePrefix = "arXiv",
    primaryClass = "hep-th",
    doi = "10.1103/PhysRevLett.129.101601",
    journal = "Phys. Rev. Lett.",
    volume = "129",
    number = "10",
    pages = "101601",
    year = "2022"
}

@article{Alday:2010vh,
    author = "Alday, Luis F. and Maldacena, Juan and Sever, Amit and Vieira, Pedro",
    title = "{Y-system for Scattering Amplitudes}",
    eprint = "1002.2459",
    archivePrefix = "arXiv",
    primaryClass = "hep-th",
    doi = "10.1088/1751-8113/43/48/485401",
    journal = "J. Phys. A",
    volume = "43",
    pages = "485401",
    year = "2010"
}

@article{Rastelli:2016nze,
    author = "Rastelli, Leonardo and Zhou, Xinan",
    title = "{Mellin amplitudes for $AdS_5\times S^5$}",
    eprint = "1608.06624",
    archivePrefix = "arXiv",
    primaryClass = "hep-th",
    doi = "10.1103/PhysRevLett.118.091602",
    journal = "Phys. Rev. Lett.",
    volume = "118",
    number = "9",
    pages = "091602",
    year = "2017"
}

@article{Dolan:2004mu,
    author = "Dolan, Francis A. and Gallot, Laurent and Sokatchev, Emery",
    title = "{On four-point functions of 1/2-BPS operators in general dimensions}",
    eprint = "hep-th/0405180",
    archivePrefix = "arXiv",
    reportNumber = "DAMTP-04-45, LAPTH-1047-04",
    doi = "10.1088/1126-6708/2004/09/056",
    journal = "JHEP",
    volume = "09",
    pages = "056",
    year = "2004"
}

@article{Woolley:2026anb,
    author = "Woolley, Mitchell",
    title = "{Reduced superblocks at next-to-next-to-extremality for half-maximally supersymmetric CFTs}",
    eprint = "2604.08673",
    archivePrefix = "arXiv",
    primaryClass = "hep-th",
    reportNumber = "QMUL-PH-26-14",
    month = "4",
    year = "2026"
}

@article{Stawinski:2023,
    author = "Stawinski, Sven F.",
    title = "{An elliptic one-loop amplitude in anti-de-Sitter space}",
    eprint = "2309.15059",
    archivePrefix = "arXiv",
    primaryClass = "hep-th",
    reportNumber = "BONN-TH-2023-10",
    doi = "10.1007/JHEP02(2024)208",
    journal = "JHEP",
    volume = "02",
    pages = "208",
    year = "2024"
}

@article{DHoker:2007a,
    author = "D'Hoker, Eric and Estes, John and Gutperle, Michael",
    title = "{Exact half-BPS Type IIB interface solutions. I. Local solution and supersymmetric Janus}",
    eprint = "0705.0022",
    archivePrefix = "arXiv",
    primaryClass = "hep-th",
    doi = "10.1088/1126-6708/2007/06/021",
    journal = "JHEP",
    volume = "06",
    pages = "021",
    year = "2007"
}

@article{DHoker:2007b,
    author = "D'Hoker, Eric and Estes, John and Gutperle, Michael",
    title = "{Exact half-BPS Type IIB interface solutions. II. Flux solutions and multi-Janus}",
    eprint = "0705.0024",
    archivePrefix = "arXiv",
    primaryClass = "hep-th",
    doi = "10.1088/1126-6708/2007/06/022",
    journal = "JHEP",
    volume = "06",
    pages = "022",
    year = "2007"
}

@article{Aharony:2011,
    author = "Aharony, Ofer and Berdichevsky, Leon and Berkooz, Micha and Shamir, Itamar",
    title = "{Near-horizon solutions for D3-branes ending on 5-branes}",
    eprint = "1106.1870",
    archivePrefix = "arXiv",
    primaryClass = "hep-th",
    doi = "10.1103/PhysRevD.84.126003",
    journal = "Phys. Rev. D",
    volume = "84",
    pages = "126003",
    year = "2011"
}

@article{Beem:2013qxa,
    author = "Beem, Christopher and Rastelli, Leonardo and van Rees, Balt C.",
    title = "{The $\mathcal{N}=4$ Superconformal Bootstrap}",
    eprint = "1304.1803",
    archivePrefix = "arXiv",
    primaryClass = "hep-th",
    doi = "10.1103/PhysRevLett.111.071601",
    journal = "Phys. Rev. Lett.",
    volume = "111",
    pages = "071601",
    year = "2013"
}

@article{Binder:2019jwn,
    author = "Binder, Damon J. and Chester, Shai M. and Pufu, Silviu S. and Wang, Yifan",
    title = "{$\mathcal{N}=4$ Super-Yang-Mills correlators at strong coupling from string theory and localization}",
    eprint = "1902.06263",
    archivePrefix = "arXiv",
    primaryClass = "hep-th",
    doi = "10.1007/JHEP12(2019)119",
    journal = "JHEP",
    volume = "12",
    pages = "119",
    year = "2019"
}

@article{Chester:2021aun,
    author = "Chester, Shai M. and Dempsey, Ross and Pufu, Silviu S.",
    title = "{Bootstrapping $\mathcal{N}=4$ super-Yang-Mills on the conformal manifold}",
    eprint = "2111.07989",
    archivePrefix = "arXiv",
    primaryClass = "hep-th",
    doi = "10.1007/JHEP01(2023)038",
    journal = "JHEP",
    volume = "01",
    pages = "038",
    year = "2023"
}

@article{Chester:2023ehi,
    author = "Chester, Shai M. and Dempsey, Ross and Pufu, Silviu S.",
    title = "{Level repulsion in $\mathcal{N}=4$ super-Yang-Mills via integrability, holography, and the bootstrap}",
    eprint = "2312.12576",
    archivePrefix = "arXiv",
    primaryClass = "hep-th",
    doi = "10.1007/JHEP07(2024)059",
    journal = "JHEP",
    volume = "07",
    pages = "059",
    year = "2024"
}

@article{Caron-Huot:2021usw,
    author = "Caron-Huot, Simon and Coronado, Frank",
    title = "{Ten dimensional symmetry of $\mathcal{N}=4$ SYM correlators}",
    eprint = "2106.03892",
    archivePrefix = "arXiv",
    primaryClass = "hep-th",
    doi = "10.1007/JHEP03(2022)151",
    journal = "JHEP",
    volume = "03",
    pages = "151",
    year = "2022"
}

@article{Henn:2010bk,
    author = "Henn, Johannes M. and Naculich, Stephen G. and Schnitzer, Howard J. and Spradlin, Marcus",
    title = "{Higgs-regularized three-loop four-gluon amplitude in $\mathcal{N}=4$ SYM: exponentiation and Regge limits}",
    eprint = "1001.1358",
    archivePrefix = "arXiv",
    primaryClass = "hep-th",
    doi = "10.1007/JHEP04(2010)038",
    journal = "JHEP",
    volume = "04",
    pages = "038",
    year = "2010"
}

@article{Fitzpatrick:2011jn,
    author = "Fitzpatrick, A. Liam and Kaplan, Jared",
    title = "{Scattering States in AdS/CFT}",
    eprint = "1104.2597",
    archivePrefix = "arXiv",
    primaryClass = "hep-th",
    year = "2011"
}

@article{Hijano:2020szl,
    author = "Hijano, Eliot and Neuenfeld, Dominik",
    title = "{Soft photon theorems from CFT Ward identites in the flat limit of AdS/CFT}",
    eprint = "2005.03667",
    archivePrefix = "arXiv",
    primaryClass = "hep-th",
    doi = "10.1007/JHEP11(2020)009",
    journal = "JHEP",
    volume = "11",
    pages = "009",
    year = "2020"
}

@article{Duary:2022pyv,
    author = "Duary, Sarthak and Hijano, Eliot and Patra, Milan",
    title = "{Towards an IR finite S-matrix in the flat limit of AdS/CFT}",
    eprint = "2211.13711",
    archivePrefix = "arXiv",
    primaryClass = "hep-th",
    year = "2022"
}

@article{Paulos:2016fap,
    author = "Paulos, Miguel F. and Penedones, Joao and Toledo, Jonathan and van Rees, Balt C. and Vieira, Pedro",
    title = "{The S-matrix bootstrap. Part I: QFT in AdS}",
    eprint = "1607.06109",
    archivePrefix = "arXiv",
    primaryClass = "hep-th",
    doi = "10.1007/JHEP11(2017)133",
    journal = "JHEP",
    volume = "11",
    pages = "133",
    year = "2017"
}

@article{Bork:2009ps,
    author = "Bork, L. V. and Kazakov, D. I. and Vartanov, G. S. and Zhiboedov, A. V.",
    title = "{Infrared Safe Observables in $\mathcal{N}=4$ Super Yang-Mills Theory}",
    eprint = "0908.0387",
    archivePrefix = "arXiv",
    primaryClass = "hep-th",
    doi = "10.1016/j.physletb.2009.10.013",
    journal = "Phys. Lett. B",
    volume = "681",
    pages = "296",
    year = "2009"
}

@article{Bork:2009nc,
    author = "Bork, L. V. and Kazakov, D. I. and Vartanov, G. S. and Zhiboedov, A. V.",
    title = "{Construction of Infrared Finite Observables in $\mathcal{N}=4$ Super Yang-Mills Theory}",
    eprint = "0911.1617",
    archivePrefix = "arXiv",
    primaryClass = "hep-th",
    doi = "10.1103/PhysRevD.81.105028",
    journal = "Phys. Rev. D",
    volume = "81",
    pages = "105028",
    year = "2010"
}

@article{Hofman:2008ar,
    author = "Hofman, Diego M. and Maldacena, Juan",
    title = "{Conformal collider physics: Energy and charge correlations}",
    eprint = "0803.1467",
    archivePrefix = "arXiv",
    primaryClass = "hep-th",
    doi = "10.1088/1126-6708/2008/05/012",
    journal = "JHEP",
    volume = "05",
    pages = "012",
    year = "2008"
}

@article{Belitsky:2013xxa,
    author = "Belitsky, A. V. and Hohenegger, S. and Korchemsky, G. P. and Sokatchev, E. and Zhiboedov, A.",
    title = "{From correlation functions to event shapes}",
    eprint = "1309.0769",
    archivePrefix = "arXiv",
    primaryClass = "hep-th",
    doi = "10.1016/j.nuclphysb.2014.04.020",
    journal = "Nucl. Phys. B",
    volume = "884",
    pages = "305",
    year = "2014"
}

@article{Belitsky:2013bja,
    author = "Belitsky, A. V. and Hohenegger, S. and Korchemsky, G. P. and Sokatchev, E. and Zhiboedov, A.",
    title = "{Event shapes in $\mathcal{N}=4$ super-Yang-Mills theory}",
    eprint = "1309.1424",
    archivePrefix = "arXiv",
    primaryClass = "hep-th",
    journal = "Nucl. Phys. B",
    volume = "884",
    pages = "206",
    year = "2014"
}

@article{DiPietro:2019hqe,
    author = "Di Pietro, Lorenzo and Gaiotto, Davide and Lauria, Edoardo and Wu, Jingxiang",
    title = "{3d Abelian Gauge Theories at the Boundary}",
    eprint = "1902.09567",
    archivePrefix = "arXiv",
    primaryClass = "hep-th",
    doi = "10.1007/JHEP05(2019)091",
    journal = "JHEP",
    volume = "05",
    pages = "091",
    year = "2019"
}

@article{Behan:2020nsf,
    author = "Behan, Connor and Di Pietro, Lorenzo and Lauria, Edoardo and van Rees, Balt C.",
    title = "{Bootstrapping boundary-localized interactions}",
    eprint = "2009.03336",
    archivePrefix = "arXiv",
    primaryClass = "hep-th",
    doi = "10.1007/JHEP12(2020)182",
    journal = "JHEP",
    volume = "12",
    pages = "182",
    year = "2020"
}

@article{Alday:2017xua,
    author = "Alday, Luis F. and Bissi, Agnese",
    title = "{Loop Corrections to Supergravity on $AdS_5 \times S^5$}",
    eprint = "1706.02388",
    archivePrefix = "arXiv",
    primaryClass = "hep-th",
    doi = "10.1103/PhysRevLett.119.171601",
    journal = "Phys. Rev. Lett.",
    volume = "119",
    pages = "171601",
    year = "2017"
}

@article{Aprile:2017bgs,
    author = "Aprile, F. and Drummond, J. M. and Heslop, P. and Paul, H.",
    title = "{Quantum Gravity from Conformal Field Theory}",
    eprint = "1706.02822",
    archivePrefix = "arXiv",
    primaryClass = "hep-th",
    doi = "10.1007/JHEP01(2018)035",
    journal = "JHEP",
    volume = "01",
    pages = "035",
    year = "2018"
}

@article{Alday:2017vkk,
    author = "Alday, Luis F. and Caron-Huot, Simon",
    title = "{Gravitational S-matrix from CFT dispersion relations}",
    eprint = "1711.02031",
    archivePrefix = "arXiv",
    primaryClass = "hep-th",
    doi = "10.1007/JHEP12(2018)017",
    journal = "JHEP",
    volume = "12",
    pages = "017",
    year = "2018"
}

@article{Beem:2016cbd,
    author = "Beem, Christopher and Peelaers, Wolfger and Rastelli, Leonardo",
    title = "{Deformation quantization and superconformal symmetry in three dimensions}",
    eprint = "1601.05378",
    archivePrefix = "arXiv",
    primaryClass = "hep-th",
    doi = "10.1007/s00220-017-2845-6",
    journal = "Commun. Math. Phys.",
    volume = "354",
    pages = "345",
    year = "2017"
}

@article{Klebanov:1999tb,
    author = "Klebanov, Igor R. and Witten, Edward",
    title = "{AdS/CFT correspondence and symmetry breaking}",
    eprint = "hep-th/9905104",
    archivePrefix = "arXiv",
    primaryClass = "hep-th",
    doi = "10.1016/S0550-3213(99)00387-9",
    journal = "Nucl. Phys. B",
    volume = "556",
    pages = "89",
    year = "1999"
}

@article{Green:1982sw,
    author = "Green, Michael B. and Schwarz, John H. and Brink, Lars",
    title = "{$N=4$ Yang-Mills and $N=8$ Supergravity as Limits of String Theories}",
    journal = "Nucl. Phys. B",
    volume = "198",
    pages = "474",
    year = "1982"
}

@article{Gardi:2009qi,
    author = "Gardi, Einan and Magnea, Lorenzo",
    title = "{Factorization constraints for soft anomalous dimensions in QCD scattering amplitudes}",
    eprint = "0901.1091",
    archivePrefix = "arXiv",
    primaryClass = "hep-ph",
    doi = "10.1088/1126-6708/2009/03/079",
    journal = "JHEP",
    volume = "03",
    pages = "079",
    year = "2009"
}

@article{Korchemsky:1987wg,
    author = "Korchemsky, G. P. and Radyushkin, A. V.",
    title = "{Renormalization of the Wilson Loops Beyond the Leading Order}",
    journal = "Nucl. Phys. B",
    volume = "283",
    pages = "342",
    year = "1987"
}

@article{Beisert:2006ez,
    author = "Beisert, Niklas and Eden, Burkhard and Staudacher, Matthias",
    title = "{Transcendentality and Crossing}",
    eprint = "hep-th/0610251",
    archivePrefix = "arXiv",
    primaryClass = "hep-th",
    doi = "10.1088/1742-5468/2007/01/P01021",
    journal = "J. Stat. Mech.",
    volume = "0701",
    pages = "P01021",
    year = "2007"
}

@article{Henn:2019swt,
    author = "Henn, Johannes M. and Korchemsky, Gregory P. and Mistlberger, Bernhard",
    title = "{The full four-loop cusp anomalous dimension in $\mathcal{N}=4$ super Yang-Mills and QCD}",
    eprint = "1911.10174",
    archivePrefix = "arXiv",
    primaryClass = "hep-th",
    doi = "10.1007/JHEP04(2020)018",
    journal = "JHEP",
    volume = "04",
    pages = "018",
    year = "2020"
}

@article{Belitsky:2022,
    author = "Belitsky, A. V. and Bork, L. V. and Pikelner, A. F. and Smirnov, V. A.",
    title = "{Exact off-shell Sudakov form factor in $\mathcal{N}=4$ SYM}",
    eprint = "2209.09263",
    archivePrefix = "arXiv",
    primaryClass = "hep-th",
    year = "2022"
}

@article{Alday:2026,
    author = {Alday, Luis F. and Armanini, Elisabetta and Belitsky, Andrei V. and H{\"a}ring, Kelian and Zhiboedov, Alexander},
    title = "{Walking Sudakov: From Cusp to Octagon}",
    eprint = "2605.16034",
    archivePrefix = "arXiv",
    primaryClass = "hep-th",
    year = "2026"
}

\end{document}